\documentclass[fleqn,usenatbib]{mnras}
\usepackage{ae,aecompl}
\usepackage[T1]{fontenc}
\usepackage[latin9]{inputenc}
\usepackage{color}
\usepackage{array}
\usepackage{refstyle}
\usepackage{booktabs}
\usepackage{multirow}
\usepackage{amsmath}
\usepackage{amssymb}
\usepackage{graphicx}
\usepackage[authoryear]{natbib}

\makeatletter

\AtBeginDocument{\providecommand\secref[1]{\ref{sec:#1}}}
\AtBeginDocument{\providecommand\figref[1]{\ref{fig:#1}}}
\AtBeginDocument{\providecommand\Eqref[1]{\ref{Eq:#1}}}
\AtBeginDocument{\providecommand\Figref[1]{\ref{Fig:#1}}}
\AtBeginDocument{\providecommand\subsecref[1]{\ref{subsec:#1}}}
\AtBeginDocument{\providecommand\eqref[1]{\ref{eq:#1}}}
\AtBeginDocument{\providecommand\tabref[1]{\ref{tab:#1}}}
\AtBeginDocument{\providecommand\Tabref[1]{\ref{Tab:#1}}}
\providecommand{\tabularnewline}{\\}
\RS@ifundefined{subsecref}
  {\newref{subsec}{name = \RSsectxt}}
  {}
\RS@ifundefined{thmref}
  {\def\RSthmtxt{theorem~}\newref{thm}{name = \RSthmtxt}}
  {}
\RS@ifundefined{lemref}
  {\def\RSlemtxt{lemma~}\newref{lem}{name = \RSlemtxt}}
  {}

\usepackage{aecompl}
\usepackage{cuted}
\usepackage{caption}

\DeclareMathOperator\atanh{atanh}
\DeclareMathOperator\Li{Li}

\title[Bayesian modelling of Emission from Blazars]{A Bayesian Approach to Modelling Multi-Messenger Emission from Blazars using Lepto-Hadronic Kinetic Equations}

\author[B. Jim\'enez-Fern\'andez et al.] {\
    Bruno Jim\'enez-Fern\'andez,$^{1}$\thanks{E-mail: B.Jimenez.Fernandez@bath.ac.uk}
H. J. van Eerten.$^{1}$
\\
$^{1}$Department of Physics, University of Bath, Claverton Down, Bath BA2 7AY, United Kingdom
}

\date{Accepted XXX\@. Received YYY; in original form ZZZ}

\pubyear{2019}

\makeatother

\begin{document}
\label{firstpage} \pagerange{\pageref{firstpage}--\pageref{lastpage}}
\maketitle
\begin{abstract}
Blazar TXS 0506+056 is the main candidate for a coincident neutrino
and gamma-ray flare event. In this paper, we present a detailed kinetic
lepto-hadronic emission model capable of producing a photon and neutrino
spectrum given a set of parameters. Our model includes a range of
large-scale geometries and both dynamical and steady-state injection
models for electrons and protons. We link this model with a Markov
Chain Monte Carlo sampler to obtain a powerful statistical tool that
allows us to both fit the Spectral Energy Distribution and study the
probability density functions and correlations of the parameters.
Assuming a fiducial neutrino flux, we demonstrate how multi-messenger
observations can be modelled jointly in a Bayesian framework. We find
the best parameters for each of the variants of the model tested and
report on their cross-correlations. Additionally, we confirm that
reproducing the neutrino flux of TXS 0506+056 requires an extreme
proton to electron ratio either in the local acceleration process
or from external injection.
\end{abstract}
\begin{keywords} BL Lacertae objects: individual: TXS 0506+056, radiation
mechanisms: non-thermal, methods: statistical \end{keywords}

\section{Introduction}

The modelling of blazar emission and blazar flares is a topic with
a long history (see for example \citealp{boettcher_progress_2019}),
yet a number of questions still remain open. Firstly, the nature of
the hot emitting plasma remains debated. Early models were mostly
leptonic \citep{jones_calculated_1968}, in the sense that only electrons
and positrons were considered to play a role in the emission. More
recently, lepto-hadronic and hadronic models have started to gain
a wide acceptance (see e.g. \citep{mastichiadis_self-consistent_1995,mastichiadis_spectral_2005,cerruti_hadronic_2015,petropoulou_betheheitler_2015,gao_direct_2017}).
Secondly, the number of relevant emitting plasma zones remains unknown.
Currently, the main models are based on the assumptions of a single
emission zone (a spherical blob) \citep{bottcher_leptonic_2013,Cerruti_2018,keivani_multimessenger_2018}
or two main emission zones (jet and sheath models) \citep{macdonald_through_2015,potter_modelling_2018},
although some multi zone emission models exist \citep{graff_multizone_2008,bottcher_timing_2010}.
Thirdly, the plasma heating mechanisms remain poorly understood. Proposed
mechanisms include Magnetic Reconnection \citep{morris_feasibility_2019,petropoulou_relativistic_2019}
and Internal Shocks \citep{spada_internal_2001,kino_hydrodynamic_2004,bottcher_timing_2010,peer_dynamical_2017}.

Recently, a mayor milestone was achieved in addressing the first problem
with the detection of directionally coincident neutrinos and gamma-rays
from blazar TXS 0506+056 \citep{eaat1378}, which have been modelled
successfully by various groups worldwide (e.g. \citet{Cerruti_2018,keivani_multimessenger_2018,murase_blazar_2018,gao_modelling_2019,xue_two-zone_2019,liu_hadronuclear_2018}),
as only models with a hadronic component are able to explain the presence
of neutrinos. However, this in turn has raised an issue with the amount
of energy required for heating and keeping the particles hot, as hadronic
models require much more energy than leptonic ones. The other questions
also remain open. In fact, we can see that at least two of the studies
explaining the emission of TXS 0506+056 use a one zone model \citep{Cerruti_2018,keivani_multimessenger_2018}
whereas others use a two zone model \citep{murase_blazar_2018,xue_two-zone_2019,liu_hadronuclear_2018},
or a nested leaky box model \citep{gao_modelling_2019}. The nature
of the plasma heating mechanism is addressed by none of the studies.
Instead, either they inject an already heated plasma or they set a
steady population of particles, whose parameters are given based on
observational and energetic constrains or determined in a multi-dimensional
parameter search.

The methods that these studies use allow them to obtain the best fit
parameters according to a grid search or a fit by eye. But they do
not allow for a full statistical analysis on the parameters, their
distributions and their correlation. This work is an attempt at achieving
this objective.

This work also builds upon a lepto-hadronic model for all the emission
processes inside the plasma with a single zone of emission. We choose
that the main particles (electrons and protons) can be in a steady
state by default or can be injected. In addition, we also investigate
the effects of the geometry of the emitting plasma on the modelling
by simulating a sphere and a disk.

On top of the emission model we add a Markov Chain Monte Carlo (MCMC)
method that allows us to quickly fit the Spectral Energy Distribution
(SED) of TXS 0506+056. This method of obtaining a fit for the SED
gives us important statistical information about the parameters and
their correlations, encoded in the posterior distribution functions
for each of the parameters and correlation plots for each pair of
parameters.

This study starts with an overview of the model, the volumes we simulate,
the kinetic equations we solve and the approach we take for the initial
particles in \secref{Method}. We continue with a brief presentation
of how we fit the models in \secref{Model-Fitting}. Subsequently,
we report our results reproducing TXS 0506+056 in \secref{Reproducing-TXS0506+056}.
And finally we draw conclusions in \secref{Conclusions}. As cosmological
parameters, we have chosen those of \citet{planck_collaboration_planck_2019}:
$H_{0}=67.66$, $\Omega_{m}=0.3111$ and $\Omega_{\Lambda}=0.6889$.

\section{Method}{\label{sec:Method}}

Our approach to simulating SEDs follows four steps. First, we define
the geometry of the emitting volume, which we will employ to set the
escape timescale and the relation between the photon and neutrino
populations and the flux an observer would receive. Second, we set
the type of population behaviour for protons and electrons, choosing
between leaving them with fixed populations, or injecting particles
from outside. Third, we simulate the evolution of the plasma in the
volume until all the particles in it reach a steady state.

As a final step, our model takes the population number density of
photons $\left(n_{\gamma}\right)$ and neutrinos $\left(n_{\nu}\right)$
and transforms it to the flux $\left(\nu F_{\nu}\right)$ that an
observer would receive.

For photons, this transformation is given by (adding a prime to variables
in the plasma frame) \citealt[equation 2.3]{boettcher_spectral_2019}:

\begin{equation}
\nu F_{\nu}=\varepsilon'^{2}n_{\gamma}'\left(\varepsilon'\right)m_{e}c^{2}\frac{\delta^{4}}{4\pi\left(1+z\right)D_{L}^{2}}\frac{V'}{\tau_{esc}'},
\end{equation}
where $\varepsilon'$ is the energy of a photon normalized to the
rest mass of an electron, $\delta$ is the Doppler boosting factor,
$D_{L}$ is the luminosity distance, $V'$ is the volume that we are
simulating, $\tau_{esc}'$ is the escape timescale for photons and
$c$ is the speed of light.

The neutrino transformation from the obtained population to the theoretically
observed flux is similar to that of photons, but we need to take into
account the flavour of the produced neutrinos and neutrino oscillations.
In order to do this, we assume that the neutrinos have had time to
be distributed equally among the three flavours.

Note that to simplify subsequent formulas, and because we usually
work in the plasma frame, we will drop the primes.

\subsection{Simulated volumes}

\begin{figure*}
\includegraphics[width=0.45\textwidth]{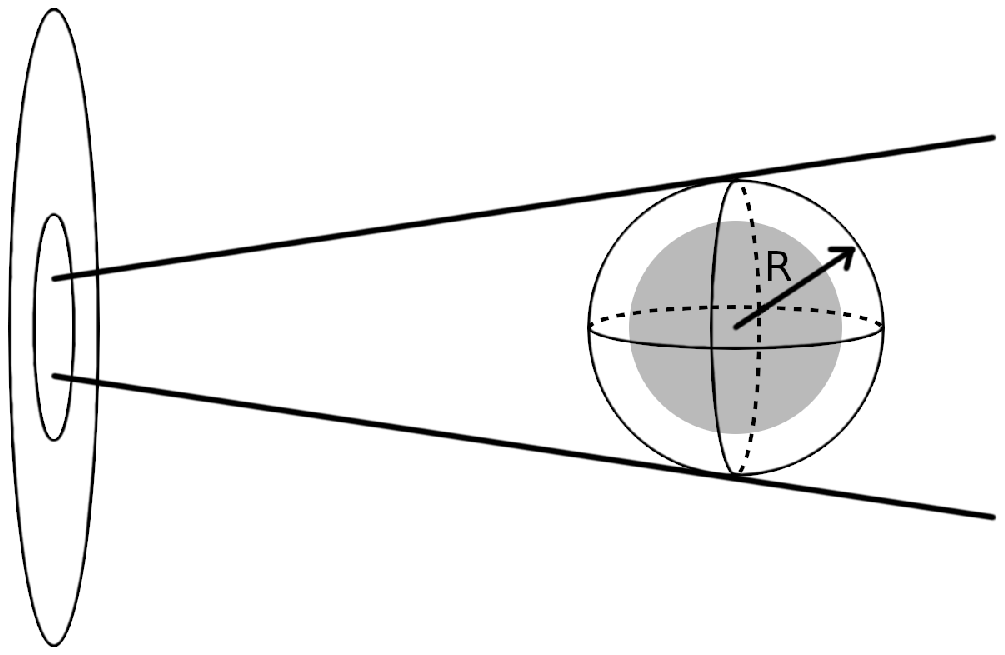}\hfill{}\includegraphics[width=0.45\textwidth]{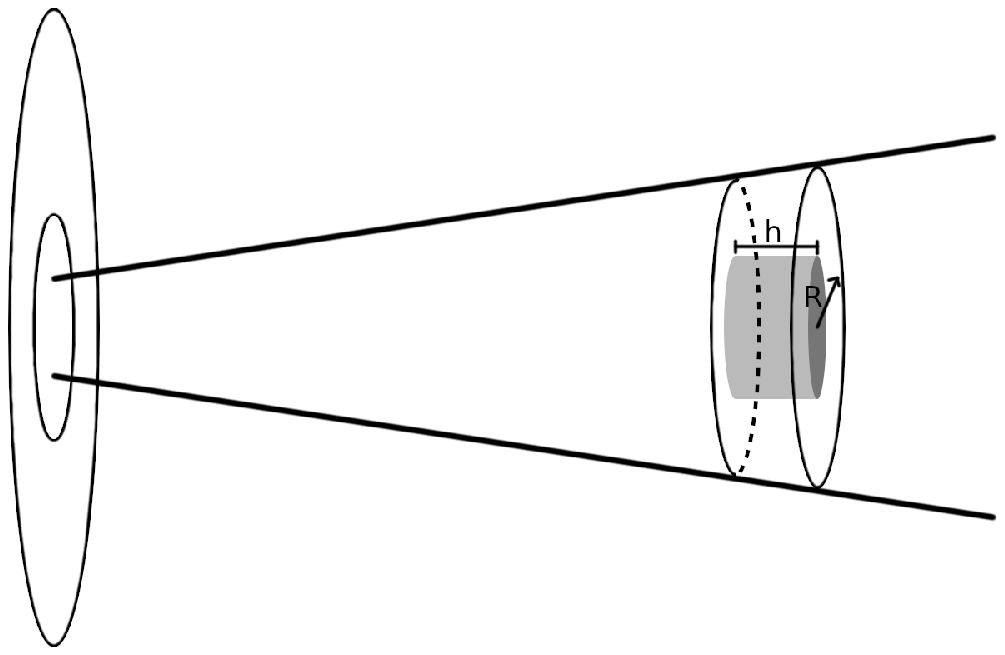}

\caption{\label{fig:Sketch-of-volumes}Sketch of both simulated volumes. Not
to scale. $R$ represents the radius of a sphere or a disk and $h$
the height of the disk. All variables are defined in the plasma frame.}

\end{figure*}

We consider two emission region geometries in our modelling. A sphere
and a thin disk. The former is a first approximation to an unknown
shape of the emitting plasma whereas the latter tries to take into
account a possible shape derived from some heating mechanisms, such
as magnetic reconnection and shock acceleration. In \figref{Sketch-of-volumes}
we show a simple sketch of both volumes.

It should be noted that one of the basic assumptions of blazar modelling
in general is that the emission zone is homogeneous and isotropic
as all the equations for particle interactions assume isotropy at
the interaction point. Although the effect is smaller for a sphere
than a disk model, any choice of geometry will introduce an inherent
anisotropy, the study of which lies beyond the scope of this work.

\subsubsection{Sphere}

The Spherical model (\figref{Sketch-of-volumes} left) is defined
by one characteristic length: the radius of the blob $R$. The free
escape timescale, that is, the time a particle that does not interact
in the volume takes to escape, $\left(\tau_{free,esc}\right)$ is
then set as the average length that a particle would have to traverse
to escape from a sphere divided by $c$:
\begin{equation}
\tau_{free,esc}=\frac{1}{c}\frac{\int_{V}\frac{\int_{\Omega}L\left(r,\Omega\right)d\Omega}{4\pi}dV}{\frac{4\pi}{3}R^{3}}=\frac{3}{4}\frac{R}{c}.\label{eq:t_esc_sphere}
\end{equation}

\subsubsection{Thin Disk}

The Thin Disk model (\figref{Sketch-of-volumes} right) is defined
by two characteristic lengths: the radius of the disk $\left(R\right)$
and the height of the disk $\left(h\right)$. The first order approximation
to the escape timescale is:
\begin{equation}
\tau_{free,esc}=\frac{1}{c}\frac{\int_{V}\frac{\int_{\Omega}L\left(r,\Omega\right)d\Omega}{4\pi}dV}{\pi R^{2}h}\approx\frac{\pi}{4}\frac{h}{c}.\label{eq:t_esc_disk}
\end{equation}

\subsection{Kinetic equation approach to emission }{\label{subsec:Kinetic-equation-approach}}

The model we employ solves the coupled kinetic equations for $12$
different species of particles: protons, neutrons, electrons, charged
and neutral pions, muons and antimuons, and electron and muon neutrinos
and antineutrinos. The general form of a kinetic equation, omitting
the linear term follows:
\begin{equation}
\frac{\partial n}{\partial t}=\mathcal{Q}_{ext}+\mathcal{Q}_{int}+\mathcal{L}-\frac{n}{\tau_{esc}}-\frac{n}{\tau_{dec}}+S\frac{\partial}{\partial\gamma}\left[\gamma^{2}n\right].\label{eq:kinetic_equation}
\end{equation}

As shown in \Eqref{kinetic_equation}, the variation of the population
of a species of particle is due to: external injection $\mathcal{Q}_{ext}$;
internal injection $\mathcal{Q}_{int}$, which represents particles
produced by processes internal to the simulated volume; loss of particles
$\mathcal{L}$ due to interactions with other particles; an escape
term where $\tau_{esc}$ represents the timescale for escaping the
simulated volume, which we take as a multiple of the free timescale
$\left(\tau_{esc}=\sigma\tau_{free,esc}\right)$; a decay term where
$\tau_{dec}$ represents the decay timescale; and a continuous loss
of energy where the energy lost is proportional to $\gamma^{2}$,
such as synchrotron losses for charged particles and the Inverse Compton
(IC) process for electrons. Lorentz factor $\gamma$ corresponds to
the energy of a particle normalized to its rest mass energy $\left(\gamma=E/mc^{2}\right)$.

Usually, the expressions for photon injection and losses are given
in terms of emission and absorption coefficients $\left(j_{\nu}\text{ and }\alpha_{\nu}\right)$.
These are related according to:
\begin{align}
\mathcal{Q}_{int} & \equiv\left|\frac{\partial n_{\gamma}\left(\varepsilon\right)}{\partial t}\right|_{inj}=\frac{4\pi}{\varepsilon m_{e}c^{2}}\frac{d\nu}{d\varepsilon}j_{\nu}\left(\nu\right)=\frac{2}{\hbar\varepsilon}j_{\nu}\left(\nu\right),\\
\mathcal{L} & \equiv\left|\frac{\partial n_{\gamma}\left(\varepsilon\right)}{\partial t}\right|_{abs}=c\alpha_{\nu}\left(\nu\right)n_{\gamma}\left(\varepsilon\right),
\end{align}
where $\hbar$ is the reduced Plank constant. For charged particle
losses, an energy loss term of arbitrary strenght $\left(S\right)$
specializes according to
\begin{align}
S\frac{\partial}{\partial\gamma}\left[\gamma^{2}n\right] & =\frac{\partial}{\partial\gamma}\left[\frac{\partial\gamma}{\partial t}n\right],
\end{align}

\bigskip{}

In our model, all the charged particles create synchrotron emission,
have synchrotron losses and cause synchrotron absorption. For this,
we model the emission as:
\begin{equation}
j_{\nu}=\frac{1}{4\pi}\int_{1}^{\infty}n\left(\gamma\right)P_{\nu}\left(\gamma\right)d\gamma,
\end{equation}
where $P_{\nu}\left(\gamma\right)$ is the energy emitted by a particle
with energy $\gamma$ at frequency $\nu$ and has the form \citep{crusius_synchrotron_1986}:
\begin{equation}
P_{\nu}\left(\gamma\right)=\frac{\sqrt{3}}{4\pi}\frac{q^{3}B}{mc^{2}}xCS\left(x\right),
\end{equation}
where $q$ and $m$ are the charge and the mass of the emitting particle,
$B$ is the magnetic field strength, $x$ is the frequency normalized
to a critical frequency $x=\nu/\nu_{c}$ with
\begin{equation}
\nu_{c}\equiv\nu_{0}\gamma^{2}\equiv\frac{3}{4\pi}\frac{qB}{mc}\gamma^{2},
\end{equation}
and $CS\left(x\right)$ is a function given in terms of Whittaker
functions:
\begin{equation}
CS\left(x\right)=W_{0,\frac{4}{3}}\left(x\right)W_{0,\frac{1}{3}}\left(x\right)-W_{\frac{1}{2},\frac{5}{6}}\left(x\right)W_{-\frac{1}{2},\frac{5}{6}}\left(x\right).
\end{equation}

The loss of energy of the emitting particles is modelled as:
\begin{equation}
\frac{\partial\gamma}{\partial t}=-\frac{4}{3}c\sigma_{T}\frac{u_{B}}{m_{e}c^{2}}\left(\frac{m_{e}}{m}\right)^{3}\gamma^{2},
\end{equation}
where $\sigma_{T}$ is the Thomson cross-section and $u_{B}$ is the
energy density of the magnetic field.

The absorption of photons is modelled as:
\begin{equation}
\alpha_{\nu}=-\frac{1}{8\pi m\nu^{2}}\int_{1}^{\infty}\gamma^{2}\frac{\partial}{\partial\gamma}\left[\frac{n}{\gamma^{2}}\right]P_{\nu}\left(\gamma\right)d\gamma.
\end{equation}

\bigskip{}

Photons interact with energetic electrons due to the inverse Compton
process, which we model as:
\begin{equation}
j_{\nu}\left(\varepsilon_{s}\right)=\frac{\hbar\varepsilon_{s}}{2}\int_{1}^{\infty}n_{e}\left(\gamma_{e}\right)\int_{0}^{\infty}n_{\gamma}\left(\varepsilon\right)g_{u,d}\left(\gamma_{e},\varepsilon,\varepsilon_{s}\right)d\varepsilon d\gamma_{e},
\end{equation}
where $\varepsilon$ is the energy of a photon normalized with respect
to the energy of a rest electron ($\varepsilon$ is the initial energy
of the photon and $\varepsilon_{s}$ is the energy after the scattering),
$n_{e}$ is the electron population, $n_{\gamma}$ is the photon population
and $g_{u,d}\left(\gamma,\varepsilon,\varepsilon_{s}\right)$ are
two functions referring to the upscattering and downscattering of
photons which where calculated by Jones \citep[equations 40 and 44]{jones_calculated_1968}:
\begin{align}
g_{u}\left(\gamma,\varepsilon,\varepsilon_{s}\right) & =\frac{3c\sigma_{T}}{4\gamma^{2}\varepsilon}\left[2q_{u}\ln q_{u}\right.\nonumber \\
 & \left.+\left(1+2q_{u}\right)\left(1-q_{u}\right)+\frac{\varepsilon_{s}^{2}}{\gamma\left(\gamma-\varepsilon_{s}\right)}\frac{1-q_{u}}{2}\right]\\
 & \quad\text{if }q_{u}\leq1\leq\frac{\varepsilon_{s}}{\varepsilon},\,\varepsilon_{s}<\gamma,\nonumber \\
g_{d}\left(\gamma,\varepsilon,\varepsilon_{s}\right) & =\frac{3c\sigma_{T}}{16\gamma^{4}\varepsilon}\left[\left(q_{d}-1\right)\left(1+\frac{2}{q_{d}}\right)-2\ln q_{d}\right]\\
 & \quad\text{if }\frac{1}{q_{d}}\leq1\leq\frac{\varepsilon}{\varepsilon_{s}},\nonumber 
\end{align}
with
\begin{align}
q_{u} & =\frac{\varepsilon_{s}}{4\varepsilon\gamma\left(\gamma-\varepsilon_{s}\right)},\\
q_{d} & =\frac{4\gamma{{}^2}\varepsilon_{s}}{\varepsilon}.
\end{align}

The loss term for photons can be obtained as:
\begin{equation}
\frac{\partial n_{\gamma}\left(\varepsilon\right)}{\partial t}=-n_{\gamma}\left(\varepsilon\right)\int_{1}^{\infty}n_{e}\left(\gamma_{e}\right)\mathcal{R}\left(\gamma_{e},\varepsilon\right)d\gamma_{e},
\end{equation}
where $\mathcal{R}\left(\gamma_{e},\varepsilon\right)$ is the reaction
rate of electrons of energy $\gamma_{e}$ with photons of energy $\varepsilon$
and can be found from integrating the Klein-Nishina cross section.

\begin{align}
\mathcal{R}\left(\gamma_{e},\varepsilon\right) & =\frac{3}{16}c\sigma_{T}\frac{1}{\beta\gamma_{e}^{2}\varepsilon^{2}}\left[-\frac{x}{4}+\frac{1}{4\left(x+1\right)}+\right.\nonumber \\
 & \phantom{=\frac{3}{16}c\sigma_{T}\frac{1}{\beta\gamma_{e}^{2}\varepsilon^{2}}}\frac{\left(x^{2}+9x+8\right)}{2x}\ln\left(1+x\right)+\\
 & \phantom{=\frac{3}{16}c\sigma_{T}\frac{1}{\beta\gamma_{e}^{2}\varepsilon^{2}}}\left.2\Li_{2}\left(-x\right)\right]_{x_{M}}^{x_{m}},\nonumber 
\end{align}
where:
\begin{align}
x_{m} & =2\varepsilon\gamma_{e}\left(1+\beta\right), & x_{M} & =2\varepsilon\gamma_{e}\left(1-\beta\right),
\end{align}
and $\Li_{2}$ the dilogarithm.

Following \citet{Bottcher_Book}, we use the simplified prescription
for the electron energy losses in inverse comptom scattering:
\begin{equation}
\frac{\partial\gamma}{\partial t}=-\frac{4}{3}c\sigma_{T}\frac{u_{\gamma}}{m_{e}c^{2}}\gamma^{2},
\end{equation}
where $u_{\gamma}$ is the energy density of the photon field:
\begin{equation}
u_{\gamma}=m_{e}c^{2}\times\int_{0}^{\infty}\varepsilon n_{\gamma}\left(\varepsilon\right)d\varepsilon.
\end{equation}

\bigskip{}

Photons can interact among themselves creating electron and positron
pairs, which we model according to \citep{bottcher_pair_1997}. Photon
losses are modelled according to the fully integrated cross section
in a similar way to the photon loss term for IC.

\begin{align}
\mathcal{R}\left(x\right) & =\frac{3\sigma_{T}}{4x^{2}}\left\{ a-2xa+\Li_{2}\left(\frac{1-a}{2}\right)-\Li_{2}\left(\frac{1+a}{2}\right)-\right.\nonumber \\
 & \phantom{=\frac{3}{4x^{2}}\quad}\left.\atanh\left(a\right)\left[-\ln\left(4x\right)-\frac{1}{x}-2x+2\right]\right\} ,
\end{align}

where $x=\varepsilon_{1}\varepsilon_{2}$, $a=\sqrt{1-\frac{1}{x}}$.

\bigskip{}

Hadrons (protons and neutrons) interact with photons producing hadronic
resonances which ultimately decay in the form of hadrons and pions.
To model this we follow the approximations of Hümmer et al \citep{hummer_simplified_2010}
in their SimB model. We diverge though in the calculations of the
loss and self-injection term for hadrons as, instead of applying a
cooling term, we model every process as a loss and where applicable,
we reinject the corresponding hadron with reduced energies.

\bigskip{}

Neutrons, charged pions and muons decay, producing ultimately protons,
electrons, positrons and neutrinos. To model these processes, we follow
Hümmer et al \citep{hummer_simplified_2010} for the decay of pions
and Lipari et al \citep{lipari_flavor_2007} for the decay of muons
and the production of neutrinos.

\bigskip{}

For electron-positron pair production and pair annihilation we follow
\citet{bottcher_pair_1997} and \citet{svensson_pair_1982} respectively.
Bethe-Heitler pair production is implemented following the model of
\citet{kelner_energy_2008}.

Finally, we do not model other processes such as proton-proton interaction,
heating of charged particles when absorbing photons and heating of
electrons when downscattering photons because their impact is negligible
relative to that of the processes included in our model.

\subsection{Steady State Models}

In our steady state model the populations of both electrons and protons
do not evolve over time. This also covers a scenario where both are
injected constantly over time with a fixed distribution that would
lead to the same steady state distribution that we here fix from the
outset.

The initial distribution of electrons is considered to be a broken
power law and for protons a simple power law:

\begin{align}
n_{e}\left(\gamma_{e}\right) & \propto\begin{cases}
A\gamma_{e}^{-p_{1}} & \text{if }\gamma_{e,min}<\gamma_{e}<\gamma_{e,break}\\
B\gamma_{e}^{-p_{2}} & \text{if }\gamma_{e,break}<\gamma_{e}<\gamma_{e,max}
\end{cases},\\
n_{p}\left(\gamma_{p}\right) & \propto C\gamma_{p}^{-p}\,\text{if }\gamma_{p,min}<\gamma_{p}<\gamma_{p,max},
\end{align}
where $A$, $B$ and $C$ are factors that normalize the distributions
and can be calculated from the conditions:
\begin{align}
\int_{\gamma_{e,min}}^{\gamma_{e,break}}A\gamma_{e}^{-p_{1}}+\int_{\gamma_{e,break}}^{\gamma_{e,max}}B\gamma_{e}^{-p_{2}}d\gamma_{e} & =1,\\
A\gamma_{e,break}^{-p_{1}} & =B\gamma_{e,break}^{-p_{2}},\\
\int_{\gamma_{p,min}}^{\gamma_{p,max}}C\gamma_{p}^{-p}d\gamma & =1,
\end{align}
and we have:
\begin{align}
A & \approx\left(p_{1}-1\right)\gamma_{e,min}^{p_{1}-1},\\
B & =A\gamma_{e,break}^{p_{2}-p_{1}},\\
C & \approx\left(p-1\right)\gamma_{p,min}^{p-1}.
\end{align}

We can assume that the ratio of protons to electrons is given by $\eta$.
This means that, given a combined mass density $\rho$ of accelerated
electrons and protons, the steady populations for electrons and protons
are:
\begin{align}
n_{e}\left(\gamma_{e}\right) & =\frac{\rho}{m_{e}+\eta\,m_{p}}A\times\nonumber \\
 & \begin{cases}
\phantom{\gamma_{e,break}^{p_{2}-p_{1}}}\gamma_{e}^{-p_{1}} & \text{if }\gamma_{e,min}<\gamma_{e}<\gamma_{e,break}\\
\gamma_{e,break}^{p_{2}-p_{1}}\gamma_{e}^{-p_{2}} & \text{if }\gamma_{e,break}<\gamma_{e}<\gamma_{e,max}
\end{cases},\\
n_{p}\left(\gamma_{p}\right) & =\frac{\eta\,\rho}{m_{e}+\eta\,m_{p}}C\times\gamma_{p}^{-p}.
\end{align}

Initially, in the simulated volume there are only electrons and protons,
so we need to evolve the plasma until it reaches a 'true' steady state
for all the particles. This is detected by calculating a statistic
related to the relative change in population of particles with time
which is then compared with a selected tolerance:
\[
\sqrt{\int\left[\frac{1}{n\left(x\right)}\frac{\partial n\left(x\right)}{\partial t}\right]^{2}dx}<\text{tol},
\]
where $n$ is the kind of particles we are interested in. In our case,
we test three particles species for convergence: photons, electrons
and neutrinos.

We consider that the system has reached the \emph{photonic steady
state} when the photon statistic has a value lower than $10^{-8}$,
and the \emph{final steady state} when all three have a value lower
than $10^{-8}$. The value $10^{-8}$ has been chosen as a good compromise
between the accuracy of the steady state and the simulation time,
given that the system only asymptotically approaches a true steady
state. We check the photon, neutrino and electron populations for
convergence for the following reasons. The photon population directly
shapes the SED, the electron population has a long covergence timescale
especially at low energies due to cooling, and neutrinos are the last
particles to be produced and therefore formally the last to achieve
steady state.

It takes time for the plasma to settle into a converged state, and
this time scale should be consistent with the variability time scale
of the conditions in the plasma as inferred from the blazar light
curve. In our application to TSX0506+056, we have verified that our
results are consistent with the 91 days time scale as experienced
by the plasma blob. We note that our modeled physical time scales
for the evolution from scratch to full convergence are an overestimate
resulting from uncertainties on both ends of the process. In reality,
there will likely be an initial distribution of energetic particles
prior to the onset of the flare, while the observed SED does not necessarily
show a completely converged state.

The steady state for protons and electrons is implemented by choosing
not to update their populations during the evolution of the system.
We can use this fact to have an a priori value for the energy density
in protons and electrons:
\begin{align}
u_{e} & =m_{e}c^{2}\times\int_{\gamma_{e,min}}^{\gamma_{e,max}}\gamma_{e}n_{e}\left(\gamma_{e}\right)d\gamma_{e},\\
u_{p} & =m_{p}c^{2}\times\int_{\gamma_{p,min}}^{\gamma_{p,max}}\gamma_{p}n_{e}\left(\gamma_{p}\right)d\gamma_{p}.
\end{align}

\textcolor{red}{}

\subsection{Dynamic Models\label{subsec:Dynamic-Models}}

Dynamic models break the assumption of a steady state population of
electrons and protons by allowing them to evolve over time. To arrive
at a steady state population of particles we have two options: we
can inject particles arriving from outside the volume with a given
distribution, or we can keep a steady population of cold particles
that are accelerated to higher energies within the volume. In this
work we focus on the first option. There are various reasons for this.
One is that acceleration introduces various parameters, the acceleration
timescales and mechanisms, which are not well understood. Second,
the accelerated particles have to come from a population of cold particles,
which we do not model. Third, acceleration requires a bit of tuning
to obtain the right slope and enough maximum energy for the particle
populations. Fourth, one can argue that acceleration is only going
to happen at the boundaries of the simulated volume (e.g. shock fronts),
and thus, its effect is to inject energetic particles with a given
distribution to the rest of the volume.

Particle injection is characterized by a luminosity ($L$) and a normalized
particle distribution $f\left(\gamma\right)$ such that:
\begin{equation}
\frac{L}{Vmc^{2}}=\int_{\gamma_{min}}^{\gamma_{max}}\gamma\mathcal{Q}_{ext,0}f\left(\gamma\right)d\gamma,
\end{equation}
where $V$ is the volume of the plasma where the injection is happening,
$m$ is the mass of the injected particle and $\mathcal{Q}_{ext,0}$
is the number density of injected particles. For a known particle
distribution, we can solve for $\mathcal{Q}_{ext,0}$ as:
\begin{equation}
\mathcal{Q}_{ext,0}=\frac{L}{\left\langle \gamma\right\rangle Vmc^{2}},
\end{equation}
where $\left\langle \gamma\right\rangle $ is the average energy of
the injected particles. In our case, we inject electrons and protons.
Assuming that the ratio of injected protons to electrons is $\eta$,
we have:
\begin{align}
\frac{L}{V} & =\mathcal{Q}_{ext,e,0}\left\langle \gamma_{e}\right\rangle m_{e}c^{2}+\eta\mathcal{Q}_{ext,e,0}\left\langle \gamma_{p}\right\rangle m_{p}c^{2},\\
Q_{e,0} & =\frac{L/V}{\left\langle \gamma_{e}\right\rangle m_{e}c^{2}+\eta\left\langle \gamma_{p}\right\rangle m_{p}c^{2}},\label{eq:Q_from_L}
\end{align}
and thus:
\begin{align}
Q_{e}\left(\gamma_{e}\right) & =\phantom{\eta}\mathcal{Q}_{ext,e,0}f_{e}\left(\gamma_{e}\right),\\
Q_{p}\left(\gamma_{p}\right) & =\eta\mathcal{Q}_{ext,e,0}f_{p}\left(\gamma_{p}\right).
\end{align}

Once a steady state is achieved and assuming that the injection follows
a broken power law, simplified versions of the kinetic equations can
be applied to obtain an approximation to the final electron and proton
populations.

For low energies, we have that the cooling time due to synchrotron
and Inverse Compton processes is longer than the escape timescale
of the particles. This means that we can approximate the population
evolution as:
\begin{equation}
\frac{\partial n}{\partial t}\approx\mathcal{Q}_{ext}-\frac{n}{\tau_{esc}}.
\end{equation}

By taking the derivative equal to $0$, we obtain that in the steady
state limit the population is:
\begin{align}
n & \approx\mathcal{Q}_{ext}\tau_{esc},\\
n_{e}\left(\gamma_{e}\right) & \approx\mathcal{Q}_{ext,e,0}\tau_{esc}A\times\nonumber \\
 & \begin{cases}
\phantom{\gamma_{e,break}^{p_{2}-p_{1}}}\gamma_{e}^{-p_{1}} & \text{if }\gamma_{e,min}<\gamma_{e}<\gamma_{e,break}\\
\gamma_{e,break}^{p_{2}-p_{1}}\gamma_{e}^{-p_{2}} & \text{if }\gamma_{e,break}<\gamma_{e}<\gamma_{e,max}
\end{cases},\\
n_{p}\left(\gamma_{p}\right) & \approx\eta Q_{e,0}\tau_{esc}C\times\gamma_{p}^{-p}.
\end{align}

If a steady state is achieved from dynamically injected protons and
electrons, the dynamic model and the steady state model relate according
to:
\begin{align}
\frac{L\tau_{esc}/V}{\left\langle \gamma_{e}\right\rangle m_{e}c^{2}+\eta\left\langle \gamma_{p}\right\rangle m_{p}c^{2}} & \approx\frac{\rho}{m_{e}+\eta\,m_{p}}.\label{eq:L_related_with_rho}
\end{align}

\textcolor{red}{}

For higher energies where we can not ignore the effects of synchrotron
cooling and Inverse Compton losses, we can approximate the population
with a power law with a given index $p$. We obtain:
\begin{align}
\frac{\partial n}{\partial t} & \approx\mathcal{Q}_{ext}+n_{0}\gamma^{-p}\left[\left(2-p\right)\gamma S-\frac{1}{\tau_{esc}}\right].
\end{align}

In the steady state, we obtain:
\begin{align}
n_{0}\gamma^{-p} & \approx\frac{\mathcal{Q}_{ext}\tau_{esc}}{1-\left(2-p\right)\gamma S\tau_{esc}}.\label{eq:pop_inj_cooling}
\end{align}

For $\left(2-p\right)\gamma S\tau_{esc}\ll1$ we recover the previous
result where we ignored the contributions of Synchrotron and Inverse
Compton cooling. When $\left(2-p\right)\gamma S\tau_{esc}\gg1$ we
obtain that the main driver of the population is cooling, and we get:
\begin{align}
n_{0}\gamma^{-p} & \approx\frac{1}{p-2}\frac{\mathcal{Q}_{ext}}{\gamma S},\\
n_{e}\left(\gamma_{e}\right) & \approx\frac{1}{p-2}\frac{\mathcal{Q}_{ext,e,0}}{S}A\times\nonumber \\
 & \begin{cases}
\phantom{\gamma_{e,break}^{p_{2}-p_{1}}}\gamma_{e}^{-p_{1}-1} & \text{if }\gamma_{e,min}<\gamma_{e}<\gamma_{e,break}\\
\gamma_{e,break}^{p_{2}-p_{1}}\gamma_{e}^{-p_{2}-1} & \text{if }\gamma_{e,break}<\gamma_{e}<\gamma_{e,max}
\end{cases},\\
n_{p}\left(\gamma_{p}\right) & \approx\frac{1}{p-2}\frac{\eta\mathcal{Q}_{ext,e,0}}{S}C\times\gamma_{p}^{-p_{inj}-1},
\end{align}
where we see that the effect that cooling produces is to increase
the index of the injected distribution of particles by $1$.

\bigskip{}

Particles with energies lower than the minimum energy of the injected
energy can appear due to cooling. If the turnover energy in a charged
particle population due to cooling lies below the level associated
with the lower injection cut-off Lorentz factor, an intermediate regime
appears in the particle spectrum. In the absence of additional physical
processes, this would lead to a power law slope of 2. Below the turnover
energy, escape takes over from cooling as the main driver, which leads
to a drop in the population.

\section{Bayesian and Markov Chain Monte Carlo analysis}{\label{sec:Model-Fitting}}

Model fitting can be done in various ways. An easy and straightforward
way of doing it is by eye \citep{keivani_multimessenger_2018}, where
the goodness of the fit is estimated by seeing if the simulated spectrum
fits the observations. The disadvantages of this approach are that
the validity of the fit can not be validated numerically and that,
depending on the parameters, finding a good fit might be a complicated
task.

Another way is by performing grid searches \citep{gao_modelling_2019},
where the parameter space is divided into a grid, and, for every point
in this grid you assign a goodness of fit value, taking the best parameters
as the final ones. This has the advantage of being fully automated
but has the disadvantage of requiring a high amount of computations,
as quite a lot of time is spent calculating the likelihood of bad
parameters.

A third method, similar to the latter, is to perform an MCMC search
over the parameter space (see for example \citet{qin_using_2018}
for a study that employed MCMC to fit simple leptonic models to blazars).
In this case, instead of dividing the parameter space into a grid
and checking every point, a series of 'walkers' traverse the parameter
space checking the likelihood of the fit at the point where they are.
Their advance is biased towards improving the likelihood so as to
guide their search towards 'good' parameters. Nevertheless, the search
is ergodic, which means that with time, the whole parameter space
will be traversed. Just like the previous method, MCMC searches are
fully automated, but they have the advantage of being focused on the
best parameters. This allows us to obtain two results besides the
best parameters. On the one hand, it allows us to obtain a measure
of the error in each of the parameters. On the other hand, a study
of the behaviour of the MCMC walkers allows us to identify correlations
between the parameters. This in turn allows us to redefine our parameters
so as to eliminate the correlations between them and improve the fits.
It should be noted that not all MCMC sampling algorithms perform worse
under correlated parameters. Nevertheless, finding good 'fit parameters'
a priori can give a very useful physical insight in the interconnectedness
of the model parameters.

Therefore, we will refer to two sets of parameters: 'model parameters',
or simply parameters, which are the input of our model; and 'fitted
parameters', which are the set that MCMC fits. Both sets are related
by invertible transformations, so it is easy to go from one to other.

MCMC requires some information from our part to perform its search.
Just as grid search needs a grid to be defined, MCMC needs information
about the distribution of the parameters it has to fit. The priors
on the fitted parameters are derived by starting from flat priors
in either linear or logarithmic space of the model parameters. The
distributions are chosen so that parameters that enter the model as
multiplicative terms are defined in logarithmic space and parameters
that enter as exponents are defined in linear space. The limits of
the distributions are taken very loosely, even to the point where
in some cases we allow unphysical values. The only hard limits are
the relations $\gamma_{e,min}<\gamma_{e,break}<\gamma_{e,max}$ and
$p_{1}<p_{2}$, and in the case of thin shells, $h<R$.

The log-likelihood of the model is calculated as:

\begin{equation}
l=-\frac{1}{2}\sum_{n}\left(\frac{y_{n}-\hat{y}_{n}}{\sigma_{n}}\right)^{2},\label{eq:loglikelihood}
\end{equation}
where $y_{n}$ are the observations, $\hat{y}_{n}$ are the predictions
of our model and $\sigma_{n}$ are the errors of the observations,
which we assume to be a few percent of the observations, depending
on the observed band. In our case, we take that for optical frequencies,
we have $\sigma_{n}=0.1\,y_{n}$; for X-rays $\sigma_{n}=0.2\,y_{n}$;
and for gamma-rays $\sigma_{n}=0.5\,y_{n}$.

The log-likelihood of the 'fitted parameters' is then defined as a
combination of the log-likelihood of the model plus the a-priori log-likelihood
of the 'fitted parameters'. Denoting the 'model parameters' as $P$
the 'fitted parameters' as $F$ and the a-priori log-likelihood of
a set of parameters as $p\left(X\right)$, we obtain:
\begin{align}
\mathcal{L}\left(F\right) & =l\left(F\right)+p\left(F\right)\nonumber \\
 & =l\left(F\right)+p\left(P\right)-\ln\left(\left|\frac{dF}{dP}\right|\right)\nonumber \\
 & =l\left(F\right)-\ln\left(\left|\frac{dF}{dP}\right|\right),\label{eq:transform_log_likelihood}
\end{align}
where the term $\left|dF/dP\right|$ represents the Jacobian of the
transformation between parameters, and $p\left(P\right)=0$ by construction.

\bigskip{}

In our implementation, we use an affine invariant sampler \citep{goodman_ensemble_2010}
implemented in the \texttt{Python} package emcee \citep{foreman-mackey_emcee:_2013}
using 128 walkers. The total numbers of samples taken differ between
runs and are reported along with the outcomes of a run.

The first data points of each walker correspond to the period of \textquoteright burn
in\textquoteright , where the walkers spread from their initial conditions
until they explore the log-likelihood function in an uncorrelated
way. During each of the runs, the data is binned and the values of
the average, median, standard deviation and 16 and 84 percentiles
are checked for trends. After they remain consistent for several bins
we consider that the run has found the final distribution and the
run is stopped.

The bins where the aforementioned values drift are considered the
'burn-in' phase and their data discarded for the final study.

As an additional consistency check for the remaining data. We rebin
the data and plot the PDFs to verify that the features of the distributions
remain essentially unchanged when drawn from subsets of the full chain.

While it is never formally possible to rule out trends with an extremely
long correlation time, we are therefore confident that the results
do not contain biases.

All the results reported in tables correspond to the median and the
16 and 84 percentiles.

\section{Applying our models in a Bayesian framework }{\label{sec:Reproducing-TXS0506+056}}

Before trying to fit a data set, we do an a priori analysis of the
physics of the emission processes in order to obtain constraints and
correlations for our parameters. These relations are then used to
find good initial parameters for the MCMC analysis and to interpret
the outcome of the fitting process.

\subsection{Initial Assumptions and Constraints\label{subsec:Constraints}}

\begin{figure*}
\begin{centering}
\includegraphics[width=1\textwidth]{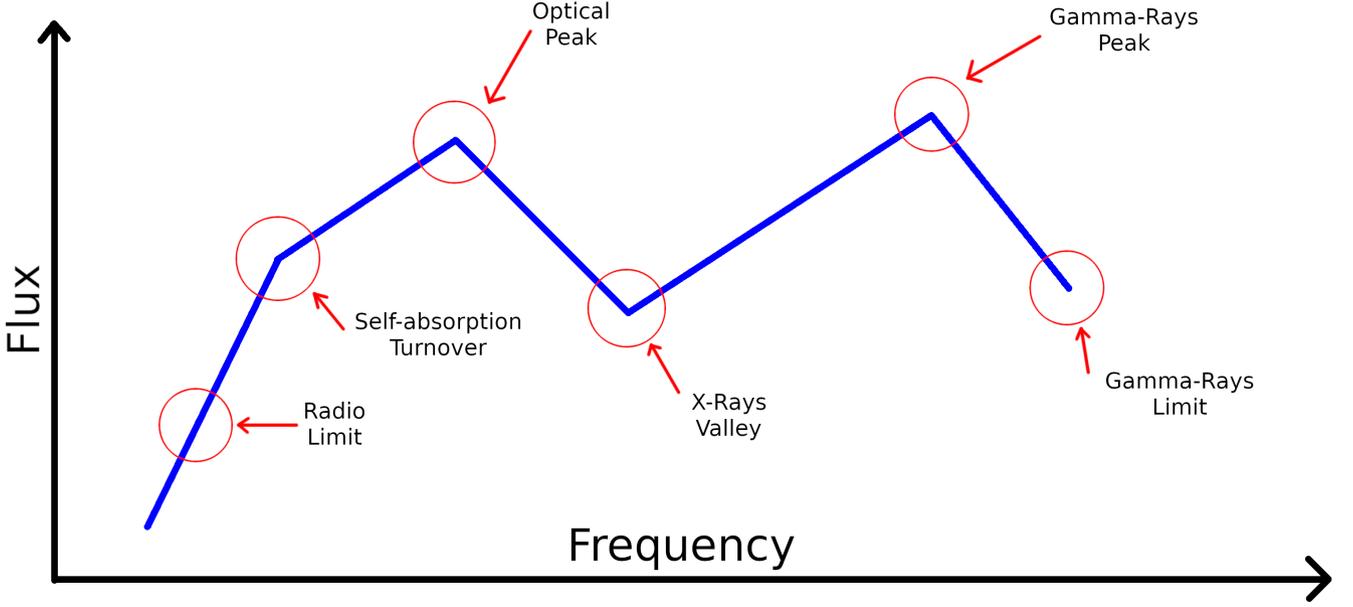}
\par\end{centering}
\caption{\label{fig:Schematic-SED}Schematic view of a blazar SED with some
interesting points marked for clarity.}
\end{figure*}

\Figref{Schematic-SED} shows a simple schematic view of an SED, which
we can use to obtain a series of constraints and first guesses for
our search:
\begin{itemize}
\item We can assume that the first peak of the SED is near the optical band,
and that it is caused by the synchrotron emission of the electrons
at the break point $\gamma_{e,break}$. This gives us the constraint:
\begin{equation}
\nu_{obs,opt}\approx\frac{3}{4\pi}\frac{qB}{m_{e}c}\gamma_{e,break}^{2}\frac{\delta}{1+z}.\label{eq:nu_break_constraint}
\end{equation}
\item We know that we have to obtain photons of a certain energy, and that
those come from the Inverse Compton process with high energy electrons.
So, we can derive a lower limit for the maximum value of $\gamma_{e,max}$:
\begin{equation}
\nu_{obs,max}\lesssim\gamma_{e,max}\frac{m_{e}c^{2}}{2\pi\hbar}\frac{\delta}{1+z}.\label{eq:gamma_max_constraint}
\end{equation}
\item A second lower limit for the value of $\gamma_{e,max}$, albeit much
less limiting, can be obtained from the X-ray valley. We can safely
assume that the emission for the low energy part of the valley comes
almost entirely from the synchrotron emission of high energy electrons.
This gives us the constraint:
\begin{equation}
\nu_{obs,xray}\lesssim\frac{3}{4\pi}\frac{qB}{m_{e}c}\gamma_{e,max}^{2}\frac{\delta}{1+z}.\label{eq:gamma_max_constraint_xray}
\end{equation}
\item We can fit a line between the optical and the X-ray emission. As it
is above the synchrotron peak, we can assume that this emission comes
from electrons with $\gamma_{e}>\gamma_{e,break}$ and, as the theoretical
slope of synchrotron emission is known, we can find the constraint:
\begin{equation}
\frac{3-p_{2}}{2}\approx\alpha_{2}.\label{eq:alpha_2_constraint}
\end{equation}
\item We can fit a line between the X-ray emission and Gamma-ray emission,
and we can assume that the emission in this range is going to be caused
by the upscattering of synchrotron photons by the Inverse Compton
process. The inverse Compton process increases the energy of photons
as $\varepsilon_{after}\approx\gamma_{e}^{2}\varepsilon_{before}$.
For photons from the synchrotron peak, upscattered by electrons at
$\gamma_{e,break}$, their energy is above the fit. So, these photons
should come from synchrotron photons emitted by electrons with $\gamma_{e}<\gamma_{e,break}$.\\
As the Inverse Compton process does not modify the slope of the emission,
we obtain:
\begin{equation}
\frac{3-p_{1}}{2}\approx\alpha_{1}.\label{eq:alpha_1_constraint}
\end{equation}
\item The radio observations can be considered as upper limits for our model,
as the flux can come from the extended emission in the jet (see for
example \citealt{Cerruti_2018}). From this fact, we can derive another
constraint. Assuming, as before, that the optical observations are
the peak of the synchrotron spectrum, and that they are caused by
electrons with energy $\gamma_{e,break}$, we can extrapolate the
observations down to the frequency associated with $\gamma_{e,min}$:
\begin{equation}
\left(\nu F_{\nu}\right)_{break}=A\nu_{break}^{\alpha_{1}},
\end{equation}
so:
\begin{equation}
\left(\nu F_{\nu}\right)_{min}=\left(\nu F_{\nu}\right)_{break}\left(\frac{\nu_{min}}{\nu_{break}}\right)^{\alpha_{1}}.
\end{equation}
At lower frequencies, this can be extrapolated further using the index
$4/3$. And we obtain:
\begin{align}
\left(\nu F_{\nu}\right)_{min} & =B\nu_{min}^{4/3},\\
B & =\left(\nu F_{\nu}\right)_{break}\frac{\nu_{min}^{\alpha_{1}-4/3}}{\nu_{break}^{\alpha_{1}}},
\end{align}
so:
\begin{equation}
\left(\nu F_{\nu}\right)_{radio}=\left(\nu F_{\nu}\right)_{break}\left(\frac{\nu_{min}}{\nu_{break}}\right)^{\alpha_{1}}\left(\frac{\nu_{radio}}{\nu_{min}}\right)^{4/3}.
\end{equation}
Now, all the frequencies can be transformed into associated gamma
factors for electrons in the fluid frame:
\begin{equation}
\left(\nu F_{\nu}\right)_{radio}=\left(\nu F_{\nu}\right)_{break}\left(\frac{\gamma_{e,min}}{\gamma_{e,break}}\right)^{2\alpha_{1}}\left(\frac{\gamma_{e,radio}}{\gamma_{e,min}}\right)^{8/3}.
\end{equation}
Note that $\gamma_{e,radio}$ is not a real value for the gamma factor
of electrons, as it can be $<1$. Also note that it might be the case
that the synchrotron self absorption frequency is higher than the
frequency associated with $\gamma_{e,min}$. In this case, by using
$\gamma_{e,min}$ we would be overestimating the approximation of
$\left(\nu F_{\nu}\right)_{radio}$. The constraint is that the observed
flux in radio has to be less than the calculated, so:
\begin{equation}
\left(\nu F_{\nu}\right)_{radio,obs}\gtrsim\left(\nu F_{\nu}\right)_{break,obs}\left(\frac{\gamma_{e,min}}{\gamma_{e,break}}\right)^{2\alpha_{1}}\left(\frac{\gamma_{e,radio}}{\gamma_{e,min}}\right)^{8/3}.\label{eq:radio_constraints}
\end{equation}
\item The energy of the photons at the optical band is enough so that the
blob can be considered thin to the radiation. In this case, the main
loss term for photons is escape. Under the simplification of a $\delta$-approximation
to synchrotron emission \citep[Chapter 3]{Bottcher_Book}\footnote{Note that Equation 3.39 follows from Equation 3.38, and that one from
Equation 3.25. But a factor $c$ was forgotten when deriving Equation
3.38 from Equation 3.25. This factor is corrected here.}, which we only employ here in order to make an estimate, we have:
\begin{equation}
j_{\nu}\left(\nu\right)=\frac{1}{6\pi}c\sigma_{T}u_{B}\sqrt{\frac{\nu}{\nu_{0}}}\frac{1}{\nu_{0}}n_{e}\left(\gamma_{e}=\sqrt{\frac{\nu}{\nu_{0}}}\right).
\end{equation}
The emission coefficient is related to the rate of injection of photons
as:
\begin{equation}
\left|\frac{\partial n_{\gamma}\left(\varepsilon\right)}{\partial t}\right|_{inj}=\frac{4\pi}{\varepsilon m_{e}c^{2}}\frac{d\nu}{d\varepsilon}j_{\nu}\left(\nu\right)=\frac{2}{\hbar\varepsilon}j_{\nu}\left(\nu\right).
\end{equation}
The $\delta$-approximation also allows us to have that $\sqrt{\nu/\nu_{0}}=\gamma_{e}$,
and thus, we can relate $\nu$, $\varepsilon$ and $\gamma_{e}$ as:
\begin{equation}
\varepsilon=\frac{2\pi\hbar\nu}{m_{e}c^{2}}=\frac{2\pi\hbar\nu_{0}}{m_{e}c^{2}}\gamma_{e}^{2},
\end{equation}
which means that:
\begin{align}
\left|\frac{\partial n_{\gamma}\left(\varepsilon\right)}{\partial t}\right|_{inj} & =\frac{m_{e}c^{2}}{\hbar^{2}\nu_{0}^{2}}\frac{1}{6}c\sigma_{T}u_{B}\frac{1}{\gamma_{e}}n_{e}\left(\gamma_{e}\right).
\end{align}
As the main losses for photons is the escape term we have:
\begin{equation}
\frac{\partial n_{\gamma}\left(\varepsilon\right)}{\partial t}=\frac{m_{e}c^{2}}{\hbar^{2}\nu_{0}^{2}}\frac{1}{6}c\sigma_{T}u_{B}\frac{1}{\gamma_{e}}n_{e}\left(\gamma_{e}\right)-\frac{n_{\gamma}\left(\varepsilon\right)}{\tau_{esc}}.
\end{equation}
In the steady state, $\partial n/\partial t=0$, so, we can solve
for the equilibrium population of photons as:
\begin{equation}
n_{\gamma}\left(\varepsilon\right)=\tau_{esc}\frac{m_{e}c^{2}}{\hbar^{2}\nu_{0}^{2}}\frac{1}{6}c\sigma_{T}u_{B}\frac{1}{\gamma_{e}}n_{e}\left(\gamma_{e}\right).
\end{equation}
The electron population is related to the density of the blob and
we have, assuming that $\gamma_{e,min}\ll\gamma_{e,break}$:
\begin{align}
n_{\gamma}\left(\varepsilon\right) & =\tau_{esc}\frac{m_{e}c^{2}}{\hbar^{2}\nu_{0}^{2}}\frac{1}{6}c\sigma_{T}u_{B}\nonumber \\
 & \phantom{=\;}\frac{1}{\gamma_{e}}\frac{\rho}{m_{e}+\eta m_{p}}\left(p_{1}-1\right)\gamma_{e,min}^{p_{1}-1}\gamma_{e}^{-p_{1}},
\end{align}
for $\gamma_{e}\le\gamma_{e,break}$.\\
The transformation from population to observed flux requires us to
multiply the previous expression by $\varepsilon^{2}$, which we do
using the relation with $\gamma_{e}$:
\begin{align}
\varepsilon^{2}n_{\gamma}\left(\varepsilon\right) & =\tau_{esc}\frac{1}{m_{e}c^{2}}\frac{c\sigma_{T}}{12\pi}B^{2}\nonumber \\
 & \phantom{=\;}\frac{\rho}{m_{e}+\eta m_{p}}\left(p_{1}-1\right)\gamma_{e,min}^{p_{1}-1}\gamma_{e}^{3-p_{1}}.
\end{align}
Doing the final transformation we have:
\begin{align}
\nu F_{\nu} & =\frac{\delta^{4}}{4\pi\left(1+z\right)D_{L}^{2}}V\frac{c\sigma_{T}}{12\pi}B^{2}\nonumber \\
 & \phantom{=\;}\frac{\rho}{m_{e}+\eta m_{p}}\left(p_{1}-1\right)\gamma_{e,min}^{p_{1}-1}\gamma_{e}^{3-p_{1}}\label{eq:optical_flux_constraints}\\
 & \propto\delta^{4}VB^{2}\rho\left(p_{1}-1\right)\gamma_{e,min}^{p_{1}-1}\gamma_{e}^{3-p_{1}}.\nonumber 
\end{align}
It is important to remember that this is only valid for energies where
the blob is not thick to the radiation, and only for $\gamma<\gamma_{e,break}$.
\item Following the previous point, we can also derive the flux for the
low energy part of the x-ray band, as we assume that those photons
are synchrotron emission from high energy electrons. The population
of high energy electrons is:
\begin{align}
n_{e}\left(\gamma_{e}>\gamma_{e,break}\right) & =\frac{\rho}{m_{e}+\eta m_{p}}\left(p_{1}-1\right)\nonumber \\
 & \phantom{=\;}\gamma_{e,min}^{p_{1}-1}\gamma_{e,break}^{p_{2}-p_{1}}\gamma_{e}^{-p_{2}}.
\end{align}
Then, the photon population for the high energy part of the synchrotron
spectrum is:
\begin{align}
n_{\gamma}\left(\varepsilon\right) & =\tau_{esc}\frac{m_{e}c^{2}}{\hbar^{2}\nu_{0}^{2}}\frac{1}{6}c\sigma_{T}u_{B}\nonumber \\
 & \phantom{=\;}\frac{1}{\gamma_{e}}\frac{\rho}{m_{e}+\eta m_{p}}\left(p_{1}-1\right)\gamma_{e,min}^{p_{1}-1}\gamma_{e,break}^{p_{2}-p_{1}}\gamma_{e}^{-p_{2}}.
\end{align}
for $\gamma_{e}>\gamma_{e,break}$. So, the X-Ray flux is:
\begin{align}
\nu F_{\nu} & =\frac{c\sigma_{T}}{48\pi^{2}}\frac{\delta^{4}V}{\left(1+z\right)D_{L}^{2}}B^{2}\nonumber \\
 & \phantom{=\;}\frac{\rho}{m_{e}+\eta m_{p}}\left(p_{1}-1\right)\gamma_{e,min}^{p_{1}-1}\gamma_{e,break}^{p_{2}-p_{1}}\gamma_{e}^{3-p_{2}}\\
 & \propto V\delta^{4}B^{2}\gamma_{e,x-ray}^{3}\rho\left(p_{1}-1\right)\gamma_{e,min}^{p_{1}-1}\gamma_{e,break}^{p_{2}-p_{1}}\gamma_{e,x-ray}^{-p_{2}}.\nonumber 
\end{align}
Following one of our previous constraints, $\gamma_{e,x-ray}\lesssim\gamma_{e,max}$,
so:
\begin{equation}
\nu F_{\nu}\gtrsim V\delta^{4}B^{2}\rho\left(p_{1}-1\right)\gamma_{e,min}^{p_{1}-1}\gamma_{e,break}^{p_{2}-p_{1}}\gamma_{e,max}^{3-p_{2}}.
\end{equation}
Dividing our previous prediction for optical flux and X-rays flux,
we have:
\begin{equation}
\frac{\left(\nu F_{\nu}\right)_{optical}}{\left(\nu F_{\nu}\right)_{x-rays}}\sim\frac{\gamma_{e,break}^{3-p_{1}}}{\gamma_{e,break}^{p_{2}-p_{1}}\gamma_{e,max}^{3-p_{2}}}\sim\left(\frac{\gamma_{e,break}}{\gamma_{e,max}}\right)^{3-p_{2}}.
\end{equation}
\item Photons from synchrotron processes get upscattered by high energy
electrons due to Inverse Compton processes. As an approximation, and
just so we can derive a functional form for the photon population
at high energies, we can say that the emission coefficient for IC
photons depends on the population of photons at low energy and the
population of electrons at high energy. As before, we assume that
the energy of a photon is increased as $\varepsilon_{after}\approx\gamma_{e}^{2}\varepsilon_{before}$.
\begin{equation}
j_{\nu,IC}\left(\varepsilon\right)=\frac{c\sigma_{T}}{4}\hbar n_{\gamma}\left(\frac{\varepsilon}{\gamma_{e}^{2}}\right)n_{e}\left(\gamma_{e}\right).
\end{equation}
Transforming to the rate of injection of photons, we obtain:
\begin{equation}
\left|\frac{\partial n_{\gamma}\left(\varepsilon\right)}{\partial t}\right|_{inj}=\frac{c\sigma_{T}}{2}\frac{1}{\varepsilon}n_{\gamma}\left(\frac{\varepsilon}{\gamma_{e}^{2}}\right)n_{e}\left(\gamma_{e}\right).
\end{equation}
Again, we assume that the main losses are the escape process, so we
can calculate the population of gamma ray photons in our blob as:
\begin{equation}
n_{\gamma}\left(\varepsilon\right)=\tau_{esc}\frac{c\sigma_{T}}{2}\frac{1}{\varepsilon}n_{\gamma}\left(\frac{\varepsilon}{\gamma_{e}^{2}}\right)n_{e}\left(\gamma_{e}\right).
\end{equation}
To compare with the optical flux, we just multiply by $\varepsilon^{2}$:
\begin{equation}
\varepsilon^{2}n_{\gamma}\left(\varepsilon\right)=\tau_{esc}\frac{c\sigma_{T}}{2}\varepsilon n_{\gamma}\left(\frac{\varepsilon}{\gamma_{e}^{2}}\right)n_{e}\left(\gamma_{e}\right).
\end{equation}
We assume that the photons that get upscattered to $\gamma$-rays
are those from the synchrotron peak, and that the electrons with $\gamma_{e,break}$
are the ones mainly responsible for the upscattering. This means that
$\varepsilon_{opt}\gamma_{e,break}^{2}=\varepsilon_{\gamma-ray}$
and $n_{\gamma}\left(\varepsilon_{opt}\right)=n_{\gamma}\left(\varepsilon_{\gamma-ray}/\gamma_{e,break}^{2}\right)$.
This means that the ratio of fluxes is:
\begin{align}
\frac{\left(\nu F_{\nu}\right)_{\gamma-rays}}{\left(\nu F_{\nu}\right)_{opt}} & =\frac{\left(\varepsilon^{2}n_{\gamma}\right)_{\gamma-rays}}{\left(\varepsilon^{2}n_{\gamma}\right)_{opt}},\nonumber \\
 & =\frac{m_{e}c^{2}}{2\pi\hbar\nu_{0}}\tau_{esc}\frac{c\sigma_{T}}{2}\nonumber \\
 & \phantom{=\;}\frac{\rho}{m_{e}+\eta m_{p}}\left(p_{1}-1\right)\gamma_{e,min}^{p_{1}-1}\gamma_{e,break}^{-p_{1}},\label{eq:g_ray_opt_constraints}\\
 & \propto\frac{1}{\delta B}\tau_{esc}\rho\left(p_{1}-1\right)\gamma_{e,min}^{p_{1}-1}\gamma_{e,break}^{-p_{1}}.\nonumber 
\end{align}
\end{itemize}

\subsection{Application to TXS0506+056: Initial Estimates}

TXS 0506+056 is a Bl Lac kind of blazar where recently a neutrino
was detected coincident with a flare, which established the importance
of a hadronic component to blazar modelling. Due to this, there is
a high amount of almost simultaneous data that we can use to test
emission models.

Following \subsecref{Constraints}, we can use the available observational
data on TXS 0506+056 to estimate good initial values for our fitting
procedure. Frequencies are given in $s^{-1}$ and fluxes in $erg \cdot s^{-1}\cdot cm^{-2}$.
We have:
\begin{align*}
\nu_{obs,radio} & \sim10^{10}, & \left(\nu F_{\nu}\right)_{obs,radio} & \sim10^{-13},\\
\nu_{obs,opt} & \sim10^{15}, & \left(\nu F_{\nu}\right)_{obs,opt} & \sim4\times10^{-11},\\
\nu_{obs,x-ray} & \sim10^{18}, & \left(\nu F_{\nu}\right)_{obs,x-ray} & \sim10^{-12},\\
\nu_{obs,\gamma-ray} & \sim4\times10^{23}, & \left(\nu F_{\nu}\right)_{obs,\gamma-ray} & \sim6\times10^{-11},\\
\nu_{obs,max} & \sim10^{26},
\end{align*}

\begin{itemize}
\item By the constraint from \eqref{nu_break_constraint} and the value
for $\nu_{obs,opt}$ we obtain:
\begin{align}
\log_{10}\left(\delta B\gamma_{e,break}^{2}\right) & \approx8.50.
\end{align}
\item By the constraint from \eqref{gamma_max_constraint} and the value
for $\nu_{obs,max}$ we obtain:
\begin{align}
\log_{10}\left(\delta\gamma_{e,max}\right) & \gtrsim6.03.
\end{align}
\item By the constraint from \eqref{gamma_max_constraint_xray} and the
value for $\nu_{obs,x-ray}$ we obtain:
\begin{align}
\log_{10}\left(\delta B\gamma_{e,max}^{2}\right) & \gtrsim11.50.
\end{align}
\item By the constraint from \eqref{alpha_2_constraint} and the values
for $\nu_{obs,opt}$, $\nu_{obs,x-ray}$, $\left(\nu F_{\nu}\right)_{obs,opt}$
and $\left(\nu F_{\nu}\right)_{obs,x-ray}$,we obtain:
\begin{align}
p_{2} & \approx4.06.
\end{align}
\item By the constraint from \eqref{alpha_1_constraint} and the values
for $\nu_{obs,x-ray}$, $\nu_{obs,\gamma-ray}$, $\left(\nu F_{\nu}\right)_{obs,x-ray}$
and $\left(\nu F_{\nu}\right)_{obs,\gamma-ray}$,we obtain:
\begin{align}
p_{1} & \approx2.36.
\end{align}
\item By the constraint from \eqref{radio_constraints} and the values for
$\nu_{obs,radio}$, $\nu_{obs,opt}$, $\left(\nu F_{\nu}\right)_{obs,radio}$
and $\left(\nu F_{\nu}\right)_{obs,opt}$,we obtain, assuming $\alpha_{1}\approx1/3$
:
\begin{equation}
\log_{10}\left(\delta B\gamma_{e,min}^{2}\right)\gtrsim4.44.
\end{equation}
\item We have obtained a form for the optical flux in \eqref{optical_flux_constraints}.
With the value for $\left(\nu F_{\nu}\right)_{obs,opt}$ we obtain:
\begin{equation}
\log_{10}\left[\delta^{4}VB^{2}\rho\left(p_{1}-1\right)\gamma_{e,min}^{p_{1}-1}\gamma_{e,break}^{3-p_{1}}\right]\approx37.83.
\end{equation}
and interestingly, combining the previous constraint with that for
$\delta B\gamma_{e,break}^{2}$ we obtain:
\begin{equation}
\log_{10}\left[\delta^{2}V\rho\left(p_{1}-1\right)\gamma_{e,min}^{p_{1}-1}\gamma_{e,break}^{-p_{1}-1}\right]\approx20.83.
\end{equation}
\item From the relation between the $\gamma$-ray and the optical flux,
\eqref{g_ray_opt_constraints} we can obtain:
\begin{equation}
\log_{10}\left[\frac{1}{\delta B}\tau_{esc}\rho\left(p_{1}-1\right)\gamma_{e,min}^{p_{1}-1}\gamma_{e,break}^{-p_{1}}\right]\approx-23.07.
\end{equation}
\item Dividing the optical flux reduced by $\left(\delta B\gamma_{e,break}^{2}\right)^{2}$
by the previous relation we obtain:
\begin{equation}
\log_{10}\left(\delta^{3}B\frac{V}{\tau_{esc}}\gamma_{e,break}^{-1}\right)\approx43.89.
\end{equation}
\item And reducing once more by $\delta B\gamma_{e,break}^{2}$:
\begin{equation}
\log_{10}\left(\delta^{2}\frac{V}{\tau_{esc}}\gamma_{e,break}^{-3}\right)\approx35.39.\label{eq:delta_g_break_R}
\end{equation}
\end{itemize}
\bigskip{}

The estimates and relations that we have derived in the previous subsection,
and the numerical estimates that we have obtained in this can then
be used in two ways. First, the relations can guide our search for
parameters that are usually together and group them into the 'fitted
parameters' for the MCMC fitting procedure. And second, the estimates
will guide our choices for the initial values and domain boundaries
of the fit parameters. At the start of our fitting process, we initialize
our walkers in a small sphere, taken to be much smaller than the allowed
range for all the parameters, around the estimates.

In total, we fit six scenarios. First, we fit the four permutations
of geometry (sphere vs disk) and particle behaviour (steady state
vs injection) as our 'default fits'. Then, guided by the results from
this analysis, we explore further (sections \ref{subsec:Interpretation-of-Posterior}
and \ref{subsec:Changing-eta}) a sphere injection scenario with a
single electron power-law and a scenario where we also allow the ratio
of protons to electrons to vary.

\subsection{Other Settings and Results }{\label{subsec:Other-Settings-and-Results}}

Our model includes several parameters that set the physics and numerics
of the emitting plasma. Some of these are fitted for, and are summarized
in \tabref{Parameters} along with their allowed ranges. Among these
parameters, the electron distribution is taken to be a broken power
law ($p_{1}$ and $p_{2}$ refer to the slopes across this break).
Other parameters are kept fixed and are discussed below.

Following other authors \citep{gao_modelling_2019,Cerruti_2018},
we assume for the accelerated proton population an injection with
a single power law in energy with decay slope of 2. The lower cut-off
value of the proton population $\gamma_{p,min}$ does not have a noticeable
impact on the final fit results and has been fixed at $10$. For the
value of $\gamma_{p,max}$ we have chosen a reasonable energy that
allows for the creation of neutrinos with the energy detected by Ice
Cube. The neutrinos that eventually result from the photomeson production
process can end up with a significiant fraction of the initial proton
energy (see e.g. \citet{kelner_energy_2008,hummer_simplified_2010}).
To allow for the creation of neutrinos at an energy range detectable
by IceCube in this manner, we have extended the upper limit of the
proton power law distribution to a value of $\gamma_{p,max}=10^{6}$,
although as we will show in \subsecref{Changing-eta}, this value
does not affect the predicted neutrino flux at $290$ TeV.

The escape time for charged particles will be longer than that for
uncharged particles. Following \citet{gao_modelling_2019}, we take
this ratio of escape times, $\sigma$ (see \subsecref{Kinetic-equation-approach}),
to be equal to $300$.

In our dynamic models where the proton and electron populations are
not initially present, we set both number densities to numerically
small values at the start.

We set the range of the numerical grid that stores the electron distribution
broad enough to safely encompass the emergent electron distribution
from injection, using a lower cut-off Lorentz factor of $2$ and an
upper cut-off value that is set to $\max\left(\gamma_{e,max},\frac{m_{p}}{m_{e}}\gamma_{p,max}\right)$,
where $\gamma_{e,max}$ and $\gamma_{p,max}$ are the upper cut-off
values respectively of the electron and proton injections. The lower
and upper boundaries of the photon population grid have also been
chosen generously to account for all possible photon production channels.
These have respectively been set at $10^{-12}$ and equal to the electron
upper grid boundary.

In principle we can sidestep the issue of charge neutrality in the
plasma by allowing the low-temperature pools of electrons and protons
to compensate for any charge imbalance in the injection terms. However,
assuming the injection to be either acceleration of these low-temperature
populations (e.g. at boundary shocks), or to originate from the same
external region, we have no obvious physical mechanism that suggests
a larger number of protons than electrons to be accelerated (if anything,
the opposite), nor for the occurence of a larger pool of protons than
electrons in the source of the injection, and our default approach
therefore assumes $\eta=1$. We explore the implications of relaxing
this constraint in \subsecref{Changing-eta}.

This still leaves us with all the parameters of \Tabref{Parameters}
to fit.

\begin{table*}
\begin{centering}
\begin{tabular}{ccc}
\toprule 
Parameter & Search Range & Physical Meaning\tabularnewline
\midrule
\midrule 
$\gamma_{e,min}$ & $10-\gamma_{e,break}$ & Lower boundary for the electron distribution\tabularnewline
\midrule 
$\gamma_{e,break}$ & $\gamma_{e,min}-\gamma_{e,max}$ & Breaking point for the electron distribution\tabularnewline
\midrule 
$\gamma_{e,max}$ & $\gamma_{e,break}-10^{10}$ & Higher boundary for the electron distribution\tabularnewline
\midrule 
$p_{1}$ & $1-p_{2}$ & First slope of the broken power law\tabularnewline
\midrule 
$p_{2}$ & $p_{1}-6$ & Second slope of the broken power law\tabularnewline
\midrule 
$B$ & $0.005-0.5$ & Magnetic field $\left(\text{G}\right)$\tabularnewline
\midrule 
$\delta$ & $2-192$ & Doppler boosting parameter\tabularnewline
\midrule 
\multirow{1}{*}{$h$} & $10^{10}-R$ & Height of the disk $\left(\text{cm}\right)$$^{\dagger}$\tabularnewline
\midrule 
\multirow{2}{*}{$R$} & $10^{10}-10^{26}$ & Radius of the sphere $\left(\text{cm}\right)$$^{*}$\tabularnewline
\cmidrule{2-3} \cmidrule{3-3} 
 & $h-10^{26}$ & Radius of the disk $\left(\text{cm}\right)$$^{\dagger}$\tabularnewline
\midrule 
$\rho$ & $10^{-26}-10^{-16}$ & Density of the fluid $\left(\text{gr}\cdot\text{cm}^{-3}\right)$$^{\ddagger}$\tabularnewline
\midrule 
$L$ & $10^{38}-10^{50}$ & Luminosity of the injected particles $\left(\text{erg}\cdot\text{s}^{-1}\right)$$^{\star}$\tabularnewline
\bottomrule
\end{tabular}
\par\end{centering}
\caption{\label{tab:Parameters}Free parameters, their search ranges and their
meanings. $\dagger$ only for disk volumes, $*$ only for spherical
volumes, $\ddagger$ only for steady state models and $\star$ only
for dynamical models.}
\end{table*}

\begin{figure*}
\begin{centering}
\includegraphics[width=1\textwidth]{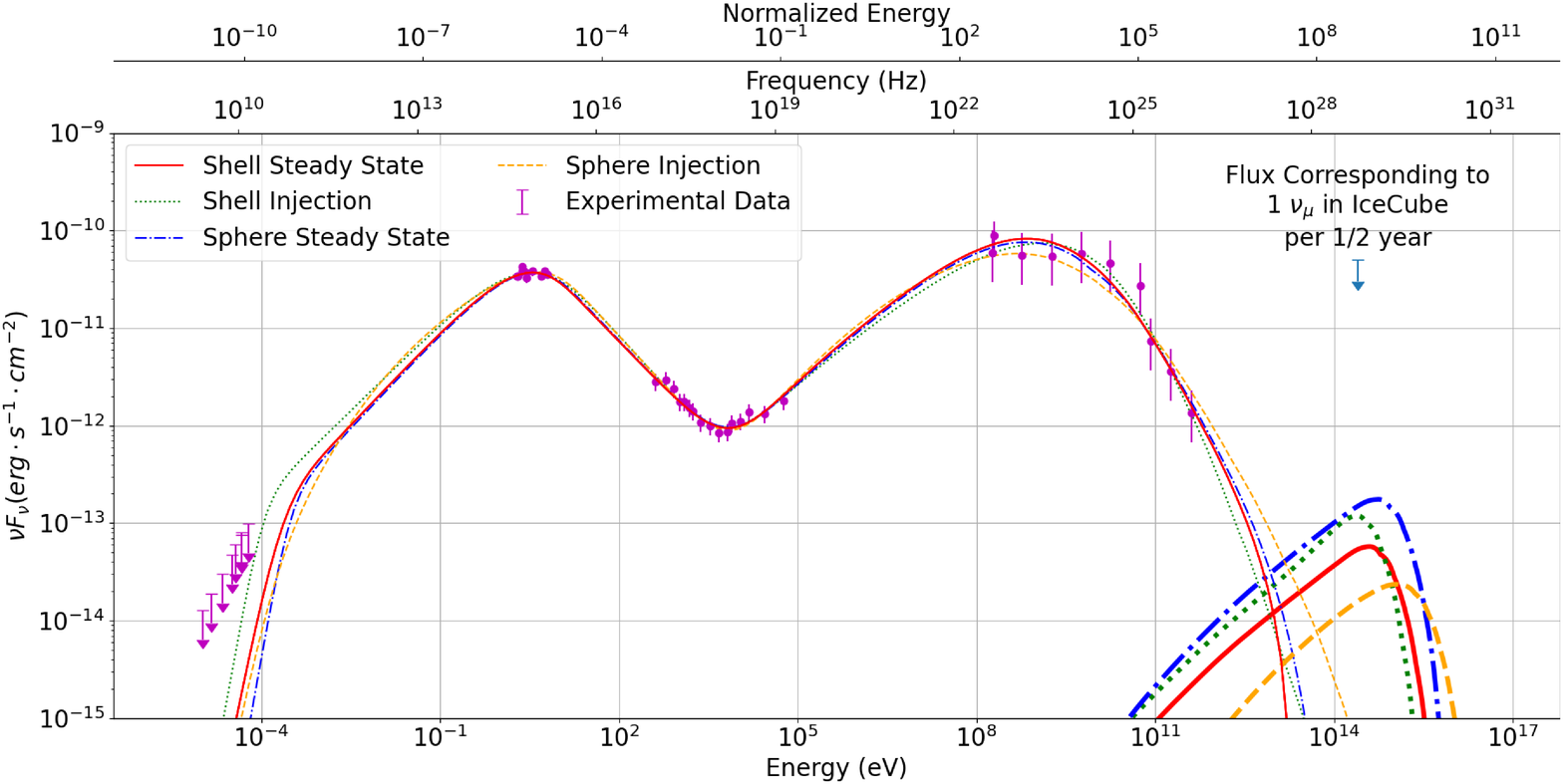}
\par\end{centering}
\caption{\label{fig:Combined-SED}Combined SED for all the models. Thick curves
correspond to the predicted neutrino flux.}
\end{figure*}

\Figref{Combined-SED} is a combination of the best SED fits (\tabref{Parameters_all_models})
that we can obtain with our model given the parameters that we set
a priori. It can be appreciated that all the models can reproduce
the SED very well, but they fall short by $\sim3$ orders of magnitude
in the neutrino flux. We return to the neutrino flux in \subsecref{Changing-eta}.

\Figref[s]{Spherical_Steady_State_Corner}, \ref{fig:Disk_Steady_State_Corner},
\ref{fig:Spherical_Injection_Corner} and \ref{fig:Disk_Injection_Corner}
show the corner plots that we obtain from the MCMC fitting. The blue
lines denote the parameters that give the best SED fit.

\subsubsection{Spherical Steady State}

The spherical steady state model has $9$ parameters: $\left(\gamma_{e,min}\text{, }\gamma_{e,break}\text{, }\gamma_{e,max}\text{, }p_{1}\text{, }p_{2}\text{, }B\text{, }\delta\text{, }R\text{ and }\rho\right)$,
but we know from our analysis of \subsecref{Constraints} that some
of them are always grouped. Thus, we choose to redefine our parameters
in the following way:
\begin{align}
F_{1} & \equiv VN_{1}\delta^{2}\nu_{br}^{2}\gamma_{e,break}^{-p_{1}-1},\nonumber \\
F_{2} & \equiv\left(\frac{\gamma_{e,max}}{\gamma_{e,break}}\right)^{3-p_{2}},\nonumber \\
F_{3} & \equiv\frac{RN_{1}}{\delta B}\gamma_{e,break}^{-p1},\nonumber \\
\delta & \equiv\delta,\nonumber \\
\delta B & \equiv\delta B,\\
\nu_{br} & \equiv\delta B\gamma_{e,break}^{2},\nonumber \\
\rho & \equiv\rho,\nonumber \\
RN_{1} & \equiv R\rho\left(p_{1}-1\right)\gamma_{e,min}^{p_{1}-1},\nonumber \\
N_{2} & \equiv\gamma_{e,break}^{p_{2}-p_{1}}.\nonumber 
\end{align}

As we showed in \eqref{transform_log_likelihood}, we need the Jacobian
of the transformation to properly calculate the log-likelihood of
the model parameters. This is given by:

\begin{align}
\left|\frac{dF}{dP}\right| & =4\gamma_{e,break}^{2}\left(p_{1}-1\right)\left(p_{2}-3\right).
\end{align}

The best SED that we can achieve is shown in \figref{Combined-SED}
where we can see that it fits quite well to the available data.

The corner plot of the fit can be seen in \figref{Spherical_Steady_State_Corner}.

\textcolor{red}{}
\begin{table*}
\begin{centering}
\begin{tabular}{ccccccccc}
\toprule 
\multirow{2}{*}{Parameter} & \multicolumn{2}{c}{Spherical Steady State} & \multicolumn{2}{c}{Disk Steady State} & \multicolumn{2}{c}{Sphere Injection} & \multicolumn{2}{c}{Disk Injection}\tabularnewline
 & Best & Median & Best & Median & Best & Median & Best & Median\tabularnewline
\midrule
\midrule 
$\log_{10}\left(\gamma_{e,min}\right)$ & $1.33$ & $1.96_{-0.65}^{+0.67}$ & $1.99$ & $2.07_{-0.73}^{+0.60}$ & $3.95$ & $4.04_{-0.20}^{+0.20}$ & $1.47$ & $1.73_{-0.51}^{+0.68}$\tabularnewline
\midrule 
$\log_{10}\left(\gamma_{e,break}\right)$ & $4.03$ & $4.10_{-0.24}^{+0.24}$ & $4.07$ & $4.13_{-0.25}^{+0.23}$ & $4.26$ & $4.40_{-0.21}^{+0.25}$ & $4.39$ & $4.31_{-0.24}^{+0.27}$\tabularnewline
\midrule 
$\log_{10}\left(\gamma_{e,max}\right)$ & $6.82$ & $7.89_{-1.48}^{+1.42}$ & $6.50$ & $7.83_{-1.42}^{+1.47}$ & $7.68$ & $7.29_{-1.00}^{+1.77}$ & $9.01$ & $7.99_{-1.38}^{+1.42}$\tabularnewline
\midrule 
$p_{1}$ & $1.69$ & $1.81_{-0.22}^{+0.23}$ & $1.71$ & $1.83_{-0.23}^{+0.25}$ & $1.22$ & $1.79_{-0.51}^{+1.00}$ & $1.82$ & $1.79_{-0.20}^{+0.17}$\tabularnewline
\midrule 
$p_{2}$ & $4.29$ & $4.29_{-0.08}^{+0.10}$ & $4.27$ & $4.29_{-0.08}^{+0.10}$ & $3.37$ & $3.43_{-0.12}^{+0.20}$ & $3.51$ & $3.55_{-0.14}^{+0.23}$\tabularnewline
\midrule 
$\log_{10}\left(B\right)$ & $-1.01$ & $-1.06_{-0.25}^{+0.24}$ & $-0.96$ & $-1.06_{-0.27}^{+0.27}$ & $-1.36$ & $-1.38_{-0.16}^{+0.18}$ & $-1.09$ & $-1.10_{-0.27}^{+0.25}$\tabularnewline
\midrule 
$\log_{10}\left(\delta\right)$ & $1.43$ & $1.37_{-0.37}^{+0.44}$ & $1.27$ & $1.32_{-0.36}^{+0.44}$ & $1.83$ & $1.71_{-0.17}^{+0.16}$ & $1.04$ & $1.22_{-0.35}^{+0.46}$\tabularnewline
\midrule 
\multirow{1}{*}{$\log_{10}\left(h\right)$} &  &  & $15.93$ & $13.49_{-2.14}^{+2.24}$ &  &  & $13.54$ & $13.69_{-0.72}^{+0.78}$\tabularnewline
\midrule 
\multirow{1}{*}{$\log_{10}\left(R\right)$} & $16.46$ & $16.62_{-0.81}^{+0.80}$ & $16.79$ & $16.82_{-0.83}^{+0.75}$ & $16.12$ & $16.43_{-0.53}^{+0.48}$ & $17.42$ & $17.08_{-0.82}^{+0.65}$\tabularnewline
\midrule 
$\log_{10}\left(\rho\right)$ & $-21.76$ & $-22.29_{-1.03}^{+0.98}$ & $-21.71$ & $-19.23_{-2.31}^{+2.15}$ &  &  &  & \tabularnewline
\midrule 
$\log_{10}\left(L\right)$ &  &  &  &  & $41.54$ & $41.82_{-0.39}^{+0.43}$ & $47.46$ & $46.37_{-1.87}^{+1.63}$\tabularnewline
\bottomrule
\end{tabular}
\par\end{centering}
\caption{\label{tab:Parameters_all_models}Parameters found for the default
fits. The number of points for each model is 309760, 496640, 116480
and 144640 respectively.}
\end{table*}

\subsubsection{Thin Disk Steady State}

The thin disk steady state model has $10$ parameters: $\left(\gamma_{e,min}\text{, }\gamma_{e,break}\text{, }\gamma_{e,max}\text{, }p_{1}\text{, }p_{2}\text{, }B\text{, }\delta\text{, }h\text{, }R\text{ and }\rho\right)$.
And again, we choose to redefine our parameters in the following way:
\begin{align}
F_{1} & \equiv VN_{1}\delta^{2}\nu_{br}^{2}\gamma_{e,break}^{-p_{1}-1},\nonumber \\
F_{2} & \equiv\left(\frac{\gamma_{e,max}}{\gamma_{e,break}}\right)^{3-p_{2}},\nonumber \\
F_{3} & \equiv\frac{hN_{1}}{\delta B}\gamma_{e,break}^{-p1},\nonumber \\
\delta & \equiv\delta,\nonumber \\
\delta B & \equiv\delta B,\\
\nu_{br} & \equiv\delta B\gamma_{e,break}^{2},\nonumber \\
\rho & \equiv\rho,\nonumber \\
N_{1} & \equiv\rho\left(p_{1}-1\right)\gamma_{e,min}^{p_{1}-1},\nonumber \\
hN_{1} & \equiv hN_{1},\nonumber \\
N_{2} & \equiv\gamma_{e,break}^{p_{2}-p_{1}}.\nonumber 
\end{align}

Again, we need the Jacobian of the transformation, which is given
by:

\begin{align}
\left|\frac{dF}{dP}\right| & =4\gamma_{e,break}^{2}\left(p_{1}-1\right)\left(p_{2}-3\right).
\end{align}

\subsubsection{Spherical Injection}

The spherical injection model has $9$ parameters: $\left(\gamma_{e,min}\text{, }\gamma_{e,break}\text{, }\gamma_{e,max}\text{, }p_{1}\text{, }p_{2}\text{, }B\text{, }\delta\text{, }R\text{ and }L\right)$.
Unfortunately, due to \eqref{pop_inj_cooling}, the parameter that
naturally would take the place of $N_{1}$ in the steady state models
depends on the value of $S$, which is not known a priori. Even if
it were known, it separates two different behaviours for the particles
in two energetic regimes. This makes obtaining simple relations for
the synchrotron, x-ray and gamma-ray fluxes a not trivial problem
that, although solvable, does not help the fitting. Because of this,
we have chosen to not redefine our parameters and fit directly the
model parameters.

\subsubsection{Thin Disk Injection}

The spherical injection model has $10$ parameters: $\left(\gamma_{e,min}\text{, }\gamma_{e,break}\text{, }\gamma_{e,max}\text{, }p_{1}\text{, }p_{2}\text{, }B\text{, }\delta\text{, }h\text{, }R\text{ and }L\right)$.
Again, for the same reason as for the case of spherical injection,
we choose to not redefine our parameters.

\subsection{First Interpretation of Results \label{subsec:Interpretation}}

\begin{figure*}
\begin{centering}
\includegraphics[width=0.25\textwidth]{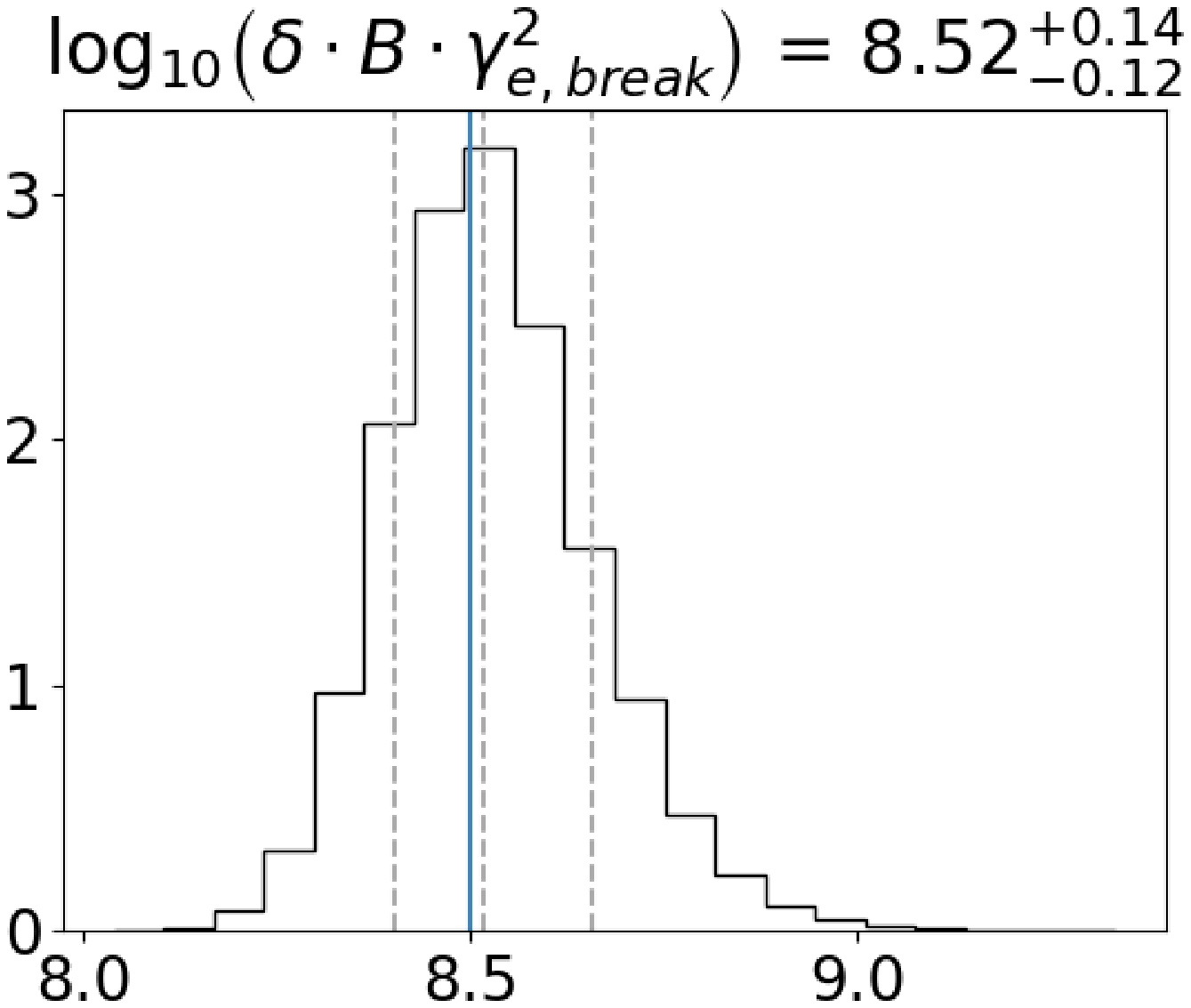}\includegraphics[width=0.25\textwidth]{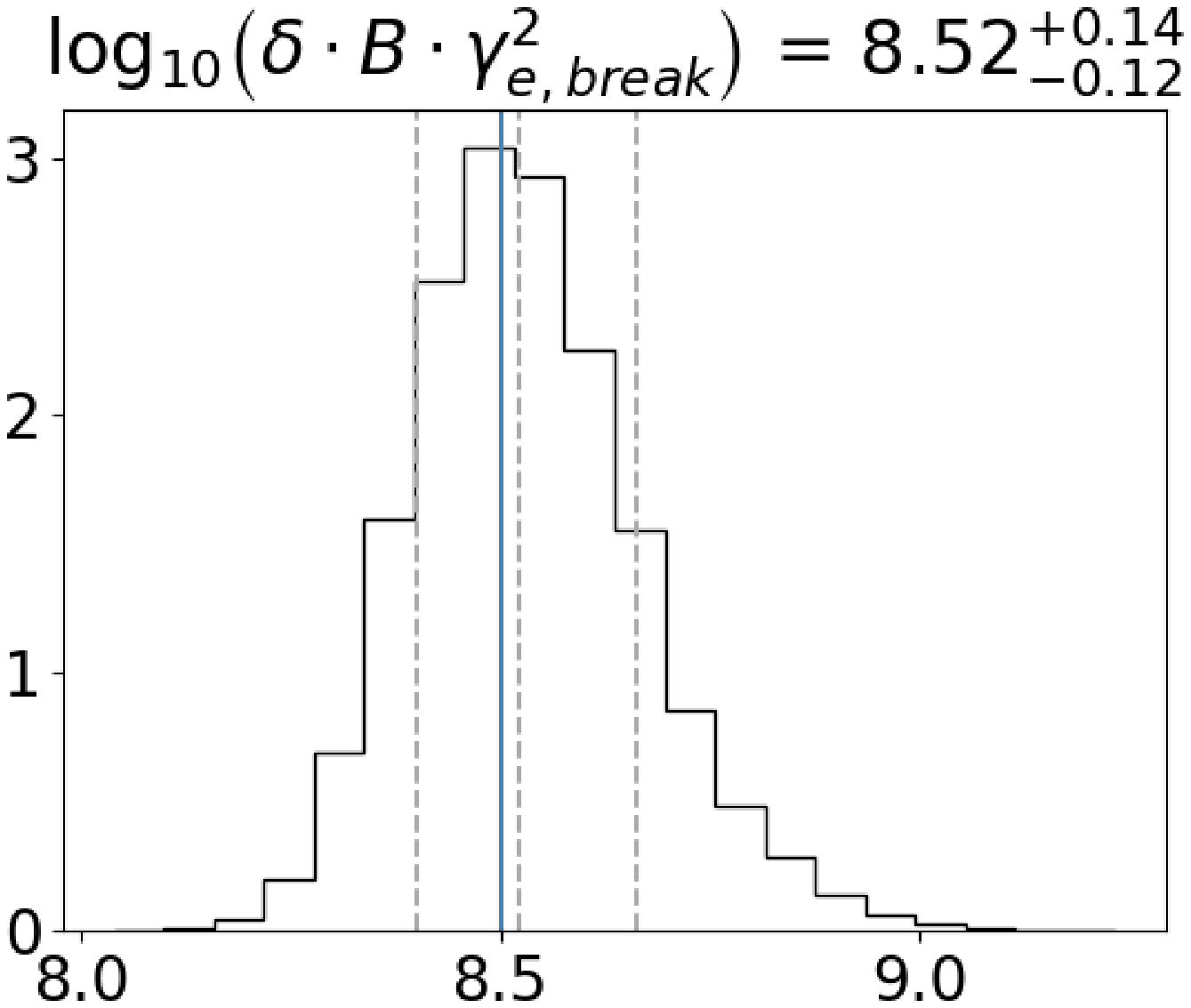}\includegraphics[width=0.25\textwidth]{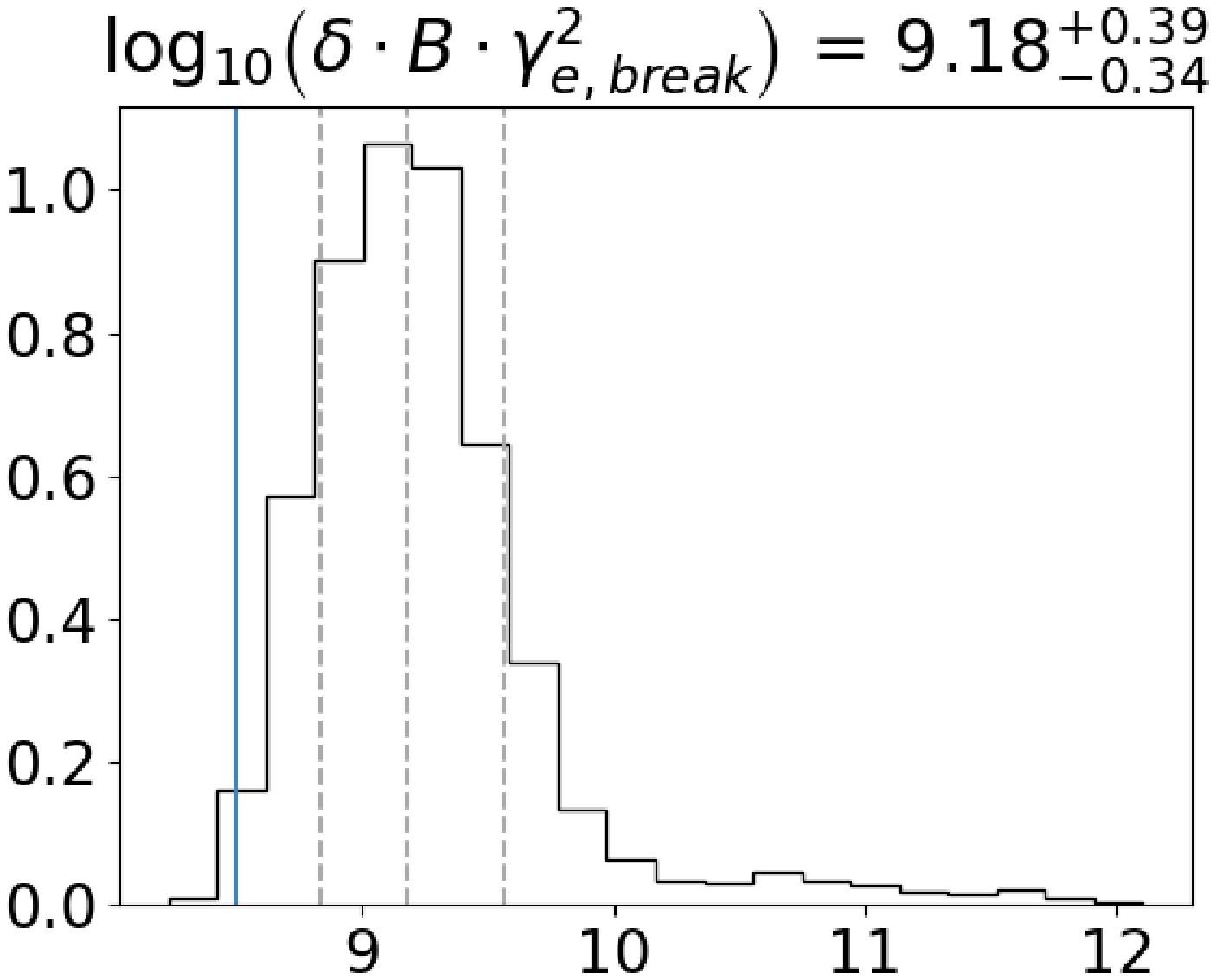}\includegraphics[width=0.25\textwidth]{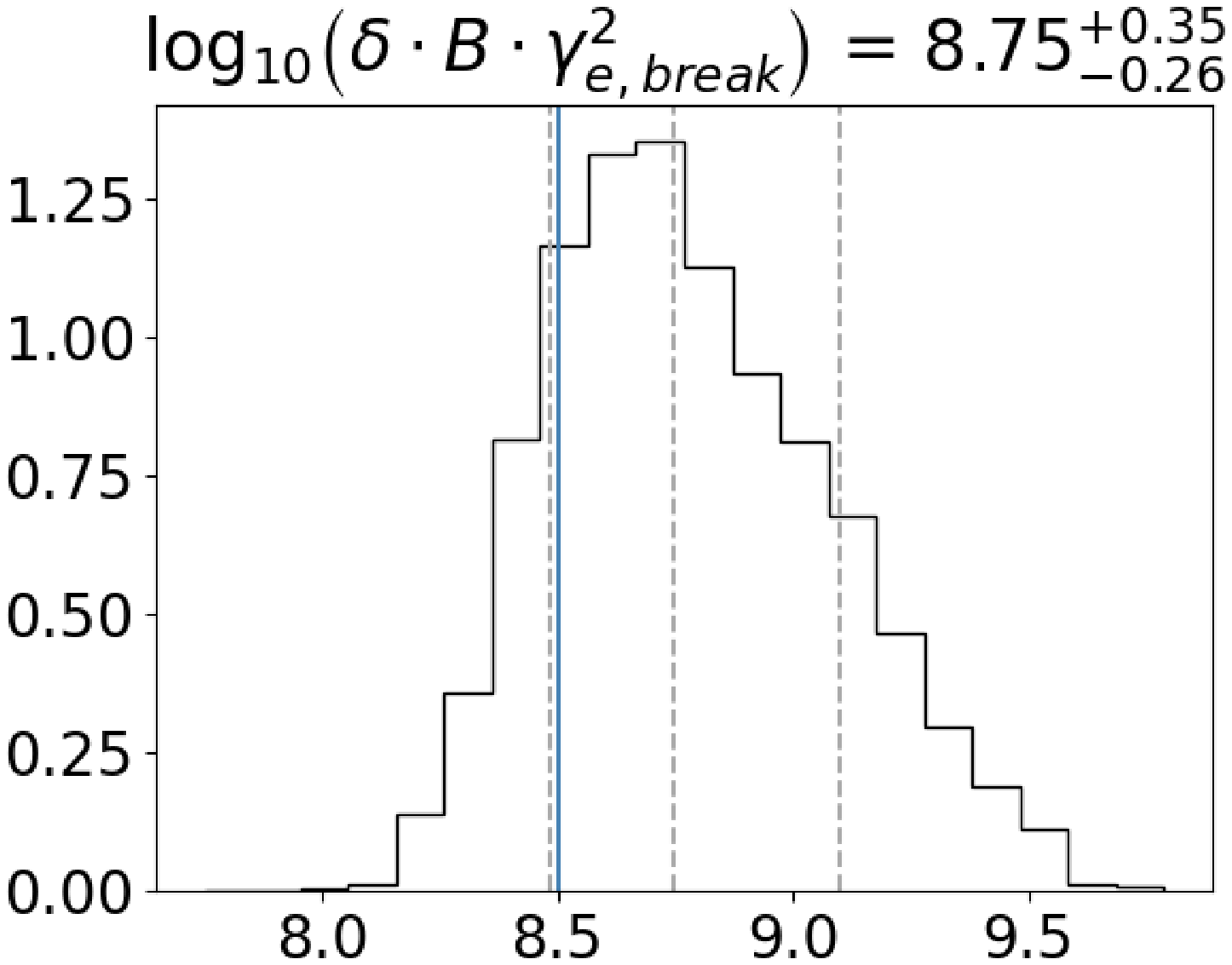}
\par\end{centering}
\caption{\label{fig:Comparison-with-nu_br}Comparison with the estimates for
$\log_{10}\left(\delta B\gamma_{e,break}^{2}\right)$, blue vertical
line. From left to right: sphere steady state, disk steady state,
sphere injection, disk injection}
\end{figure*}

First, we check the validity of some of our initial estimates for
the initial fitted parameters. For the value of $\log_{10}\left(\delta B\gamma_{e,break}^{2}\right)$
we had estimated $8.50$. \Figref{Comparison-with-nu_br} reflects
that, with the exception of the sphere injection model, where we obtain
a value $\sim3$ times bigger, our estimate was correct. A possible
explanation for this minor discrepancy lies in \eqref{pop_inj_cooling}.
This equation sets the energy which separates the regimes where particles
mainly escape from the regime where particles mainly cool. Spherical
models have typically higher values of $R$ than discs for $h$, which
in turns means higher values for $\tau_{esc}$. Then, everything else
being equal, we know that $\gamma_{e,break,sphere}<\gamma_{e,break,disk}$.

\bigskip{}

\begin{figure*}
\begin{centering}
\includegraphics[width=0.25\textwidth]{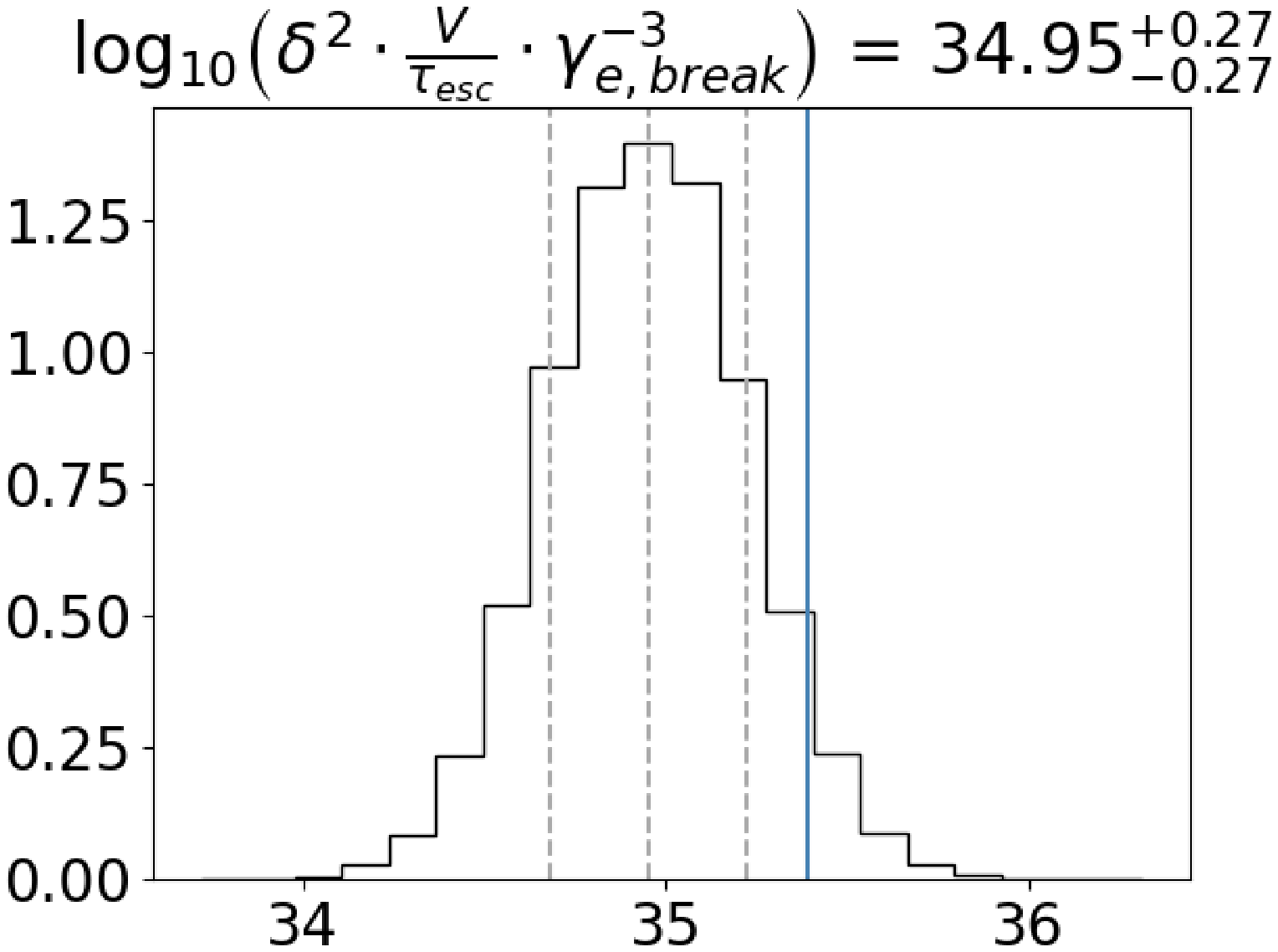}\includegraphics[width=0.25\textwidth]{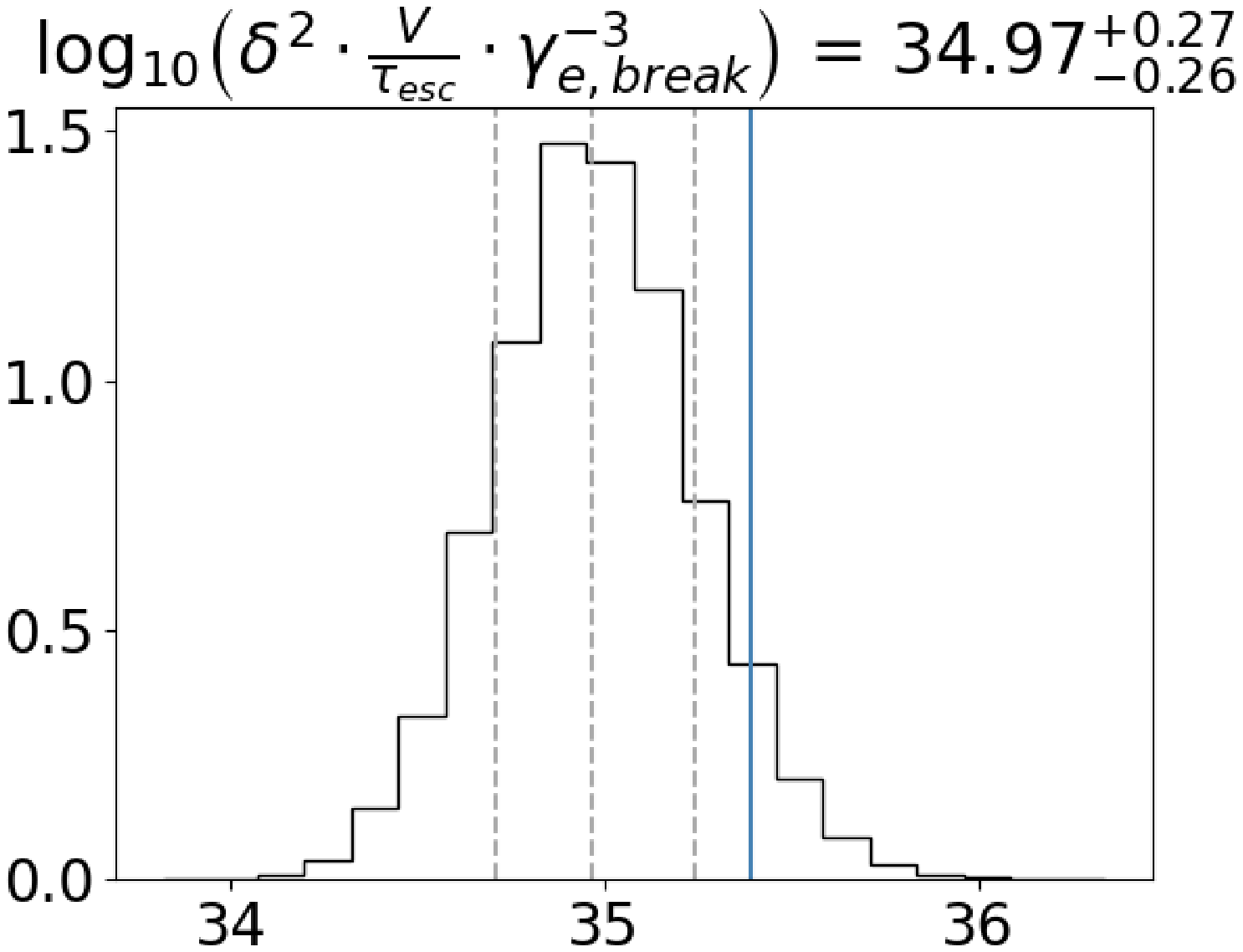}\includegraphics[width=0.25\textwidth]{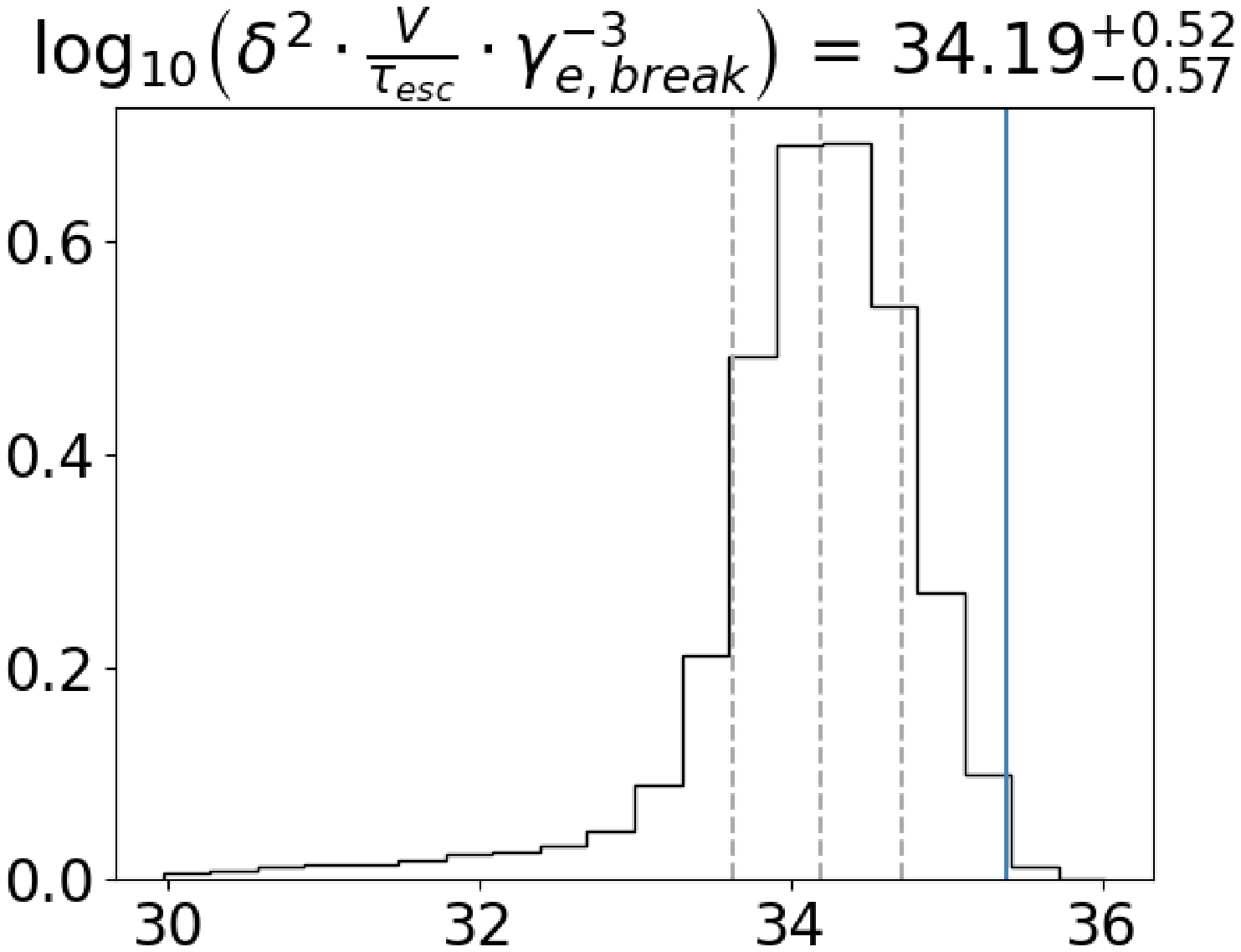}\includegraphics[width=0.25\textwidth]{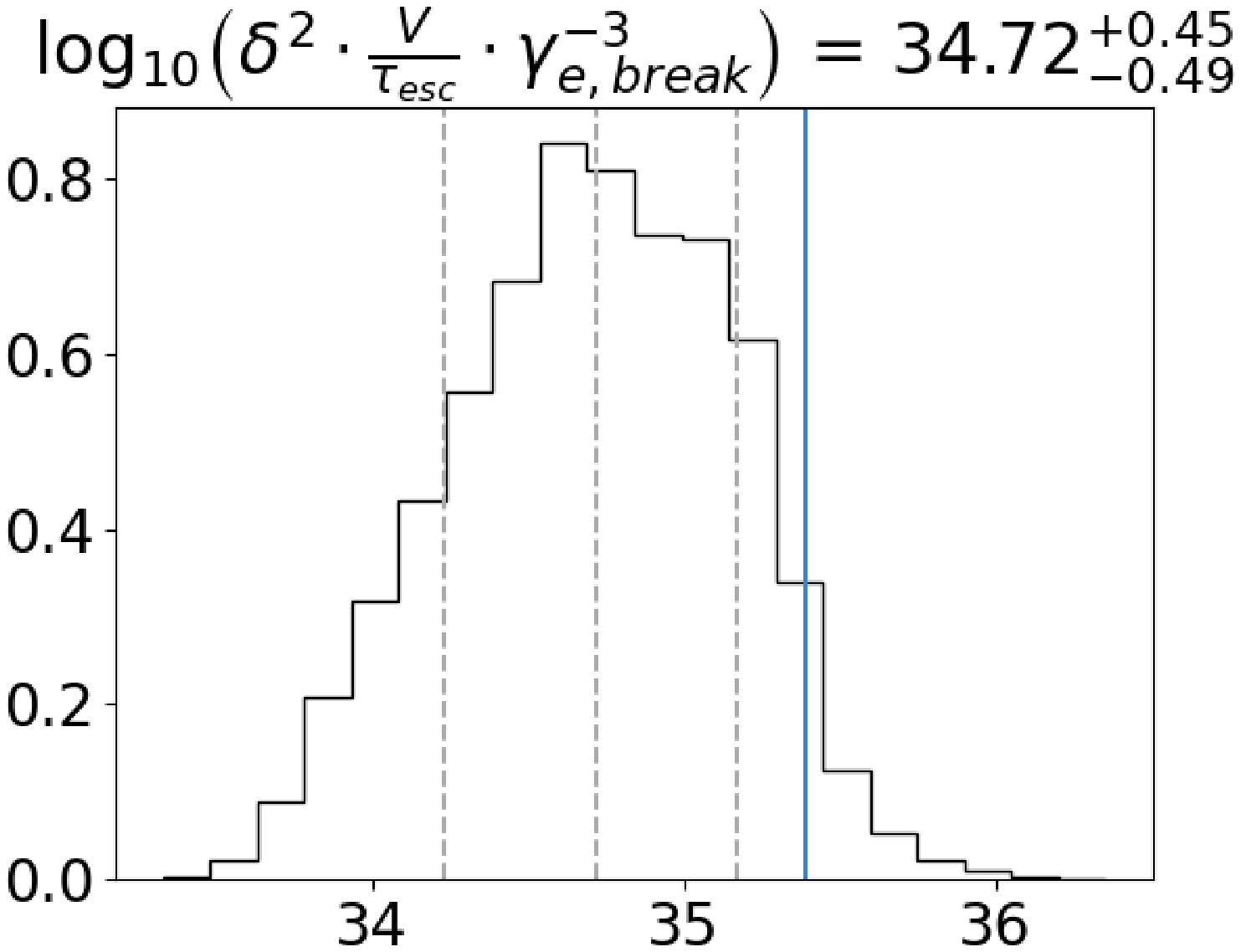}
\par\end{centering}
\caption{\label{fig:Comparison-with-thing}Comparison with the initial estimates
for $\log_{10}\left(\delta^{2}\frac{V}{\tau_{esc}}\gamma_{e,break}^{-3}\right)$,
blue vertical line. From left to right: sphere steady state, disk
steady state, sphere injection, disk injection}
\end{figure*}

Second, we check the validity of our initial estimate for $\log_{10}\left(\delta^{2}\frac{V}{\tau_{esc}}\gamma_{e,break}^{-3}\right)$,
where we had obtained a value of $35.39$. Note that the factor $V/\tau_{esc}$
depends only on $R^{2}$ regardless of the geometry. \Figref{Comparison-with-thing}
shows that we systematically obtain results below the estimate by
a factor $\sim3$. Part of this discrepancy can be explained by looking
at the gamma-ray peak of \figref{Combined-SED}. We can see that our
model obtains a gamma-ray flux a bit above the data points, although
well within the error bars. So, the modelled flux is above that that
we used for the estimate, and thus its value should be reduced. The
other part comes from the physics used in the estimation, which is
based on a crude approximation of inverse Compton upscattering.

\bigskip{}

\begin{figure}
\begin{centering}
\includegraphics[width=0.25\textwidth]{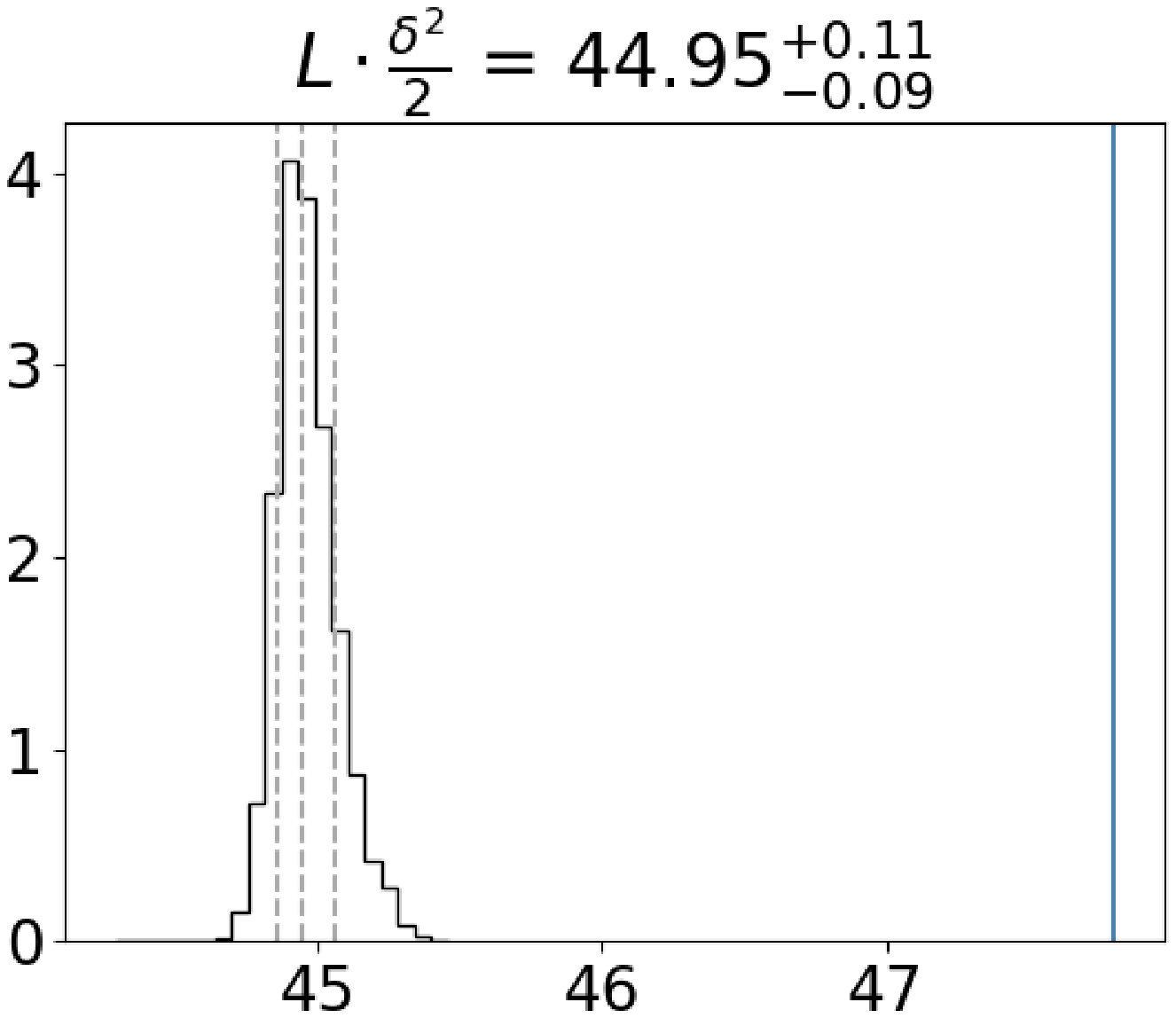}\includegraphics[width=0.25\textwidth]{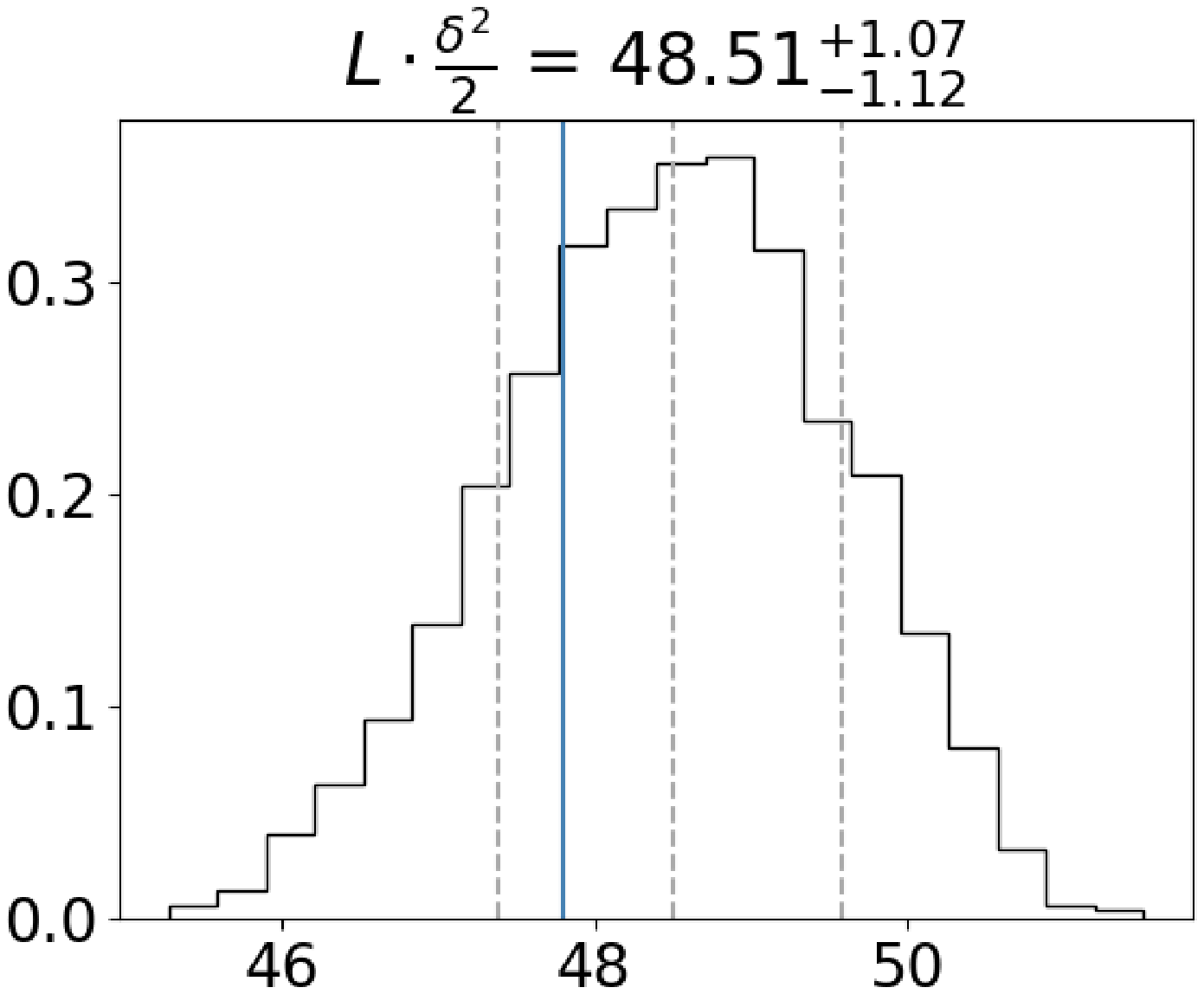}
\par\end{centering}
\caption{\label{fig:Comparison-with-L}Comparison with \citet{gao_modelling_2019}'s
estimate for the Eddington luminosity of TXS 0506+056 assuming that
the central black hole has a mass similar to that of M87. Left: sphere
injection, right: disk injection}
\end{figure}

\citet{gao_modelling_2019} give an estimate for the Eddington luminosity
of the black hole at the center of TXS 0506+056 assuming a mass similar
to that of M87: $10^{47.8}\,\text{erg}\cdot\text{s}^{-1}$. As one
of our parameters for the injection models is precisely injected luminosity,
we can directly compare with this value. \Figref{Comparison-with-L}
shows that the distribution of values we obtain for spherical injection
are well below the estimated one, whereas the values for disk injection
are above it.

The cause for the discrepancy is due to the different behaviour of
the particles with low energies in the sphere and disk cases. For
the Sphere, the escape timescale is much larger than the cooling timescale,
so the high-energy particles have enough time to cool before escaping
from the volume and can accumulate, producing a sizable population
at low energy. By contrast, the escape timescale for disks are much
lower, which allows the low-energy electrons to escape before having
enough time to cool.

This is confirmed by the steady-state models that show the need for
a low-energy population of electrons through their low values for
$\gamma_{e,min}$ ($\log_{10}$ of $1.96$ for the sphere steady-state
model, versus $4.04$ for injection). As mentioned above, although
the sphere injection model can obtain a low-energy population through
cooling, the disk injection model needs to include this in the injection
directly ($\log_{10}\gamma_{e,min}$ of $1.73$ for disk injection,
versus the aforementioned $4.04$ for disk injection).

From the results of the spherical steady state model, we can integrate
the amount of particles between $\gamma_{e,min}$ and $\gamma_{e,break}$
to find that about $98\%$ of the electrons are in this range of energies.
By using \eqref{L_related_with_rho} we can calculate an rough estimate
of the luminosity that would be required to obtain a population of
electrons similar to that of the steady state case in a pure injection
and escape case, obtaining a value of $L_{estimated}\sim10^{46}$,
which is in line with the estimate for the Eddington Luminosity (after
a frame transformation).

\subsection{Interpretation of Posterior Distribution Results and Cross-Correlations}{\label{subsec:Interpretation-of-Posterior}}

As discussed in \ref{sec:Model-Fitting} when we presented the procedure
of fitting with MCMC, one of the products that we obtain is the posterior
density function for each of the parameters. And that, in turn, can
provide interesting information, for example about how individual
parameters impact a model or whether or not bimodalities occur across
the range of good fits. One example of the former are the posterior
density functions for $\gamma_{e,max}$. We have not found examples
of bimodal distributions.

The four $\gamma_{e,max}$ posterior distributions clearly show a
lower cut off at low energies and a flat shape at the rest of the
energies in the explored parameter space. The lower cut-off can be
explained by the constraints of \eqref[s]{gamma_max_constraint} and
(\ref{eq:gamma_max_constraint_xray}), whereas the flat shape can
be explained by realizing that our model is not constraining at this
range. In fact, an increase of $\gamma_{e,max}$ requires a number
of particles so small that it can barely modify $\rho$ or $L$.

\bigskip{}

Another of the products of using MCMC are the correlation plots between
the parameters, where we can clearly see the relations between the
parameters and how changing one affects others. Good examples are
the triad $\gamma_{e,break}\text{, }\delta\text{, }R$ or the correlation
between $\rho$ and $R$ or $h$ depending on the geometry of the
emitting volume for steady state models.

The correlations between $\gamma_{e,break}\text{, }\delta\text{ and }R$
can be explained by \eqref{delta_g_break_R}. $V/\tau_{esc}$ is a
function of $R^{2}$ regardless of the geometry of the model, which
is the cause of the cross-correlations.

The correlation between $\rho$ and $R$ or $h$ involves the escape
timescale $\tau_{esc}$. As stated in the previous paragraph, $V/\tau_{esc}$
is only a function of $R^{2}$, which means that when transforming
from the number of photons in the blob to the received flux the only
geometric factor is $R^{2}$. But, from our exploration of both synchrotron
emission and inverse Compton emission, we found that the population
of photons in the rest frame of the fluid depends on $\tau_{esc}$.
This means that the population depends directly on $R$ or $h$ depending
on whether the geometry is a sphere or a disk.

For a posterior distribution of $h$ peaking around $10^{13}$ cm,
we note that the thin disk model is in tension with the assumption
that the plasma is able to contain the upper high-energy part at $\gamma\sim10^{6}$
of a population of protons. At magnetic fields $B$ of $0.1$ G, their
gyroradius becomes of an order comparable to $h$, leaving little
room for diffusion (accounted for through our $\sigma$ parameter).
Sphere models are therefore arguably preferable to disk models for
explaining a high-energy proton population responsible for a measurable
neutrino flux.

\bigskip{}

\begin{figure}
\begin{centering}
\includegraphics[width=0.45\textwidth]{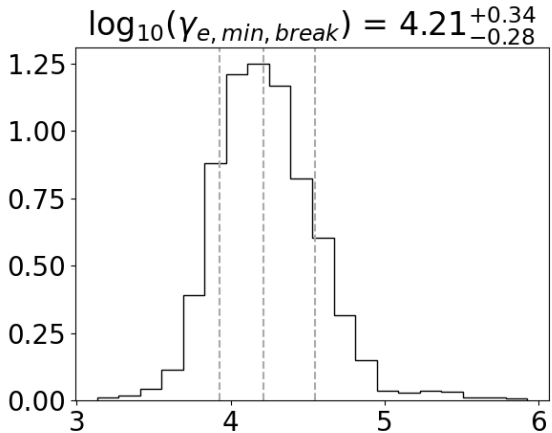}
\par\end{centering}
\caption{\label{fig:SI_g_min_break}Probability density function for $\gamma_{e,min}$
and $\gamma_{e,break}$ together for the Spherical Injection model}
\end{figure}

A more detailed analysis of the posterior density functions for the
case of the spherical injection model reveals something interesting.
In the case of the distribution for $p_{1}$, we find that its distribution
is much more extended than the ones for the other three models. Moreover,
looking at the distributions for $\gamma_{e,min}$ and $\gamma_{e,break}$
we find that their ranges overlap and, as can be seen in \figref{SI_g_min_break},
if we plot the distribution for both combined we find that we obtain
a single peak. We infer that the first part of the broken power law
is not relevant for the fit, as it is very narrow and its slope can
vary wildly.

To check the validity of this reasoning. We have chosen to also test
the case of the spherical injection but with a simple power law.

\subsubsection{Spherical Injection with a Power Law}

The spherical injection with a power law model has $7$ parameters:
$\left(\gamma_{e,min}\text{, }\gamma_{e,max}\text{, }p\text{, }B\text{, }\delta\text{, }R\text{ and }L\right)$
where we again have chosen to not redefine them.

\begin{figure*}
\begin{centering}
\includegraphics[width=1\textwidth]{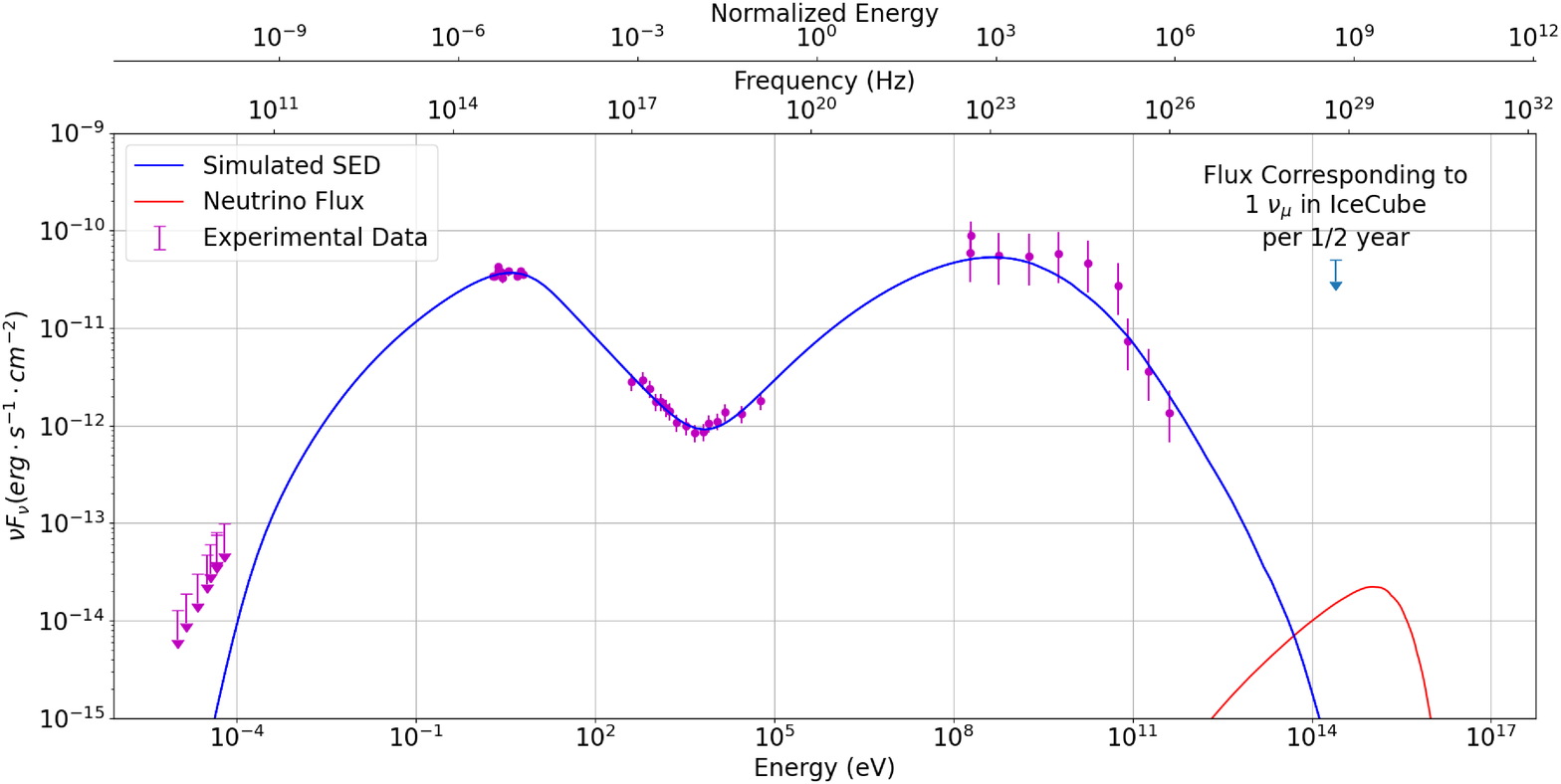}
\par\end{centering}
\caption{\label{fig:best_SI_PL}SED resulting from the best parameters for
the Sphere Injection with a Power Law.}
\end{figure*}

\textcolor{red}{}
\begin{table*}
\begin{centering}
\begin{tabular}{ccccccc}
\toprule 
\multirow{3}{*}{Parameter} & \multicolumn{2}{c}{Spherical Injection} & \multicolumn{2}{c}{Spherical Injection} & \multicolumn{2}{c}{Spherical Injection}\tabularnewline
 &  &  & \multicolumn{2}{c}{Simple Power Law} & \multicolumn{2}{c}{Simple Power Law varying $\eta$}\tabularnewline
 & Best & Median & Best & Median & Best & Median\tabularnewline
\midrule 
$\log_{10}\left(\gamma_{e,min}\right)$ & $3.95$ & $4.04_{-0.20}^{+0.20}$ & $4.04$ & $4.12_{-0.22}^{+0.21}$ & $4.03$ & $4.04_{-0.19}^{+0.19}$\tabularnewline
\midrule 
$\log_{10}\left(\gamma_{e,break}\right)$ & $4.26$ & $4.40_{-0.21}^{+0.25}$ &  &  &  & \tabularnewline
\midrule 
$\log_{10}\left(\gamma_{e.max}\right)$ & $7.68$ & $7.29_{-1.00}^{+1.77}$ & $7.31$ & $7.83_{-1.37}^{+1.43}$ & $7.53$ & $7.69_{-1.51}^{+1.55}$\tabularnewline
\midrule 
$p_{1}$ & $1.22$ & $1.79_{-0.51}^{+1.00}$ &  &  &  & \tabularnewline
\midrule 
$p_{2}$ & $3.37$ & $3.43_{-0.12}^{+0.20}$ & $3.31$ & $3.35_{-0.10}^{+0.14}$ & $3.37$ & $3.40_{-0.11}^{+0.18}$\tabularnewline
\midrule 
$\log_{10}\left(B\right)$ & $-1.36$ & $-1.38_{-0.16}^{+0.18}$ & $-1.32$ & $-1.36_{-0.19}^{+0.20}$ & $-1.33$ & $-1.30_{-0.20}^{+0.19}$\tabularnewline
\midrule 
$\log_{10}\left(\delta\right)$ & $1.83$ & $1.71_{-0.17}^{+0.16}$ & $1.82$ & $1.74_{-0.22}^{+0.21}$ & $1.85$ & $1.85_{-0.17}^{+0.16}$\tabularnewline
\midrule 
$\log_{10}\left(R\right)$ & $16.12$ & $16.43_{-0.53}^{+0.48}$ & $16.09$ & $16.32_{-0.66}^{+0.65}$ & $16.03$ & $16.00_{-0.53}^{+0.55}$\tabularnewline
\midrule 
$\log_{10}\left(\eta\right)$ &  &  &  &  & $3.53$ & $3.48_{-0.28}^{+0.26}$\tabularnewline
\midrule 
$\log_{10}\left(L\right)$ & $41.54$ & $41.82_{-0.39}^{+0.43}$ & $41.53$ & $41.75_{-0.50}^{+0.58}$ & $44.97$ & $44.91_{-0.61}^{+0.64}$\tabularnewline
\bottomrule
\end{tabular}
\par\end{centering}
\textcolor{red}{\caption{\label{tab:Spherical_Injection_variations_Parameters}Parameters of
the Sphere Injection with a Power Law Model and a the same case allowing
the ratio of injected protons to electrons $\left(\eta\right)$ to
change, with 234240 and 168960 points, respectively. The leftmost
spherical injection columns repeat the base case of Spherical Injection
for comparison purposes.}
}
\end{table*}

\Tabref{Spherical_Injection_variations_Parameters} (central two columns)
corresponds to the parameter values that we have found, \figref{best_SI_PL}
shows the SED corresponding to the best parameters, and \figref{Spherical_Injection_PL_Corner}
shows the corresponding corner plot of the model.

As we can see in \figref{best_SI_PL}, the fit that we obtain is very
similar to that of the broken power law model. We conclude that a
simple power law is indeed sufficient to reproduce the observed SED
for an injection model in spherical geometry.

\subsection{Large proton to electron ratio injection}{\label{subsec:Changing-eta}}

\Figref{Combined-SED} shows the comparison between the SED we obtained
with the four models and the observed fluxes. For $\eta=1$ (equal
injection numbers for protons and electrons), the SED can be produced
very well, but the neutrino flux remains orders of magnitude below
that allowed by the tentative neutrino detection. Below we discuss
two mechanisms to increase the neutrino flux.

One option is to shift the value of gamma\_p,max. As discussed in
\subsecref{Other-Settings-and-Results}, a value of $10^{6}$ is enough
to produce neutrinos with the appropriate energy through more than
one channel. Nevertheless, higher energies allow for a greater energy
range for protons, pions and muons to produce neutrinos with $\sim290$
TeV. Additionally, an increase of the maximum energy for protons also
increases the total number of protons that have energies above the
threshold to produce neutrinos at $290$ TeV, thus leading to an increase
of the total neutrino flux. For example \citet{keivani_multimessenger_2018}
uses $\gamma_{p,max}=1.6\times10^{6}-3\times10^{9}$ and \citet{Cerruti_2018}
has $\gamma_{p,max}=6\times10^{7}-2.5\times10^{9}$.

The second option is to increase the ratio of protons to electrons,
both in the steady state population and in the injected population.
The main consequence that a higher value of $\eta$ would have is
an increase of the number of protons at all energies, including those
with the appropriate one to produce neutrinos. Thus, we will obtain
a higher neutrino flux at all energies. For example \citet{gao_modelling_2019}
uses $\eta\approx8606$ (as we show in Appendix \ref{sec:reproducing_gao})
and \citet{Cerruti_2018} has $\eta\approx200$ Appendix \ref{sec:reproducing_cerruti}.

\begin{figure*}
\begin{centering}
\includegraphics[width=1\textwidth]{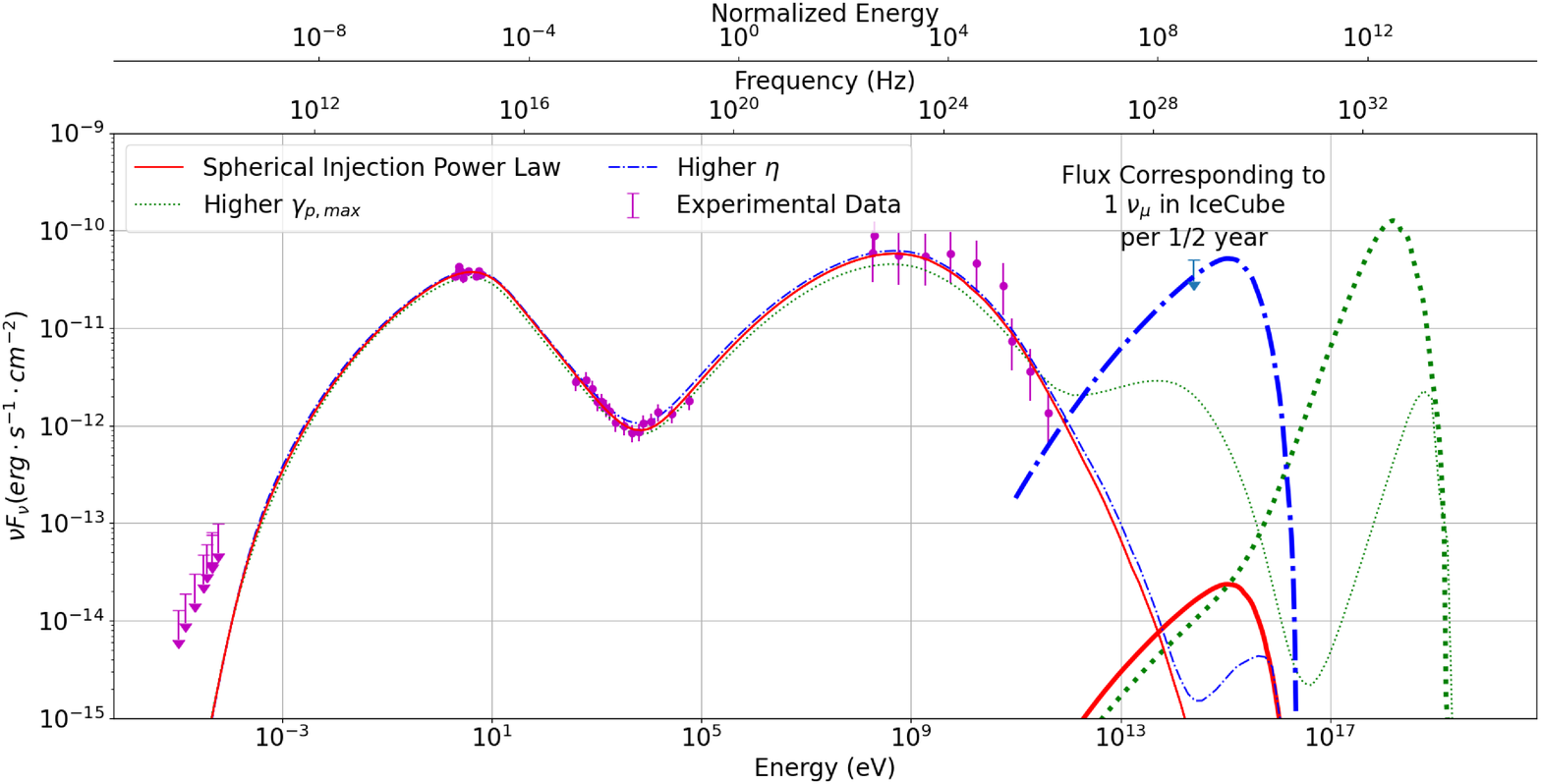}
\par\end{centering}
\caption{\label{fig:Combined-Neutrino-tests}Combined neutrino tests. The red
line corresponds to the reference of the best result for the spherical
injection with a simple power law model, the green dots correspond
to the same model but increasing $\gamma_{p,max}$ to $10^{9}$ and
the blue dot dashed line corresponds to the same model but increasing
$\eta$ by $1750$. Thick lines correspond to the predicted neutrino
flux.}
\end{figure*}

\Figref{Combined-Neutrino-tests} shows a simple test of these two
hypotheses with our model. As a base we have taken the best parameters
for the spherical injection model. Then, to check the first hypothesis
we have increased the value of $\gamma_{p,max}$ to $10^{9}$ and
to check the second we have increased the value of $\eta$ by a factor
1750 and modified the value of $L$ according to \eqref{Q_from_L}
so that the value of injected electrons is the same in the two models.
As we predicted in the previous paragraphs, in the first case we see
an increase in the total flux of neutrinos, specially those with energies
higher than $500$ TeV, whereas in the second case the increase is
restricted only to those of energies around $290$ TeV.

Following \figref{Combined-Neutrino-tests}, and combined with the
simplicity of the spherical injection with a single power law model,
we choose to go further in our modelling and allow the value of $\eta$
to vary. The neutrino flux is an upper limit, so we do not have an
actual data point to compare our model results, but for illustrative
purposes, we have chosen to act as if the upper limit was an actual
detection and use its value in the calculation of the log-likelihood.

The model that we fit has 8 parameters: $\left(\gamma_{e,min}\text{, }\gamma_{e,max}\text{, }p\text{, }B\text{, }\delta\text{, }R\text{, }\eta\text{ and }L\right)$

\begin{figure*}
\begin{centering}
\includegraphics[width=1\textwidth]{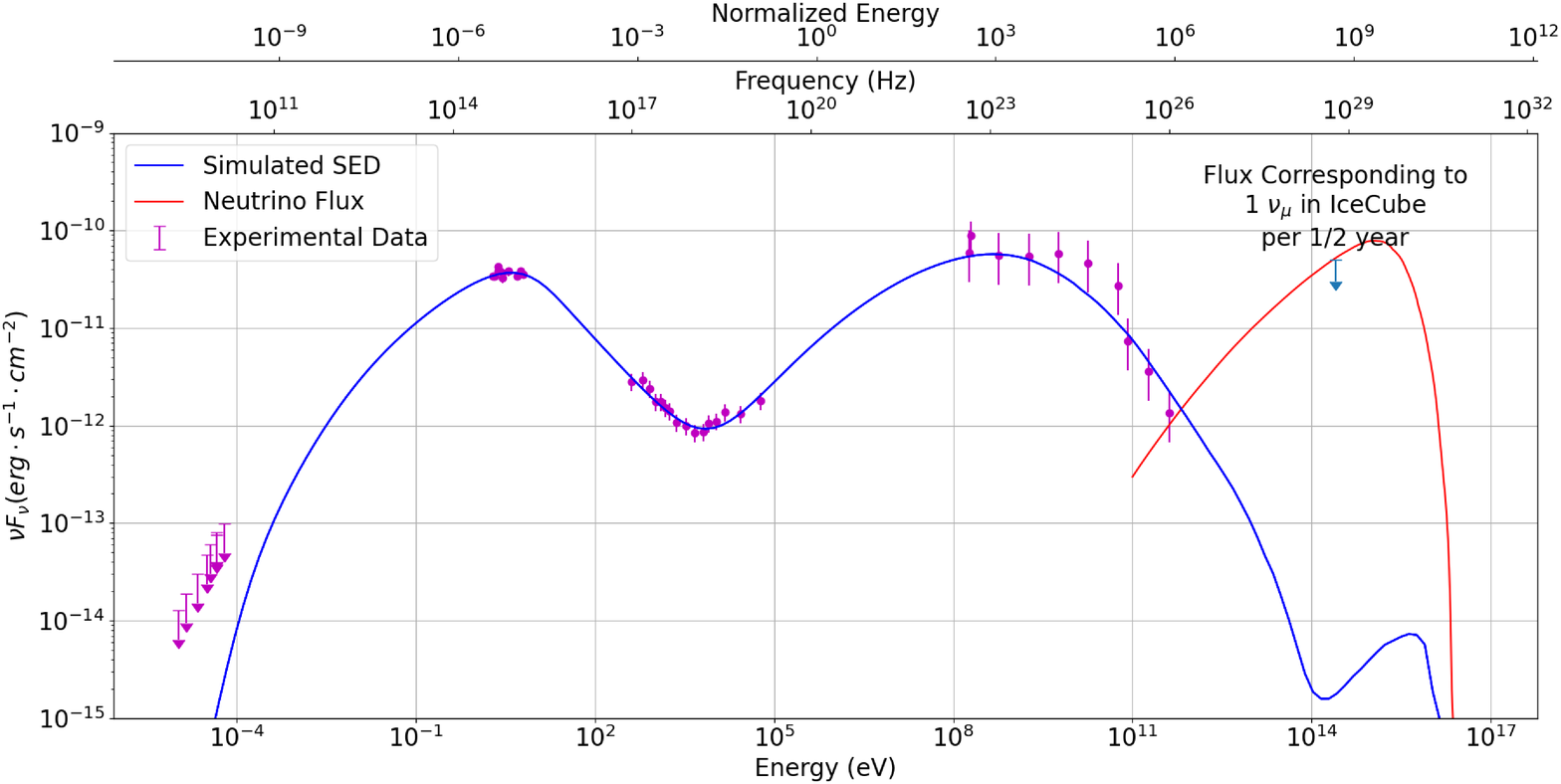}
\par\end{centering}
\caption{\label{fig:Best_eta}SED for the best parameters of the Spherical
Injection with a Power Law allowing $\eta$ to vary.}
\end{figure*}

\begin{figure}
\begin{centering}
\includegraphics[width=0.45\textwidth]{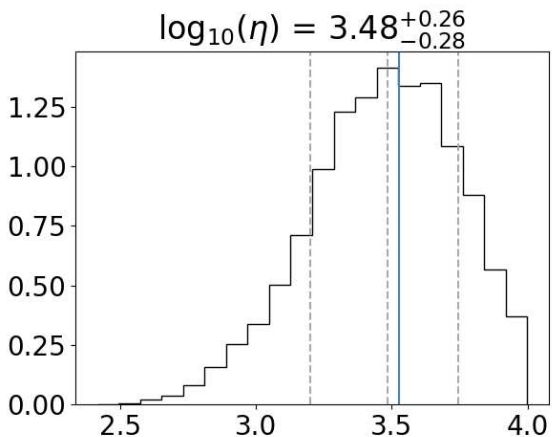}
\par\end{centering}
\caption{\label{fig:eta_histogram}Probability density function for $\eta$.}
\end{figure}

\Tabref{Spherical_Injection_variations_Parameters} (two rightmost
columns) corresponds to the parameter values that we have found, \figref{Best_eta}
shows the fit that we obtain with the best parameters and \figref{Spherical_Injection_PL_eta_Corner}
shows the corresponding corner plot of the model. \Figref{eta_histogram}
plots the resulting probability density function for $\eta$, showing
a broad distribution peaking around $\log_{10}\eta=3.48$. Below $\eta$
of $\sim10^{3}$, the neutrino flux will be underpredicted, whereas
for eta approaching $10^{4}$ it becomes difficult to avoid overpredicting
the X-ray flux in the SED valley.

It can be seen from \figref{Best_eta} that both the SED and the neutrino
flux can be fitted at the same time without overwhelming the x-ray
flux. But the best-fitting curve in the figure uses $\eta\approx3388$
(i.e. the value indicated by a blue line in \figref{eta_histogram})
for the ratio of protons to electrons, which implies a discrepancy
of orders of magnitude in the numbers of accelerated protons and electrons
injected in the plasma.

\section{Summary and Conclusions}{\label{sec:Conclusions}}

We present a detailed kinetic lepto-hadronic emission model capable
of producing photon and neutrino spectra for a range of parameters,
plasma geometries and electrons and protons either injected or at
steady state. This model has been developed independently of other
approaches (e.g. \citet{cerruti_hadronic_2015} or \citet{gao_direct_2017}),
and can be used as a cross-check on their work. We have coupled our
model to a Markov-Chain Monte Carlo method, and discuss the characteristics
of the parameter space that this allows us to probe in a novel manner.
The method is applied to the multi-messenger case of TXS 0506+056.

MCMC sampling reveals the full probability distribution of the fit
parameters as well as their cross-correlations and degeneracies. We
demonstrate that the six iterations of our model (between sphere and
disk geometry, steady state and injection, simple power-law and varying
$\eta$) are all able to fit TXS 0506+056. The MCMC method shows that
a single power law injection of electrons is sufficient in spherical
geometry, that $\gamma_{e,max}$ is poorly constrained and that some
of the fitted variables are correlated.

To reproduce a high level of neutrino flux comparable to the peak
value allowed by the tentative detection by IceCube requires a large
proton to electron ratio. Our MCMC method quantifies the probability
distribution of this ratio eta, and we find $\log_{10}\eta=3.48_{-0.28}^{+0.26}$.
If the accelerated protons and electrons are not taken to be from
different origins altogether, this poses a challenge to models of
charged particle acceleration to explain the origin of an overwhelmingly
larger number of accelerated protons than electrons.

\section*{Acknowledgements}

The authors would like to thank Matteo Cerruti and Anatoli Fedynitch
for the valuable discussions about their models and Petar Mimica for
hosting them during their meetings at the University of Valencia.
B.J.F. would also like to thank the late Cosmo the Cat for his support.
B.J.F. acknowledges funding from STFC through the joint STFC-IAC scholarship
program. The authors gratefully acknowledge the anonymous referee
for a constructive and thorough process that has significantly helped
to improve the quality of the paper.

\section*{Data availability}

The observational data used in this paper was taken from \citet{eaat1378}.
All data produced in this analysis is available upon request. The
software used in this study is available upon request, but will be
made open-source in the near future (Jimenez-Fernandez \& van Eerten,
in prep).

\bibliographystyle{mnras}
\bibliography{bibliography/biblio}

\begin{thebibliography}{}
\makeatletter
\relax
\def\mn@urlcharsother{\let\do\@makeother \do\$\do\&\do\#\do\^\do\_\do\%\do\~}
\def\mn@doi{\begingroup\mn@urlcharsother \@ifnextchar [ {\mn@doi@}
  {\mn@doi@[]}}
\def\mn@doi@[#1]#2{\def\@tempa{#1}\ifx\@tempa\@empty \href
  {http://dx.doi.org/#2} {doi:#2}\else \href {http://dx.doi.org/#2} {#1}\fi
  \endgroup}
\def\mn@eprint#1#2{\mn@eprint@#1:#2::\@nil}
\def\mn@eprint@arXiv#1{\href {http://arxiv.org/abs/#1} {{\tt arXiv:#1}}}
\def\mn@eprint@dblp#1{\href {http://dblp.uni-trier.de/rec/bibtex/#1.xml}
  {dblp:#1}}
\def\mn@eprint@#1:#2:#3:#4\@nil{\def\@tempa {#1}\def\@tempb {#2}\def\@tempc
  {#3}\ifx \@tempc \@empty \let \@tempc \@tempb \let \@tempb \@tempa \fi \ifx
  \@tempb \@empty \def\@tempb {arXiv}\fi \@ifundefined
  {mn@eprint@\@tempb}{\@tempb:\@tempc}{\expandafter \expandafter \csname
  mn@eprint@\@tempb\endcsname \expandafter{\@tempc}}}

\bibitem[\protect\citeauthoryear{B\"ottcher}{B\"ottcher}{2019}]{boettcher_progress_2019}
B\"ottcher M.,  2019, arXiv:1901.04178 [astro-ph]

\bibitem[\protect\citeauthoryear{B\"ottcher \& Baring}{B\"ottcher \&
  Baring}{2019}]{boettcher_spectral_2019}
B\"ottcher M.,  Baring M.~G.,  2019, arXiv:1903.12381 [astro-ph]

\bibitem[\protect\citeauthoryear{B\"ottcher \& Dermer}{B\"ottcher \&
  Dermer}{2010}]{bottcher_timing_2010}
B\"ottcher M.,  Dermer C.~D.,  2010, \mn@doi [The Astrophysical Journal]
  {10.1088/0004-637X/711/1/445}, 711, 445

\bibitem[\protect\citeauthoryear{B\"ottcher \& Schlickeiser}{B\"ottcher \&
  Schlickeiser}{1997}]{bottcher_pair_1997}
B\"ottcher M.,  Schlickeiser R.,  1997, Astronomy \& Astrophysics, p.~5

\bibitem[\protect\citeauthoryear{{B\"ottcher}, {Harris}  \&
  {Krawczynski}}{{B\"ottcher} et~al.}{2012}]{Bottcher_Book}
{B\"ottcher} M.,  {Harris} D.~E.,   {Krawczynski} H.,  2012, {Relativistic Jets
  from Active Galactic Nuclei}

\bibitem[\protect\citeauthoryear{B\"ottcher, Reimer, Sweeney  \&
  Prakash}{B\"ottcher et~al.}{2013}]{bottcher_leptonic_2013}
B\"ottcher M.,  Reimer A.,  Sweeney K.,   Prakash A.,  2013, \mn@doi [The
  Astrophysical Journal] {10.1088/0004-637X/768/1/54}, 768, 54

\bibitem[\protect\citeauthoryear{Cerruti, Zech, Boisson  \& Inoue}{Cerruti
  et~al.}{2015}]{cerruti_hadronic_2015}
Cerruti M.,  Zech A.,  Boisson C.,   Inoue S.,  2015, \mn@doi [Monthly Notices
  of the Royal Astronomical Society] {10.1093/mnras/stu2691}, 448, 910

\bibitem[\protect\citeauthoryear{{Cerruti}, {Zech}, {Boisson}, {Emery}, {Inoue}
   \& {Lenain}}{{Cerruti} et~al.}{2018}]{Cerruti_2018}
{Cerruti} M.,  {Zech} A.,  {Boisson} C.,  {Emery} G.,  {Inoue} S.,   {Lenain}
  J.-P.,  2018, preprint, \href
  {http://adsabs.harvard.edu/abs/2018arXiv180704335C} {} (\mn@eprint {arXiv}
  {1807.04335})

\bibitem[\protect\citeauthoryear{Collaboration et~al.,}{Collaboration
  et~al.}{2019}]{planck_collaboration_planck_2019}
Collaboration P.,  et~al., 2019, arXiv:1807.06209 [astro-ph]

\bibitem[\protect\citeauthoryear{Crusius \& Schlickeiser}{Crusius \&
  Schlickeiser}{1986}]{crusius_synchrotron_1986}
Crusius A.,  Schlickeiser R.,  1986, Astronomy \& Astrophysics, 164, L16

\bibitem[\protect\citeauthoryear{Foreman-Mackey, Hogg, Lang  \&
  Goodman}{Foreman-Mackey et~al.}{2013}]{foreman-mackey_emcee:_2013}
Foreman-Mackey D.,  Hogg D.~W.,  Lang D.,   Goodman J.,  2013, \mn@doi
  [Publications of the Astronomical Society of the Pacific] {10.1086/670067},
  125, 306

\bibitem[\protect\citeauthoryear{Gao, Pohl  \& Winter}{Gao
  et~al.}{2017}]{gao_direct_2017}
Gao S.,  Pohl M.,   Winter W.,  2017, \mn@doi [The Astrophysical Journal]
  {10.3847/1538-4357/aa7754}, 843, 109

\bibitem[\protect\citeauthoryear{Gao, Fedynitch, Winter  \& Pohl}{Gao
  et~al.}{2019}]{gao_modelling_2019}
Gao S.,  Fedynitch A.,  Winter W.,   Pohl M.,  2019, \mn@doi [Nature Astronomy]
  {10.1038/s41550-018-0610-1}, 3, 88

\bibitem[\protect\citeauthoryear{Goodman \& Weare}{Goodman \&
  Weare}{2010}]{goodman_ensemble_2010}
Goodman J.,  Weare J.,  2010, \mn@doi [Communications in Applied Mathematics
  and Computational Science] {10.2140/camcos.2010.5.65}, 5, 65

\bibitem[\protect\citeauthoryear{Graff, Georganopoulos, Perlman  \&
  Kazanas}{Graff et~al.}{2008}]{graff_multizone_2008}
Graff P.~B.,  Georganopoulos M.,  Perlman E.~S.,   Kazanas D.,  2008, \mn@doi
  [The Astrophysical Journal] {10.1086/592427}, 689, 68

\bibitem[\protect\citeauthoryear{H\"ummer, R\"uger, Spanier  \&
  Winter}{H\"ummer et~al.}{2010}]{hummer_simplified_2010}
H\"ummer S.,  R\"uger M.,  Spanier F.,   Winter W.,  2010, \mn@doi [The
  Astrophysical Journal] {10.1088/0004-637X/721/1/630}, 721, 630

\bibitem[\protect\citeauthoryear{IceCube et~al.,}{IceCube
  et~al.}{2018}]{eaat1378}
IceCube T.,  et~al., 2018, \mn@doi [Science] {10.1126/science.aat1378}, 361

\bibitem[\protect\citeauthoryear{Jones}{Jones}{1968}]{jones_calculated_1968}
Jones F.~C.,  1968, \mn@doi [Physical Review] {10.1103/PhysRev.167.1159}, 167,
  1159

\bibitem[\protect\citeauthoryear{Keivani et~al.,}{Keivani
  et~al.}{2018}]{keivani_multimessenger_2018}
Keivani A.,  et~al., 2018, \mn@doi [The Astrophysical Journal]
  {10.3847/1538-4357/aad59a}, 864, 84

\bibitem[\protect\citeauthoryear{Kelner \& Aharonian}{Kelner \&
  Aharonian}{2008}]{kelner_energy_2008}
Kelner S.~R.,  Aharonian F.~A.,  2008, \mn@doi [Physical Review D]
  {10.1103/PhysRevD.78.034013}, 78

\bibitem[\protect\citeauthoryear{Kino, Mizuta  \& Yamada}{Kino
  et~al.}{2004}]{kino_hydrodynamic_2004}
Kino M.,  Mizuta A.,   Yamada S.,  2004, \mn@doi [The Astrophysical Journal]
  {10.1086/422305}, 611, 1021

\bibitem[\protect\citeauthoryear{Lipari, Lusignoli  \& Meloni}{Lipari
  et~al.}{2007}]{lipari_flavor_2007}
Lipari P.,  Lusignoli M.,   Meloni D.,  2007, \mn@doi [Physical Review D]
  {10.1103/PhysRevD.75.123005}, 75

\bibitem[\protect\citeauthoryear{Liu, Wang, Xue, Taylor, Wang, Li  \& Yan}{Liu
  et~al.}{2018}]{liu_hadronuclear_2018}
Liu R.-Y.,  Wang K.,  Xue R.,  Taylor A.~M.,  Wang X.-Y.,  Li Z.,   Yan H.,
  2018, arXiv:1807.05113 [astro-ph]

\bibitem[\protect\citeauthoryear{MacDonald, Marscher, Jorstad  \&
  Joshi}{MacDonald et~al.}{2015}]{macdonald_through_2015}
MacDonald N.~R.,  Marscher A.~P.,  Jorstad S.~G.,   Joshi M.,  2015, \mn@doi
  [The Astrophysical Journal] {10.1088/0004-637X/804/2/111}, 804, 111

\bibitem[\protect\citeauthoryear{Mastichiadis \& Kirk}{Mastichiadis \&
  Kirk}{1995}]{mastichiadis_self-consistent_1995}
Mastichiadis A.,  Kirk J.~G.,  1995, Astronomy \& Astrophysics, 295, 613

\bibitem[\protect\citeauthoryear{Mastichiadis, Protheroe  \& Kirk}{Mastichiadis
  et~al.}{2005}]{mastichiadis_spectral_2005}
Mastichiadis A.,  Protheroe R.~J.,   Kirk J.~G.,  2005, \mn@doi [Astronomy \&
  Astrophysics] {10.1051/0004-6361:20042161}, 433, 765

\bibitem[\protect\citeauthoryear{Morris, Potter  \& Cotter}{Morris
  et~al.}{2019}]{morris_feasibility_2019}
Morris P.~J.,  Potter W.~J.,   Cotter G.,  2019, \mn@doi [Monthly Notices of
  the Royal Astronomical Society] {10.1093/mnras/stz920}, 486, 1548

\bibitem[\protect\citeauthoryear{Murase, Oikonomou  \& Petropoulou}{Murase
  et~al.}{2018}]{murase_blazar_2018}
Murase K.,  Oikonomou F.,   Petropoulou M.,  2018, \mn@doi [The Astrophysical
  Journal] {10.3847/1538-4357/aada00}, 865, 124

\bibitem[\protect\citeauthoryear{Pe'er, Long  \& Casella}{Pe'er
  et~al.}{2017}]{peer_dynamical_2017}
Pe'er A.,  Long K.,   Casella P.,  2017, \mn@doi [The Astrophysical Journal]
  {10.3847/1538-4357/aa80df}, 846, 54

\bibitem[\protect\citeauthoryear{Petropoulou \& Mastichiadis}{Petropoulou \&
  Mastichiadis}{2015}]{petropoulou_betheheitler_2015}
Petropoulou M.,  Mastichiadis A.,  2015, \mn@doi [Monthly Notices of the Royal
  Astronomical Society] {10.1093/mnras/stu2364}, 447, 36

\bibitem[\protect\citeauthoryear{Petropoulou, Sironi, Spitkovsky  \&
  Giannios}{Petropoulou et~al.}{2019}]{petropoulou_relativistic_2019}
Petropoulou M.,  Sironi L.,  Spitkovsky A.,   Giannios D.,  2019,
  arXiv:1906.03297 [astro-ph]

\bibitem[\protect\citeauthoryear{Potter}{Potter}{2018}]{potter_modelling_2018}
Potter W.~J.,  2018, \mn@doi [Monthly Notices of the Royal Astronomical
  Society] {10.1093/mnras/stx2371}, 473, 4107

\bibitem[\protect\citeauthoryear{Qin, Wang, Yang, Yuan, Kang  \& Mao}{Qin
  et~al.}{2018}]{qin_using_2018}
Qin L.,  Wang J.,  Yang C.,  Yuan Z.,  Kang S.,   Mao J.,  2018, \mn@doi
  [Publications of the Astronomical Society of Japan] {10.1093/pasj/psx150}, 70

\bibitem[\protect\citeauthoryear{Spada, Ghisellini, Lazzati  \& Celotti}{Spada
  et~al.}{2001}]{spada_internal_2001}
Spada M.,  Ghisellini G.,  Lazzati D.,   Celotti A.,  2001, \mn@doi [Monthly
  Notices of the Royal Astronomical Society]
  {10.1046/j.1365-8711.2001.04557.x}, 325, 1559

\bibitem[\protect\citeauthoryear{Svensson}{Svensson}{1982}]{svensson_pair_1982}
Svensson R.,  1982, \mn@doi [The Astrophysical Journal] {10.1086/160081}, 258,
  321

\bibitem[\protect\citeauthoryear{Xue, Liu, Petropoulou, Oikonomou, Wang, Wang
  \& Wang}{Xue et~al.}{2019}]{xue_two-zone_2019}
Xue R.,  Liu R.-Y.,  Petropoulou M.,  Oikonomou F.,  Wang Z.-R.,  Wang K.,
  Wang X.-Y.,  2019, arXiv:1908.10190 [astro-ph]

\makeatother
\end{thebibliography}

\appendix

\section{Notes on reproducing the work of Cerruti et al.}{\label{sec:reproducing_cerruti}}

As a way of checking our model, we try to reproduce the results of
\citet{Cerruti_2018}. Their study has both hadronic and lepto-hadronic
models and their parameters and the goodness of the fit are available
in the extra materials. We choose to reproduce the parameters for
a lepto hadronic model that give the best $\chi^{2}$ value.

Cerruti et al.'s model is based upon a stationary particle distribution
with the shape of a broken power-law with an exponential drop which
produces the primary photon emission. Although for this work, they
set $\gamma_{e,break}=\gamma_{e,max}$, obtaining a simple power law.
In addition, they calculate the lepton population that would result
from photon-photon pair production and bethe-heitler cascades and
subsequently add their own emission to the primary emission. For the
neutrino emission, they calculate the resulting emission from pion
and muon decays.

This hybrid approach to the calculation of the lepton population poses
a problem for our model as we can either have a single steady state
population or a fully dynamic one. Moreover, as we will see when we
derive the corresponding parameters for our model, these can easily
cause that one of our assumptions, that of the lepton injection due
to pair production and bethe-heitler processes being small, to be
broken. With this in mind, we choose to employ the spherical steady
state version of our model, knowing that it will not be able to fully
reproduce the shape of the observed SED.

As said, their model defines the steady populations for electrons
and protons as:

\begin{equation}
N_{e,p}\left(\gamma\right)=K_{e,p}'\gamma^{-\alpha_{e,p}}\exp\left(-\frac{\gamma}{\gamma_{e,p,max}}\right).
\end{equation}

Their population parameter $K_{e,p}'$ is given in term of the normalizations
for the electron distribution at $\gamma=1$ and the ratio with protons
$\left(r\right)$. This means that:
\begin{align}
N_{e}\left(1\right)=K_{e} & =K_{e}'\exp\left(-\frac{1}{\gamma_{e,max}}\right),\\
N_{p}\left(1\right)=K_{p} & =rK_{e}.
\end{align}
So:
\begin{align}
K_{e}' & =\phantom{r}K_{e}\exp\left(\frac{1}{\gamma_{e,max}}\right),\\
K_{p}' & =rK_{e}\exp\left(\frac{1}{\gamma_{p,max}}\right).
\end{align}

Our model defines the population parameters differently but we can
transform between the two as:
\begin{align}
\frac{\rho}{m_{e}+\eta m_{p}}A & =\phantom{r}K_{e}\exp\left(\frac{1}{\gamma_{e,max}}\right),\\
\frac{\eta\rho}{m_{e}+\eta m_{p}}C & =rK_{e}\exp\left(\frac{1}{\gamma_{p,max}}\right).
\end{align}
So:
\begin{align}
\eta & =\frac{A}{C}r\frac{\exp\left(\frac{1}{\gamma_{p,max}}\right)}{\exp\left(\frac{1}{\gamma_{e,max}}\right)},\\
\rho & =\frac{K_{e}}{A}\exp\left(\frac{1}{\gamma_{e,max}}\right)\left(m_{e}+\eta m_{p}\right),
\end{align}
where $A$ and $C$ are normalization constants for the particles
distributions.

\Tabref{cerruti_our_parameters} shows the results of the translation
of the parameters between Cerruti et al.'s model and ours. There are
three parameters that will have a very marked effect on the resulting
SED: $\gamma_{e,max}$, $\gamma_{p,max}$ and $\eta$. The first $\left(\gamma_{e,max}\right)$
effectively sets the maximum energy of synchrotron and Inverse Compton
photons, and as it is two orders of magnitude lower than ours, we
can predict that we will not be able to fit neither the high energy
end of the synchrotron peak, nor that of the Inverse Compton peak.
The second $\left(\gamma_{p,max}\right)$ contributes to increasing
the amount of protons available to interact with photons in both,
pion production and bethe-heitler pair production. This will cause
the amount of injected leptons and neutrinos to increase, and their
energies will be bigger. The third $\left(\eta\right)$ has a similar
effect to $\gamma_{p,max}$, but increases the number of protons at
all energies. This means that not only the pion and lepton production
will be higher, but also the synchrotron emission.

\begin{table*}
\begin{centering}
\begin{tabular}{cccc}
\toprule 
Parameter & Description & Cerruti & Ours\tabularnewline
\midrule
\midrule 
$B$ & Magnetic Field $\left(G\right)$ & $0.376356$ & $0.376356$\tabularnewline
\midrule 
$R$ & Blob Radius $\left(cm\right)$ & $4.2638\times10^{15}$ & $4.2638\times10^{15}$\tabularnewline
\midrule 
$\delta$ & Doppler Factor & $40$ & $40$\tabularnewline
\midrule 
$\alpha_{e}$ & $e^{-}$ spectral index & $2$ & $2$\tabularnewline
\midrule 
$\gamma_{e,min}$ & Minimal $e^{-}$ Lorentz factor & $500$ & $500$\tabularnewline
\midrule 
$\gamma_{e,max}$ & Maximal $e^{-}$ Lorentz factor & $10926.6$ & $10926.6$\tabularnewline
\midrule 
$\alpha_{p}$ & $p^{+}$ spectral index & $2$ & $2$\tabularnewline
\midrule 
$\gamma_{p,min}$ & Minimal $p^{+}$ Lorentz factor & $1$ & $1$\tabularnewline
\midrule 
$\gamma_{p,max}$ & Maximal $p^{+}$ Lorentz factor & $1.02547\times10^{8}$ & $1.02547\times10^{8}$\tabularnewline
\midrule 
$K_{e}$ & Normalization constant & $60659.7$ & -\tabularnewline
\midrule 
$K_{p}/K_{e}$ & Ratio of the normalization & $0.381661$ & -\tabularnewline
\midrule 
$\rho$ & Density & - & $3.872355422482072\times10^{-20}$\tabularnewline
\midrule 
$\eta$ & $p^{+}$ to $e^{-}$ ratio & - & $200$\tabularnewline
\bottomrule
\end{tabular}
\par\end{centering}
\caption{\label{tab:cerruti_our_parameters}List of parameters for reproducing
the SED of TXS 0506+056 by Cerruti et al. and the adaptation to our
model.}
\end{table*}

\begin{figure*}
\begin{centering}
\includegraphics[width=1\textwidth]{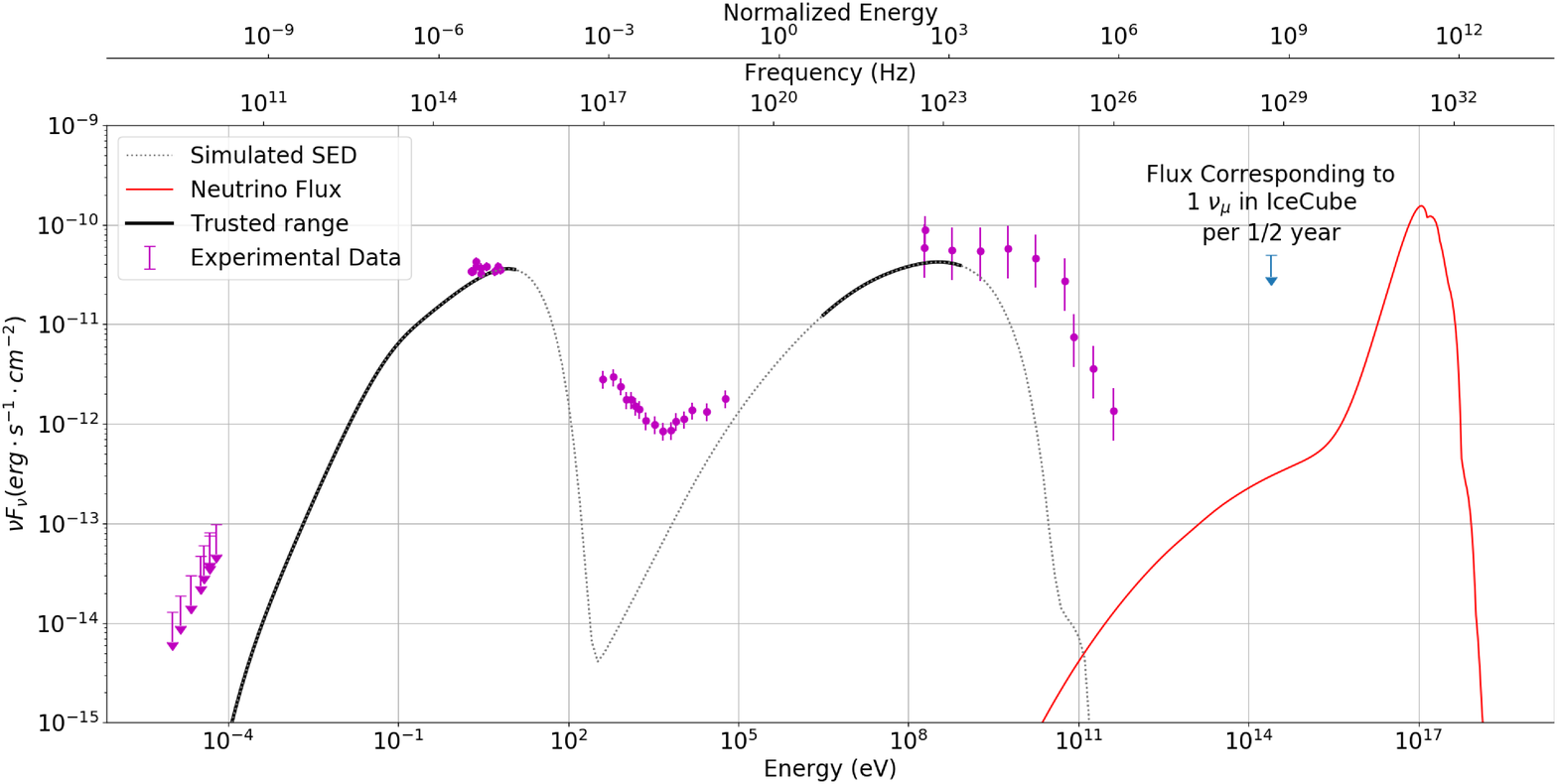}
\par\end{centering}
\caption{\label{fig:Cerruti_reproduction}SED and neutrino flux obtained with
the parameters from \citet{Cerruti_2018}}
\end{figure*}

\Figref{Cerruti_reproduction} shows the results obtained by our model
using Cerruti et al.'s parameters. We have chosen to separate the
simulated SED in the total range and a trusted range because, as we
have explained, we can not be sure of the emission that would come
from electrons with energies above that of injection. Still, for the
trusted range, we obtain an acceptable fit with the observed flux
levels. Although the neutrino flux is still lower than the estimated
by Ice Cube at the relevant energies.

\section{Notes on reproducing the work of Gao et al. }{\label{sec:reproducing_gao}}

A second check that we can do for our model is to try to reproduce
the SED of TXS 0506+056 employing the parameters of the lepto-hadronic
model of Gao et al.'s hybrid model. It must be noted that in their
model, they separate the luminosities of protons and electrons, whereas
in our model we combine them into one luminosity and regulate the
particle populations by means of the ratio of protons to electrons
parameter.

In their model:
\begin{equation}
L_{e,p}=m_{e,p}c^{2}\times\int\gamma K_{e,p}\gamma^{-\alpha_{e,p}}d\gamma.
\end{equation}
By separating $K_{e,p}\gamma^{-\alpha_{e,p}}$into a normalized distribution
and a constant, as we did in \subsecref{Dynamic-Models} we easily
find:
\begin{equation}
Q_{e,p,0}=\frac{L_{e,p}}{\left\langle \gamma_{e,p}\right\rangle m_{e,p}c^{2}}.
\end{equation}
The ratio of injected protons to electrons is then:
\begin{equation}
\eta=\frac{Q_{p,0}}{Q_{e,0}}=\frac{L_{p}}{\left\langle \gamma_{p}\right\rangle }\frac{\left\langle \gamma_{e}\right\rangle }{L_{e}}\frac{m_{e}}{m_{p}}.
\end{equation}
Using the parameters of Gao et al. \citep{gao_modelling_2019} we
obtain $\eta=8606$.

Another difference between our model and theirs is that they take
that the charged particles do not escape in the free escape timescale,
as they assume that protons, electrons and positrons escape with a
velocity of $c/300$. We have chosen to simulate this by setting $\sigma=300$
which has the same effect. Another difference lies in the definition
of the free escape timescales, where we have included a factor $3/4$
due to geometrical reasons (\ref{eq:t_esc_sphere}):
\begin{equation}
t_{esc,our}=\frac{3}{4}t_{esc,gao}.
\end{equation}

In \tabref{gao_our_parameters} we list the original parameters and
our choices.

\begin{table*}
\begin{centering}
\begin{tabular}{cccc}
\toprule 
Parameter & Description & Gao & Ours\tabularnewline
\midrule
\midrule 
$B$ & Magnetic Field $\left(G\right)$ & $0.14$ & $0.14$\tabularnewline
\midrule 
$R$ & Blob Radius $\left(cm\right)$ & $10^{16}$ & $10^{16}$\tabularnewline
\midrule 
$\delta$ & Doppler Factor & $28.0$ & $28.0$\tabularnewline
\midrule 
$L_{e}$ & $e^{-}$ injection luminosity $\left(erg/s\right)$ & $10^{40.9}$ & -\tabularnewline
\midrule 
$\alpha_{e}$ & $e^{-}$ spectral index & $-3.5$ & $-3.5$\tabularnewline
\midrule 
$\gamma_{e,min}$ & Minimal $e^{-}$ Lorentz factor & $10^{4.2}$ & $10^{4.2}$\tabularnewline
\midrule 
$\gamma_{e,max}$ & Maximal $e^{-}$ Lorentz factor & $10^{5.1}$ & $10^{5.1}$\tabularnewline
\midrule 
$L_{p}$ & $p^{+}$ injection luminosity $\left(erg/s\right)$ & $10^{45.7}$ & -\tabularnewline
\midrule 
$\alpha_{p}$ & $p^{+}$ spectral index & $-2.0$ & $-2.0$\tabularnewline
\midrule 
$\gamma_{p,min}$ & Minimal $p^{+}$ Lorentz factor & $10.0$ & $10.0$\tabularnewline
\midrule 
$\gamma_{p,max}$ & Maximal $p^{+}$ Lorentz factor & $10^{5.4}$ & $10^{5.4}$\tabularnewline
\midrule 
$L$ & Injection luminosity $\left(erg/s\right)$ & - & $10^{45.7}$\tabularnewline
\midrule 
$\eta$ & $p^{+}$ to $e^{-}$ ratio & - & $8606$\tabularnewline
\midrule 
$\eta_{esc}$ & escape velocity of $e^{\pm}$ and $p^{+}$ & $c/300$ & $c/300$\tabularnewline
\bottomrule
\end{tabular}
\par\end{centering}
\caption{\label{tab:gao_our_parameters}List of parameters for reproducing
the SED of TXS 0506+056 by Gao et al and the adaptation to our model.
It should be noted that both injections are taken as being a simple
power law and the volume a sphere.}
\end{table*}

\begin{figure*}
\begin{centering}
\includegraphics[width=1\textwidth]{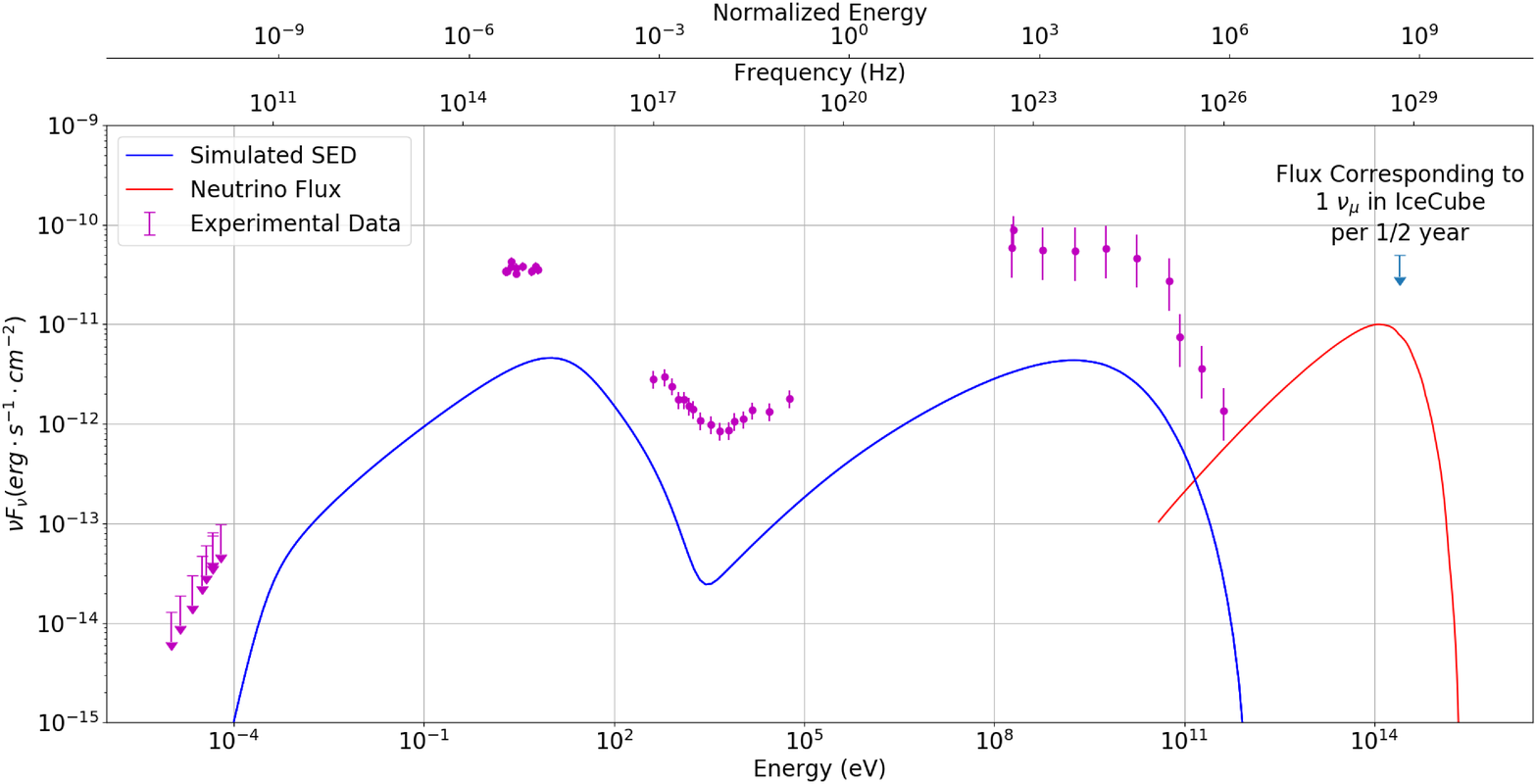}
\par\end{centering}
\caption{\label{fig:Gao_reproduction}SED and neutrino flux obtained with the
parameters from \citet{gao_modelling_2019}}
\end{figure*}

\ref{fig:Gao_reproduction} shows the resulting SED and neutrino flux
that we obtain applying our model with the parameters of \citep{gao_modelling_2019}.
It is clear that, although the shape is correct, we are not able to
correctly reproduce the flux level, being our prediction lower by
$\sim1$ order of magnitude. It should be noted that their model is,
in effect, a two zone model, consisting of two concentric spheres,
and that we are modelling the emission of the inner sphere for the
purpose of our comparison. Our analysis of the injection terms for
electrons and positrons from the Bethe-Heitler process suggests that
the photon field is not sufficient to produce the amount of high-energy
electrons needed to give rise to the required gamma-ray flux. The
main hypothesis that we have is that the external sphere that the
model of Gao et al include injects enough low energy photons to the
inner sphere to trigger a pair cascade, and that subsequently, these
high-energy pairs are responsible for the upscattering of photons
up to gamma-ray energies.

\section{Corner Plots}

\begin{figure*}
\begin{centering}
\includegraphics[width=0.9\paperwidth]{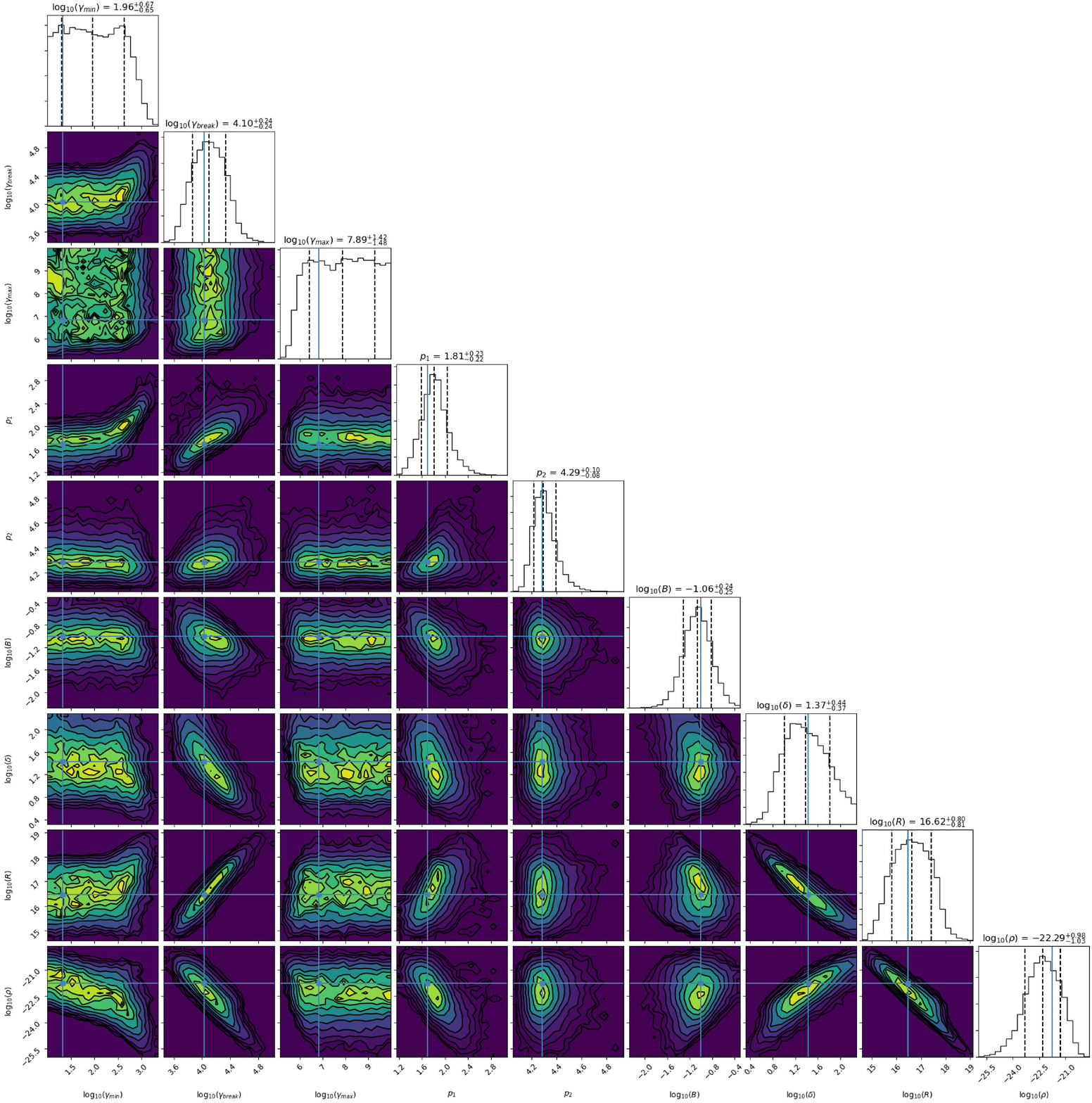}
\par\end{centering}
\caption{\label{fig:Spherical_Steady_State_Corner}Corner plot of the fit to
the Sphere Steady State Model}
\end{figure*}

\begin{figure*}
\begin{centering}
\includegraphics[width=0.9\paperwidth]{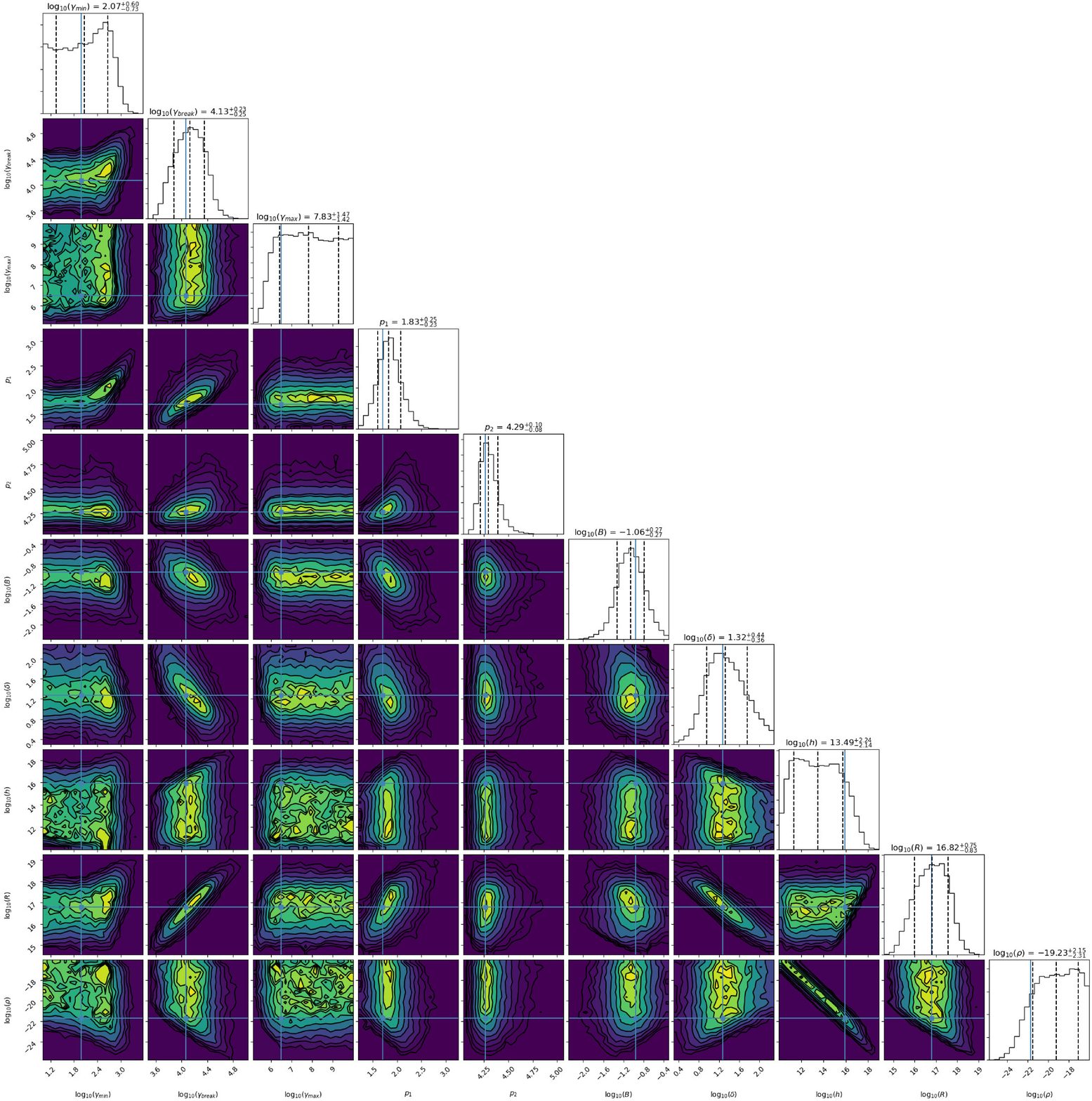}
\par\end{centering}
\caption{\label{fig:Disk_Steady_State_Corner}Corner plot of the fit to the
Disk Steady State Model}
\end{figure*}

\begin{figure*}
\begin{centering}
\includegraphics[width=0.9\paperwidth]{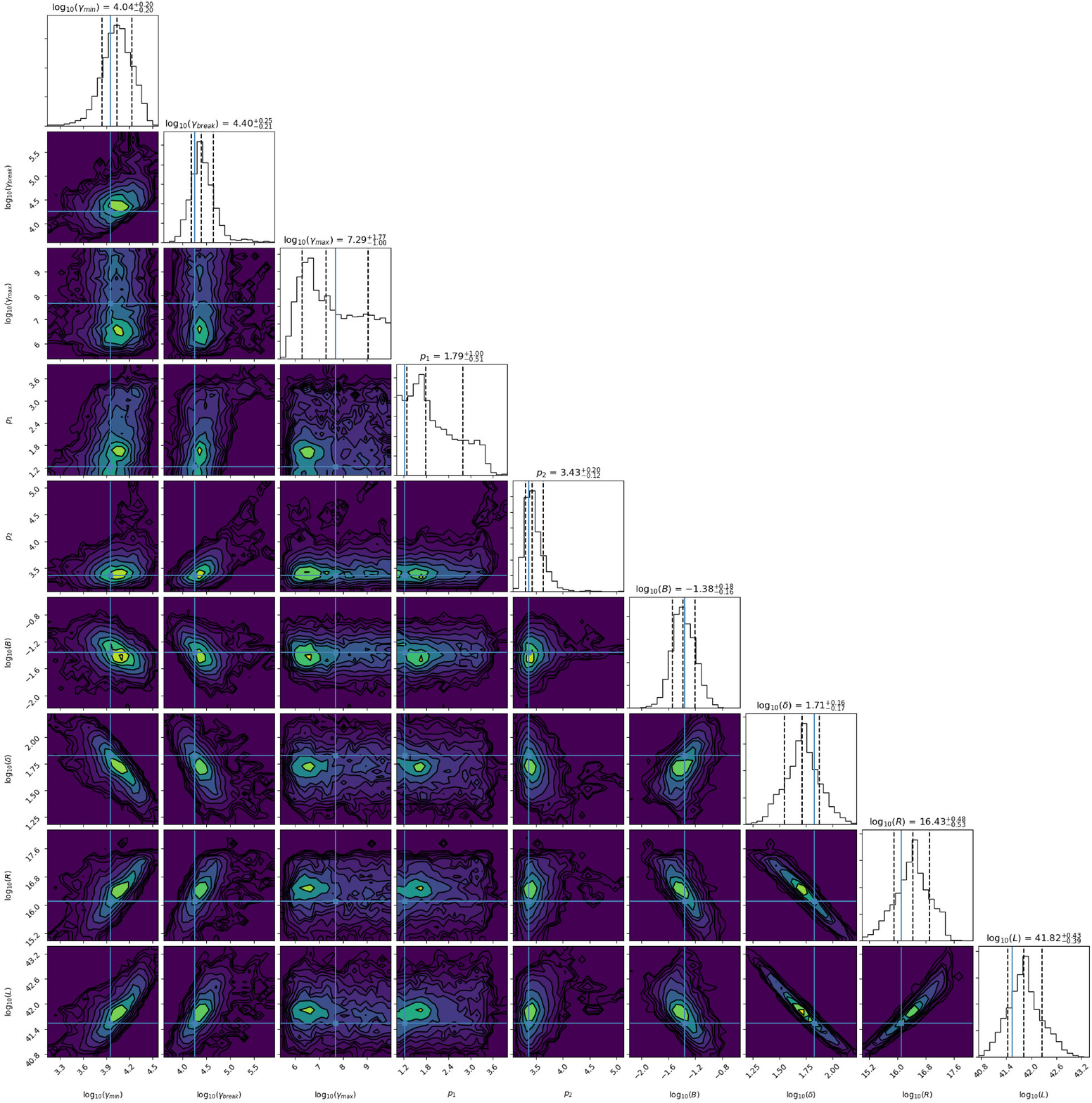}
\par\end{centering}
\caption{\label{fig:Spherical_Injection_Corner}Corner plot of the fit to the
Sphere Injection Model}
\end{figure*}

\begin{figure*}
\begin{centering}
\includegraphics[width=0.9\paperwidth]{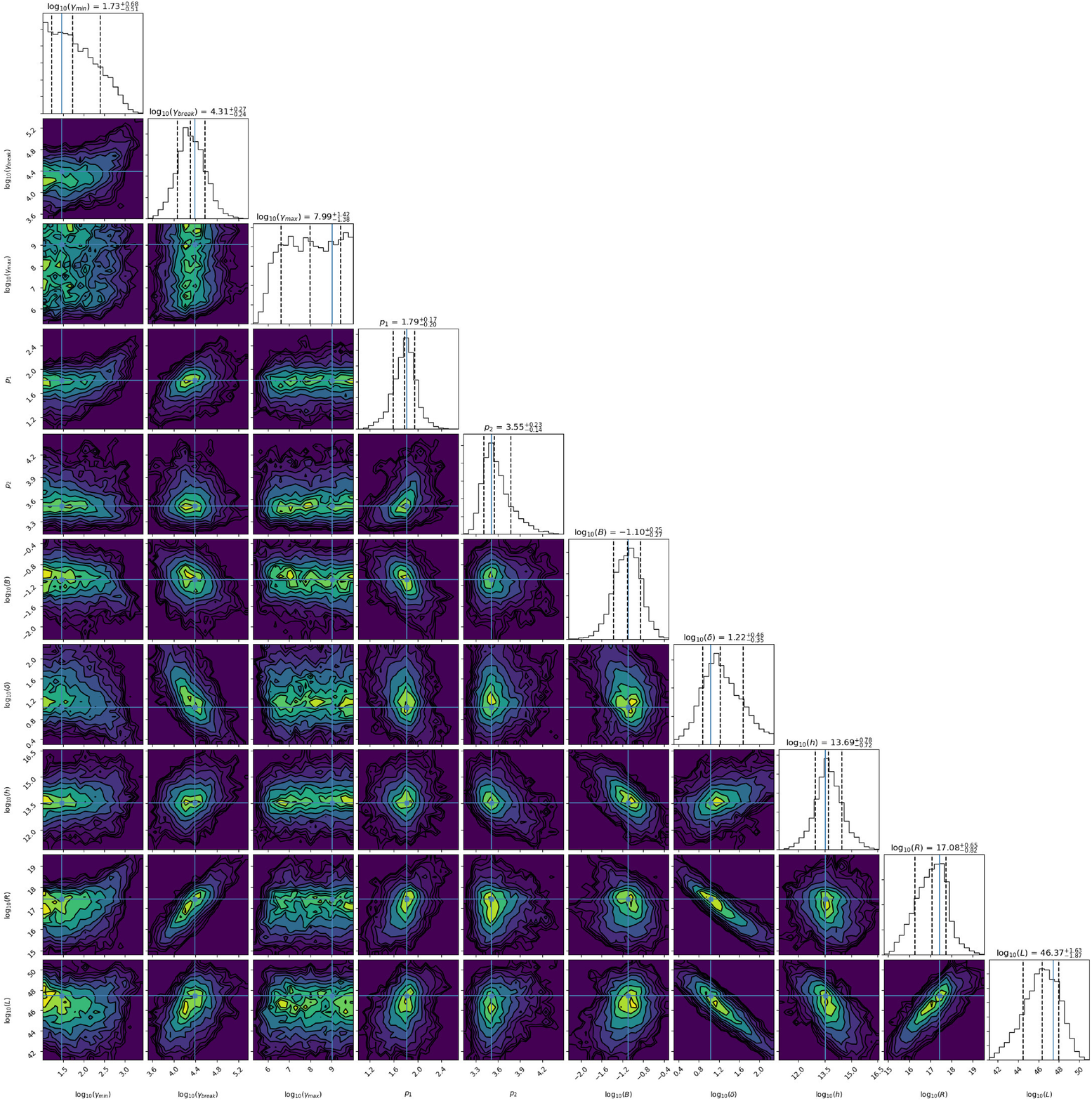}
\par\end{centering}
\caption{\label{fig:Disk_Injection_Corner}Corner plot of the fit to the Disk
Injection Model}
\end{figure*}

\begin{figure*}
\begin{centering}
\includegraphics[width=0.9\paperwidth]{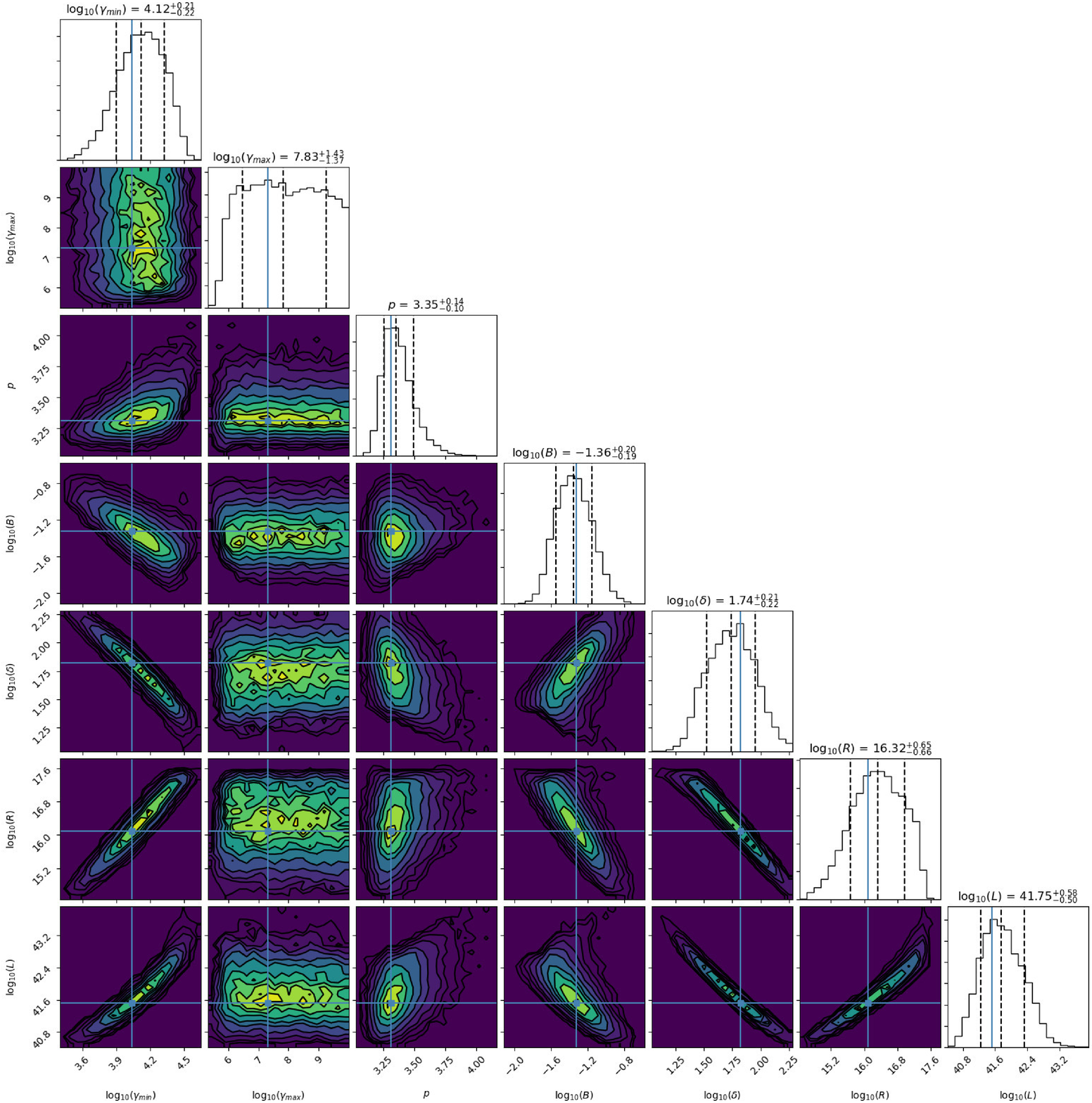}
\par\end{centering}
\caption{\label{fig:Spherical_Injection_PL_Corner}Corner plot of the fit to
the Sphere Injection with a Power Law Model}
\end{figure*}

\begin{figure*}
\begin{centering}
\includegraphics[width=0.9\paperwidth]{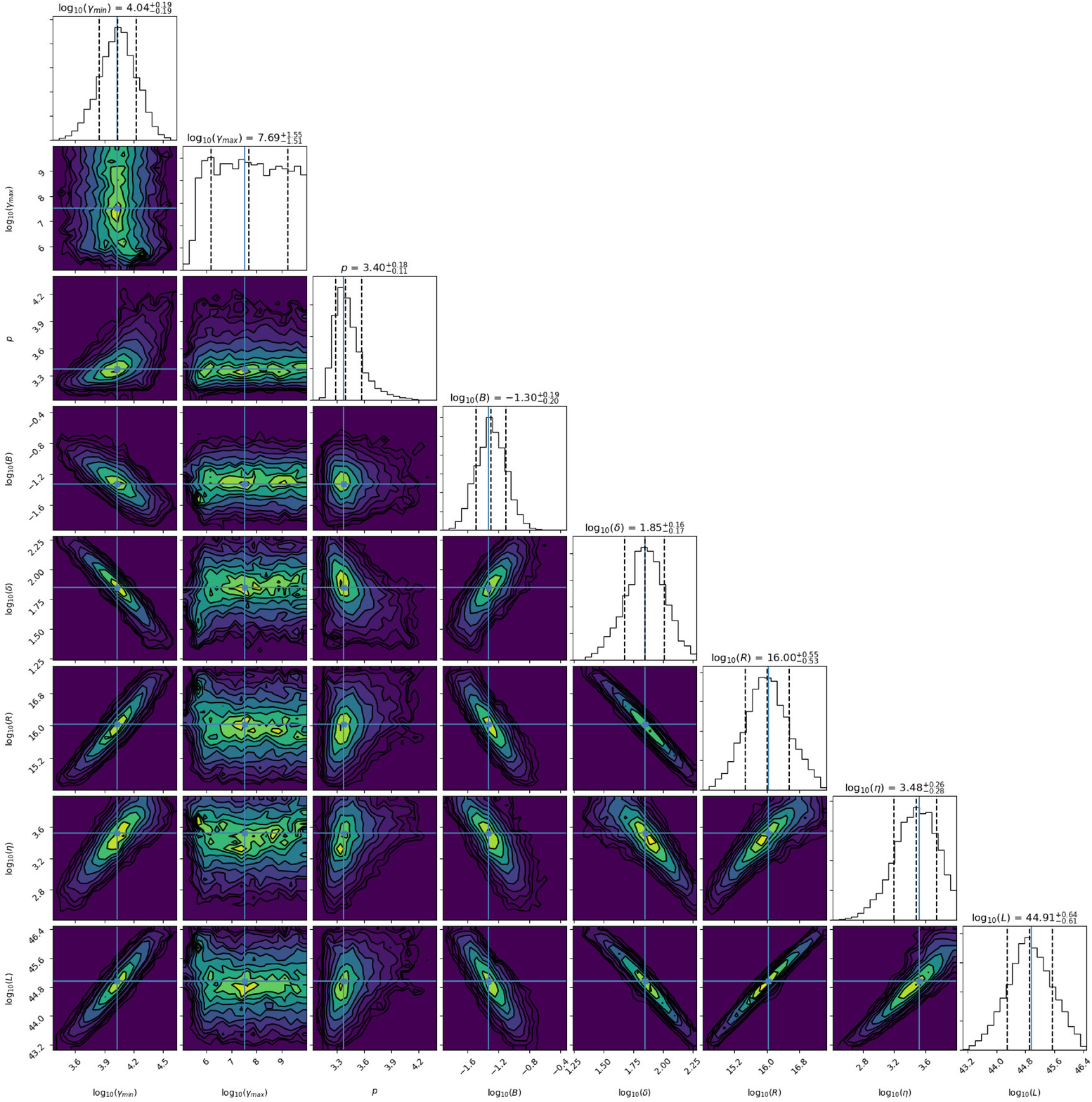}
\par\end{centering}
\caption{\label{fig:Spherical_Injection_PL_eta_Corner}Corner plot of the fit
to the Sphere Injection with a Power Law Model allowing $\eta$ to
vary.}
\end{figure*}

\bsp 

\label{lastpage} 
\end{document}